\documentclass[12pt]{article}
\usepackage[pdftex]{graphicx}
\usepackage{amsmath}
\usepackage{amssymb}
\usepackage{amsfonts}
\usepackage{latexsym}
\usepackage{caption}
\def\bea{\begin{eqnarray}}
\def\eea{\end{eqnarray}}
\def\beq{\begin{equation}}
\def\eeq{\end{equation}}
\def\ba{\begin{eqnarray}}
\def\ea{\end{eqnarray}}
\def\be{\begin{equation}}
\def\ee{\end{equation}}
\newcommand{\muf}{\mu_{\rm\sss F}}
\newcommand{\mur}{\mu_{\rm\sss R}}
\renewcommand{\S}{\mathcal{S}}
\newcommand{\Lum}{\mathcal{L}}
\newcommand{\Lrap}{L_{q\bar q}}

\newcommand{\abs}[1]{\left|\,#1\,\right|}
\newcommand{\plus}[1]{\left[ #1 \right]_+}
\newcommand{\plust}[1]{\left( #1 \right)_+}

\newcommand{\gammae}{\gamma_{\scriptscriptstyle E}}
\newcommand{\Ord}{\mathcal{O}}

\newcommand{\gsim}{\gtrsim}

\newcommand{\sss}{\scriptscriptstyle\rm}
\newcommand{\as}{\alpha_{\sss S}}
\def\toinf#1{\mathrel{\mathop{\sim}\limits_{\scriptscriptstyle{#1\rightarrow\infty }}}}
\def\({\left(}
\def\){\right)}
\def\[{\left[}
\def\]{\right]}
\def\ab{\bar\alpha}
\def    \hepph  #1 {{\tt hep-ph/#1}}
\def    \hepex  #1 {{\tt hep-ex/#1}}

\numberwithin{equation}{section}
\usepackage[pdftex]{hyperref}

\begin{document}

\pagestyle{empty}

\begin{flushright}

IFUM-964-FT\\ GeF/TH/3-10\\
\end{flushright}

\begin{center}
\vspace*{0.5cm}
{\Large \bf Soft gluon resummation\\ of Drell-Yan rapidity
  distributions:\\

\medskip

theory and phenomenology}\\
\vspace*{1.cm}
Marco Bonvini,$^{a}$ Stefano~Forte$^{b}$ and Giovanni~Ridolfi$^{a}$
\\
\vspace{1.cm}  {\it
{}$^a$Dipartimento di Fisica, Universit\`a di Genova and
INFN, Sezione di Genova,\\
Via Dodecaneso 33, I-16146 Genova, Italy\\ 
\medskip
{}$^b$Dipartimento di Fisica, Universit\`a di Milano and
INFN, Sezione di Milano,\\
Via Celoria 16, I-20133 Milano, Italy}\\
\vspace*{1.5cm}

{\bf Abstract}
\end{center}
\bigskip

We examine critically the theoretical underpinnings and
phenomenological implications of soft gluon (threshold) resummation
of rapidity distributions at a hadron collider, taking
Drell-Yan production at the Tevatron and the LHC as a reference test case.
First, we show that in perturbative QCD soft gluon resummation is
necessary whenever the partonic (rather the hadronic) center-of-mass 
energy is close enough to threshold, and we provide tools to
assess when resummation is relevant for a given process. Then, we
compare different prescriptions for handling the divergent nature of the series
of resummed perturbative corrections, specifically the minimal and
Borel prescriptions. We assess the intrinsic ambiguities of resummed
results,
both due to the asymptotic nature of their
perturbative expansion, and to the treatment of subleading terms.
Turning to phenomenology, we introduce a fast and
accurate method for the implementation of resummation with the minimal
and Borel prescriptions using an expansion on a basis of Chebyshev
polynomials. We then present results for $W$ and $Z$ production as well
as both high- and low-mass dilepton pairs at the LHC, and show that
soft gluon resummation effects are generally comparable in size to
NNLO corrections, but sometimes affected by substantial ambiguities.

\noindent

\vspace*{1cm}

\vfill
\noindent

\begin{flushleft} September 2010 \end{flushleft}
\eject

\setcounter{page}{1} \pagestyle{plain}

\eject

\section{Introduction}
\label{sec:intro}
Drell-Yan rapidity distributions  are likely to be the standard candle
which is both theoretically calculable and experimentally 
measurable with highest accuracy at hadron colliders, in particular
the LHC. The current QCD theoretical accuracy for this process is
NNLO~\cite{admp},  and small effects such as those related to the
coupling of the gauge boson to final-state leptons have been studied
recently~\cite{Catani:2010en}. Whereas the resummation of contributions related
to the emission of soft gluons are routinely included in the
computation of Drell-Yan transverse-momentum
distributions~\cite{Grazzini:2009nd}, where they are mandatory in order to
stabilize the behaviour of the cross-section at low $p_T$, their
impact on rapidity distributions has received so far only a moderate
amount of attention. This is partly due to the fact that even in
fixed-target Drell-Yan production experiments, such as the Tevatron
E866~\cite{e866}, let alone LHC experiments, the available center-of-mass
energy is much larger than the mass of typical final states, thereby
suggesting that threshold resummation is not relevant. 

However, it has been pointed out since long~\cite{Appell:1988ie} that
because hadronic cross-sections are found convoluting a hard
cross-section with a parton luminosity, the effect of resummation may be
relevant even relatively far from the hadronic threshold.  Indeed, in
Ref.~\cite{bolz} threshold resummation has been claimed to affect
significantly Drell-Yan production for E866 kinematics, though
somewhat different results have been found in
Ref.~\cite{Becher:2007ty}.  It is important to observe that Drell-Yan
data from E866 and related experiments play a crucial role in the
precision determination of parton distributions~\cite{Ball:2010de}, so
their accurate treatment is crucial for precise LHC
phenomenology. Furthermore, threshold resummation is known~\cite{Grazzini:2010zc} to
affect in a non-negligible way standard Higgs production in
gluon-gluon fusion at the LHC, even though the process is clearly
very far from threshold.

A systematic assessment of the relevance of soft-gluon resummation
for rapidity distributions at hadron colliders is called for: it is
the purpose of the present paper. We will take the Drell-Yan process
as a test case, but our formalism and results can be applied to other
collider processes such as heavy quark production or Higgs production.

The main ingredients of this assessment are the following. First, we
determine when and why threshold resummation is relevant. Then, we
discuss how the resummed contribution is defined, and specifically we
deal with the divergence of the series of resummed terms. Next, we
address the issue of combining resummed and fixed-order
results. Finally, we turn to phenomenology and assess the size and
theoretical uncertainty of resummation at the Tevatron and LHC. Each
of these steps turns out to be nontrivial, as we now briefly sketch.

The standard physical argument to explain why resummation may be
relevant even when the hadronic process is relatively far from
threshold goes as follows~\cite{catani}. The quantity which is
resummed in perturbative QCD is the hard partonic cross-section, which
depends on the partonic center-of-mass energy and the dimensionless
ratio of the latter to the final state invariant mass. Therefore,
resummation is relevant when it is the partonic subprocess that is
close to threshold. The partonic center-of-mass energy in turn can
take any value from threshold up to the hadronic center-of-mass
energy, and its mean value is determined by the shape of the PDFs:
therefore, one expects threshold resummation to be more important if
the average partonic center-of-mass energy is small, i.e.~if the
relevant PDFs are peaked at small $x$ (such as gluons or sea quarks,
as opposed to valence quarks). This for instance explains why
threshold resummation is especially relevant for Higgs production in
gluon-gluon fusion.

We will show that this can be made quantitative using a saddle-point
argument in Mellin space: for any given value of the hadronic
dimensionless variable $\tau$ the dominant contribution to the cross
section comes from a narrow range of the variable $N$ which is
conjugate to $\tau$ upon Mellin transform. In Mellin space the
cross-section is the product of parton distributions (PDFs) and a
hard coefficient, but it turns out that the position of the saddle is
mostly determined by the PDFs. Moreover, the result is quite
insensitive to the non-perturbative (low-scale) shape of the parton
distribution and mostly determined by its scale dependence,
specifically by the low--$x$ (or low--$N$) behaviour of the relevant
Altarelli-Parisi splitting functions: the faster the small--$x$ growth
of the splitting function, the smaller the average partonic
center-of-mass energy, the farther from the hadronic threshold the
resummation is relevant. This is reassuring, because it means that the
region of applicability of threshold resummation is controlled by
perturbative physics. 
The issue of the persistence of sizable
soft gluon  emission terms even far from threshold was also addressed,
using methods of soft-collinear effective theory,
in Ref.~\cite{Becher:2007ty}, where in the large $\tau\gsim 0.2$ region 
it was related to the (non-perturbative)
shape of parton distributions, though it was also observed for smaller 
$\tau$ values.~\footnote{Shortly after this paper appeared in preprint form, 
the treatment of this issue in soft-collinear effective
theory was revisited in a quantitative way in
Ref.~\cite{Bauer:2010jv}, where it was related to a parameter
determined by the shape of parton distributions.}

Having established the region in which threshold resummation is
relevant, we have to face the fact that resummation sums an infinite
series of contributions to the expansion of the hard partonic
cross-section in powers of the perturbative strong coupling $\as(Q^2)$
(with $Q^2$ the hard scale of the process) which diverges if the
resummation is performed at any given logarithmic order.  This
divergence can be treated in various ways: here we will consider the
minimal prescription (MP)~\cite{cmnt} and the Borel prescription
(BP)~\cite{frru,afr}, in both of which the resummed result is a
function of $\as(Q^2)$ to which the perturbative expansion in
powers of $\as(Q^2)$ is asymptotic. Both the BP and MP can be
obtained by adding to the perturbative series a contribution which
removes the divergence, which is power-suppressed in $Q^2$ for the
BP, and it has support in an unphysical kinematic region (below
threshold) for the MP.

In practice, however, different resummation prescriptions differ not
only because of the way the high-order divergence of the expansion is
handled, but also because if they are applied to a low-order
truncation of the divergent series they differ by subleading terms. We
will show that in fact this difference turns out to be by far
phenomenologically the most significant, unless one is very close to
the hadronic threshold, which is in practice a very rare
occurrence. We will thus assess the ambiguity that subleading terms
induce on resummed results, and to which extent they can be optimized
in order to ensure stability of the resummed expansion once resummed
results are matched to the standard fixed-order ones.

With one (or more), possibly optimal, resummation prescriptions at
hand, we turn to the resummation of rapidity distributions.
It turns out that this can be performed  by relating the
resummed expressions to those of the inclusive cross-section. This
entails some further ambiguities in the treatment of subleading terms,
though 
we shall see that these are in practice phenomenologically very
small. 
Finally, we will implement the resummation up to the
next-to-next-to-leading log (NNLL) level  combined with the
next-to-next-to-leading fixed-order (NNLO) result. We will show that
this can be done efficiently by projecting resummed results on a basis
of Chebyshev polynomials: with this approach, resummed results can be
easily obtained using any external set of parton distributions.
This will enable us to obtain predictions at the Tevatron and LHC. We
will see that even at the LHC the impact of the resummation is not
negligible, and comparable to the size of the corresponding fixed-order
corrections, especially in the central rapidity region.

The structure of this paper is the following: in Section~2 we will
present in detail the arguments which allow one to determine the
relevant partonic center-of-mass energy for given hadronic
kinematics and parton distribution, and thus to
assess the relevance of threshold resummation. In Section~3 we will
discuss and compare the minimal and Borel prescription for
resummation, the way they can be matched to fixed-order expressions,
  the way subleading terms are treated with the
various prescriptions, and the associate ambiguities. In Section~4 we
will present the general formalism for the construction of resummed
expressions for rapidity distributions from their inclusive
counterparts, and the numerical implementation of resummed results. 
In Section~5 we will turn to phenomenology for the production of
neutral and charged Drell-Yan pairs, with different values of their
invariant mass at Tevatron and LHC energies.
We will compare theoretical predictions both to E866 and recent CDF data.

\section{When is threshold resummation relevant?}
\label{sec:saddlepoint}

As we have mentioned in the introduction, both simple physical
arguments and evidence from explicit
calculations~\cite{catani} suggest that threshold resummation may be
relevant even quite far from the hadronic threshold, provided the
partonic average center-of-mass energy is sizably smaller than the
available hadronic center-of-mass energy. Here we derive this
conclusion from a quantitative argument, which will allow us to assess
the relevance of threshold resummation for a given process. We
assess the impact of parton distributions by means of a Mellin-space
argument. For a given process and a given value of $\tau=Q^2/s$, we
determine the region of the variable $N$, Mellin-conjugate to $\tau$,
which provides the dominant contribution to the cross-section. First,
we show that the saddle point is mostly determined by the small--$x$
behaviour of the PDFs, which in turn is driven by perturbative
evolution. Next, we determine the impact of the inclusion of PDFs for
the Drell-Yan process. Finally, with specific reference to the
Drell-Yan process, we assess the $N$ region where threshold
resummation is relevant.

\subsection{The impact of PDFs: $N$--space vs.~$\tau$--space}
\label{sec:ipnvtau}

The cross-section for a hadronic process with scale $Q^2$ and
center-of-mass energy $s=Q^2/\tau$ can be written as a sum of
contributions of the form
\be
\label{fact}
\sigma(\tau,Q^2)=\int_\tau^1\frac{dz}{z} \;
{\cal L}(z) \;\hat\sigma\left(\frac{\tau}{z},\as(Q^2)\right) 
\ee
in terms of a partonic cross-section $\hat \sigma$ and a parton
luminosity, in turn determined in terms of parton distributions $f_i(x_i)$
as
\be
\label{lumi}
{\cal L}(z)=\int_z^1\frac{dx_1}{x_1} f_1(x_1)
f_2\left(\frac{z}{x_1}\right).
\ee
Here we denote generically by $\sigma$
a suitable quantity (in general, process-dependent) which has 
the property of factorizing as in Eq.~\eqref{fact}.
Such quantities are usually related in a simple way to
cross-sections or distributions; for example, in the case of the
invariant mass distribution of Drell-Yan pairs, $\sigma$ is in
fact $\frac{1}{\tau}\frac{d\sigma}{dQ^2}$.

In general, the cross-section gets a contribution
like Eq.~\eqref{fact} from all parton channels which contribute to the given
process at the given order, but this is inessential for our argument,
so we concentrate on one such contribution.

In Eq.~\eqref{fact},
the partonic cross-section, which is computed in perturbation theory,
is evaluated as a function of the partonic center-of-mass energy
\be
\label{shatdef}
\hat s= \frac{Q^2}{\tau/z}=x_1 x_2 s,
\ee
where $x_1$ and $x_2\equiv z/x_1$ are the momentum fractions of the
two partons.
Therefore, the threshold region, where resummation is relevant, is the 
region in which $\hat s$ is not much larger than $Q^2$. However, all
values of $x_1,x_2$ between $\tau$ and $1$ are accessible, so whether or not
resummation is relevant depends on which region gives the dominant
contribution to the convolution integrals
Eqs.~(\ref{fact},~\ref{lumi}). This dominant region can be determined
using a Mellin-space argument.

The Mellin transform of $\sigma(\tau,Q^2)$ is 
\be
\sigma(N,Q^2)=\int_0^1d\tau\,\tau^{N-1}\, \sigma(\tau,Q^2),
\label{mt}
\ee
with inverse
\be
\sigma(\tau,Q^2)=\frac{1}{2\pi i}\int_{\bar N-i\infty}
^{\bar N+i\infty}dN\,\tau^{-N}\,\sigma(N,Q^2)
=\frac{1}{2\pi i}\int_{\bar N-i\infty}^{\bar N+i\infty}dN\,e^{E(\tau,N;Q^2)},
\label{imt}
\ee
where $\bar N$ is larger than the real part of the rightmost singularity
of $\sigma(N,Q^2)$ (by slight abuse of notation we denote with
$\sigma$ both the function and its transform), and in the last step we
have defined
\be
\label{melexp}
E(\tau,N;Q^2)\equiv N\ln\frac{1}{\tau}+\ln \sigma(N,Q^2).
\ee
The function $\sigma(N,Q^2)$ has a singularity on the real positive axis
because of the parton luminosity; to the right of this singularity,
it is a decreasing function of $N$, because the area below the curve
$\tau^{N-1} \sigma(\tau,Q^2)$ obviously decreases as $N$ increases.
As a consequence, $E(\tau,N;Q^2)$ always
has a minimum on the real positive $N$ axis at some $N=N_0$.
Hence, the
inversion integral is dominated by the region of $N$ around $N_0$, and
can be approximated by saddle-point, 
expanding $E(\tau,N;Q^2)$ around $N_0$.

When $\tau\to 1$, the slope of the straight line
$N\ln\frac{1}{\tau}$ decreases, and the position $N_0$ of the minimum is
pushed to larger values, so the large--$\tau$  behaviour of $\sigma(\tau,Q^2)$
is determined by the large--$N$ behaviour of
$\sigma(N,Q^2)$, as it easy to show using the saddle point
approximation: in fact, for typical parton distributions
in this  limit the saddle
point approximation becomes exact. This is shown 
in Appendix~\ref{app:saddle}, where some  properties of the
saddle point approximation to Mellin transforms are collected. 

\begin{figure}[htb]
\begin{center}
\includegraphics[width=0.75\textwidth]{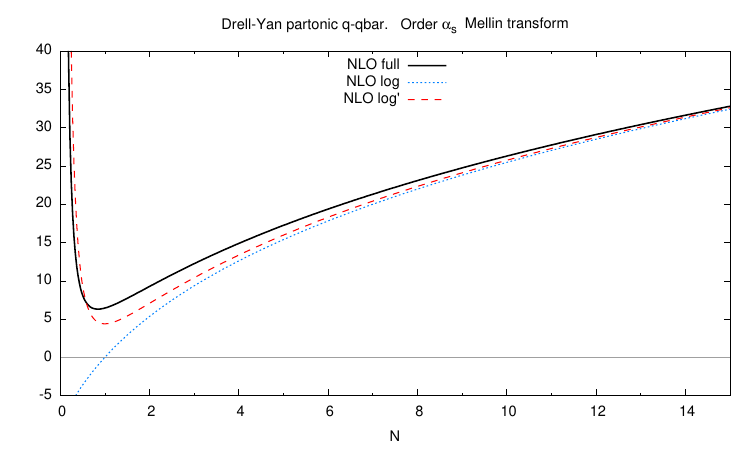}
\caption{The $\Ord(\as)$ neutral current Drell-Yan partonic coefficient
$C_1(N)$, Eq.~\eqref{eq:c1}, plotted as a function of $N$ (solid curve).  
The logarithmic approximations
$C_1^{\rm log}(N)$ Eq.~(\ref{eq:Mpsi}) and 
$C_1^{\rm log'}(N)$ Eq.~(\ref{eq:Mpsi1}) are also shown (dotted and dashed
curves, respectively).}
\label{fig:sd4}
\end{center}
\end{figure}

The position of the saddle point $N_0$ is strongly influenced by the 
rate of decrease
of the cross-section $\sigma(N,Q^2)$ as  $N$ grows.  Indeed, in Mellin
space, the cross-section Eq.~(\ref{fact}) factorizes:
\beq
\label{factN}
\sigma(N,Q^2)={\cal L}(N,Q^2)\;\hat\sigma(N,\as(Q^2)).
\eeq
It is then easy to see that the decrease of $\sigma(N,Q^2)$ with $N$
is driven by the parton luminosity ${\cal L}(N,Q^2)$: in fact, it
turns out that, for large $N$, $\hat\sigma(N,\as(Q^2))$ is an {\it
increasing} function of $N$. Indeed, it turns out (see
Appendix~\ref{app:saddle}) that a distribution, unlike an ordinary
function, need not be a decreasing function of $N$. However,
the parton
luminosity always offsets this increase if the convolution integral
exists, because the cross-section $\sigma(\tau,Q^2)$ is an
ordinary  function. For example, the $\Ord(\as)$ Drell-Yan partonic
cross-section is shown in Fig.~\ref{fig:sd4}, where it is clear that
even though at small $N$ the cross-section decreases with $N$, at
large $N$ it increases.

As a consequence, when $\tau$ is large, the position of the
saddle point $N_0$ is completely controlled by the drop of the parton
luminosity: indeed, in the absence of parton luminosity, the saddle
point would be very close to the minimum of the partonic
cross-section, which is around $N\simeq 1$.  When $\tau$ is smaller,
even without PDF the location of the saddle is controlled by the
partonic cross-section, which in this region is a decreasing function
of $N$. However, this
decrease is much stronger in the presence of a luminosity, so the
location of the saddle is substantially larger. 
Hence, in the large $\tau$ region the effect of the resummation
is made much stronger by the luminosity, while for medium-small $\tau$  if
the luminosity decreases fast enough, $N_0$ may be quite large even if
$\tau\ll 1$, i.e.~far from the hadronic threshold, thereby extending
the region in which resummation is relevant.

The position of the saddle point  in the various regions 
can be simply estimated on
the basis of general considerations.
At the leading log level, parton densities can be written as linear
combinations of terms of the form
\be
f_i(N,Q^2)=\exp\left[\frac{\gamma_i(N)}{\beta_0}
\ln\frac{\as(Q_0^2)}{\as(Q^2)}\right]f^{(0)}_i(N),
\label{pdfform}
\ee
where $\gamma_i(N)$ are eigenvalues of the leading-order anomalous
dimension matrix:
\be
\frac{\partial\ln f_i(N,Q^2)}{\partial\ln Q^2}=\as(Q^2)\gamma_i(N)
+\Ord(\as^2);\qquad
\frac{\partial\ln\as(Q^2)}{\partial\ln Q^2}=-\beta_0\as(Q^2)+\Ord(\as^2),
\ee
and  $f^{(0)}_i(N)=f_i(N,Q_0^2)$ are initial
conditions given at some reference scale $Q_0^2$. The cross-section is
correspondingly decomposed into a sum of contributions, each of which
has the form of Eq.~(\ref{imt}), with 
\beq
E(\tau,N;Q^2)= N\ln\frac{1}{\tau}+\frac{\gamma_i(N)+\gamma_j(N)}{\beta_0}
\ln\frac{\as(Q_0^2)}{\as(Q^2)}
+\ln f^{(0)}_i(N)+\ln f^{(0)}_j(N) +\ln\hat\sigma(N,\as(Q^2)).
\label{appexp}
\eeq
At large $N$, this expression is dominated by the first term, 
which grows linearly with $N$, while at small $N$ the behaviour
of $E(\tau,N;Q^2)$ is determined by the singularities of the anomalous
dimensions, which are stronger than those of the initial conditions if
$Q^2>Q_0^2$, given that low-scale physics is both expected
theoretically from Regge theory~\cite{collins,Abarbanel:1969eh} 
and known phenomenologically from PDF
fits~\cite{Ball:2008by} to produce at most poles but not essential singularities such as
those obtained exponentiating the anomalous dimensions. Indeed, assuming
a power behaviour for $f^{(0)}_i(z)$ both at small and large $z$,
\be
f^{(0)}_i(z)=z^{\alpha_i}(1-z)^{\beta_i},
\label{inpdfform}
\ee
so that
\beq
f^{(0)}_i(N)=\frac{\Gamma(N+\alpha_i)\Gamma(\beta_i+1)}
{\Gamma(N+\alpha_i+\beta_i+1)},
\label{mellinpdfform}
\eeq
$\ln f^{(0)}_i(N)$ behaves as $\ln N$ both at large and small $N$, and is
thus subdominant in comparison to either the $\tau$ dependent term or the
anomalous dimension contribution  in Eq.~\eqref{appexp}.
A similar argument holds for the partonic cross-section term
$\ln\hat\sigma(N)$.

The position of the minimum is therefore  determined by the transition
from the leading small--$N$ drop due to the anomalous dimension
term and the leading large--$N$ rise due to the $\tau$--dependent
term, up to a correction due to the other contributions to
Eq.~(\ref{appexp}). When $\tau$ is large, the rise in the first term
is slow, and it only sets in for rather large $N$ so the correction
due to the other contributions may be substantial. This is the
region in which resummation is surely relevant because the hadronic
$\tau$ is large.  But when $\tau$ is not so large, the rise sets in
more rapidly, in the region 
where the second term is dominant and the correction from
the  initial PDFs and the partonic cross-section is negligible. 

In order to provide an estimate of the value of $N_0$ in either
region, we note that
the leading-order Altarelli-Parisi anomalous dimensions 
at small $N$ behave as
\beq
\gamma_{i}\sim\frac{1}{N-N_p}+\hbox{less singular},
\label{polapprox}
\eeq
where ``less singular'' denotes terms whose singularity has a smaller
real part, and
$N_p=0$ for $\gamma_{qq},\gamma_{qg}$, and
$N_p=1$ for $\gamma_{gq},\gamma_{gg}$. This implies that one of the two singlet 
anomalous dimension eigenvectors has a leading singularity at $N=1$,
while the other has a singularity at $N=0$ and it is thus suppressed
at small $N$, like the nonsinglet anomalous dimension which is
thus also suppressed. This pattern persists to all perturbative
orders. It follows that singlet quark and gluon distributions have a
steeper small--$N$ and thus small--$z$ behaviour.

Expanding the
anomalous dimension about its rightmost singularity
at leading order we have 
\be
\gamma_+(N)=\frac{\gamma_+^{(0)}}{N-1}\left[1+\Ord(N-1)\right];\qquad
\gamma_{\rm ns}(N)=\frac{\gamma_{\rm ns}^{(0)} }{N}\left[1+\Ord(N)\right],
\label{leadandims}
\ee
where $\gamma_+$ and $\gamma_{\rm ns}$ are respectively 
the dominant small--$N$ singlet eigenvalue and nonsinglet anomalous
dimension, and
\be
\gamma_+^{(0)}=\frac{N_c}{\pi};\qquad 
\gamma_{\rm ns}^{(0)}=\frac{C_F}{2\pi}.
\label{loadvals}
\ee
We note that $\gamma_+(2)=0$ (because of momentum conservation), while
$\gamma_{\rm ns}(1)=0$ (because of baryon number conservation): hence,
we expect the small $N$ approximation
Eqs.~(\ref{leadandims},~\ref{loadvals}) 
to break down around $N\approx 2-k_i$, with 
\be
\label{kvals}
k_{\rm +}=0;\quad k_{\rm ns}=1.
\ee

We can then consider three cases, according to whether $\gamma_i$,
$\gamma_j$ in Eq.~(\ref{appexp}) are both singlet, both nonsinglet, or
one singlet and one nonsinglet. The three cases correspond respectively to the
leading behaviour of, for instance, Higgs production in gluon fusion,
Drell-Yan production at the Tevatron, and Drell-Yan production at the
LHC.
Substituting Eq.~(\ref{leadandims})  in Eq.~(\ref{appexp}) and
neglecting the last three terms the saddle is found to be at 
\beq
N^0_{ij}=1-k_{i}k_{j}+\sqrt{
\frac{\gamma^{(0)}_{ij}}{\beta_0\ln\frac{1}{\tau}}\ln\frac{\as(Q_0^2)}{\as(Q^2)}
}
\label{n0app}
\eeq
where $i,\>j$ take the values ${\rm ns}$ and $+$, $k_i$ are given by
Eq.~(\ref{kvals}), and
\be
\gamma^{(0)}_{\rm ns \> ns}=2 \gamma^{(0)}_{\rm ns};
\qquad \gamma^{(0)}_{+ +}=2 \gamma^{(0)}_+;
\qquad \gamma^{(0)}_{+ \> {\rm ns}}= \gamma^{(0)}_+.
\label{gamvals}
\ee
We expect Eq.~(\ref{n0app}) to provide a good approximation to the
position of the saddle point in the region $1- k_ik_j<N\ll 2-k_ik_j$, while
for larger values of $N$ the last three terms on the right-hand side
of Eq.~(\ref{appexp}) provide a correction of increasing size. This
correction cannot be estimated in a universal way, in that it will
generally depend on the shape of the initial parton distributions
$f_i^{(0)}$. However, note that at large $N$ both the nonsinglet and
singlet anomalous dimension drop: they are negative, and their modulus
grows as $\ln N$. Furthermore, $\frac{\Gamma(N)}{\Gamma(N+\eta)}=
N^{-\eta}(1+\Ord(1/N))$, which, together with
Eqs.~(\ref{inpdfform},~\ref{mellinpdfform}) implies that also the
contributions from $f_i^{(0)}$ to Eq.~(\ref{appexp}) drop
logarithmically as $N\to\infty$.  This drop is a generic feature of
the initial PDFs: it is necessary in order for the convolution
integral which gives the physical cross-section to exist (see
Appendix~\ref{app:saddle}); likewise, the drop of the anomalous
dimension (at most as a power of $\ln N$) is a generic feature of
perturbative evolution, related to momentum conservation of gluon
radiation, and it persists to all orders.

It follows that at large $N$ the $1/N$ drop Eq.~(\ref{leadandims}) is
replaced by a slower logarithmic drop, with a coefficient determined
by both the large--$N$ behaviour of the anomalous dimension and by the
initial PDFs. This leads to a rather larger value of the saddle point
$N_0$ than given by Eq.~(\ref{n0app}).
\begin{figure}[htb]
\begin{center}
\includegraphics[width=0.8\textwidth]{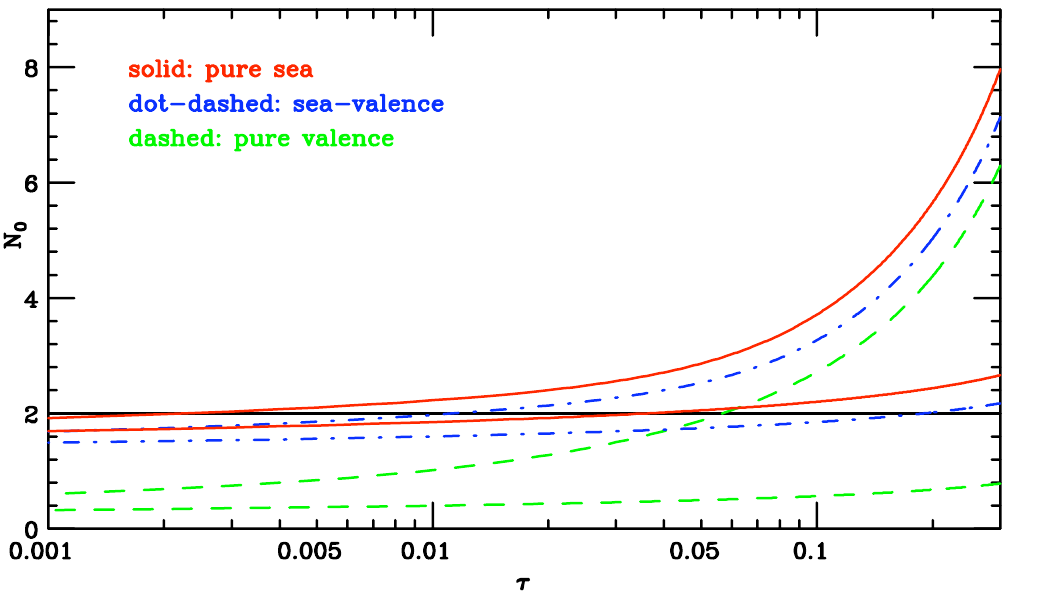}
\caption{The position of the saddle point $N_0$ for the Mellin inversion
integral Eq.~(\ref{imt}) with the exponent Eq.~(\ref{appexp})
as a function of $\tau$ determined neglecting $\hat \sigma$ and using
toy initial PDFs Eqs.~(\ref{inpdfform},~\ref{pdfexps}) and
leading-order anomalous dimensions. Upper curves: exact value;
lower curves: the approximation Eq.~(\ref{n0app}). The curve shown is obtained
with $Q_0=1$~GeV, $Q=100$~GeV.}
\label{fig:n0app}
\end{center}
\end{figure}
The value of $N^0_{ij}$ as a function of $\tau$ given by Eq.~\eqref{n0app}
is plotted in Fig.~\ref{fig:n0app} for the three cases of
Eq.~\eqref{gamvals}: in each case, we show both the result obtained
using the approximation
Eq.~\eqref{n0app}, and that found by exact numerical minimization
of Eq.~\eqref{appexp}, with the full leading-order expression of the
pertinent anomalous dimensions $\gamma_i$, $\gamma_j$, and initial
PDFs of the form Eq.(\ref{inpdfform}), with 
\be
\alpha_{\rm ns}=\frac{1}{2};\quad\beta_{\rm ns}=3;
\qquad\alpha_+=0;\quad\beta_+=4,
\label{pdfexps}
\ee
but still neglecting the $\ln \hat\sigma(N,\as(Q^2))$.
We see that indeed the pole approximation Eq.~(\ref{n0app}) holds
reasonably well for $1- k_ik_j<N\ll 2-k_ik_j$.

We can get a feeling for the implication of this for the $\tau$ region
in which resummation is relevant by 
very roughly taking
the
value $N=2$ as that
of the transition from the small--$N$ region to the large--$N$ 
region. Indeed, this value corresponds to the energy-momentum operator
and it thus sets the transition between the large $N$ region in which
parton distributions drop (while growing in modulus) and the small $N$
region in which they grow as the scale is raised.
This rough estimate is borne out by a 
more quantitative assessment of the transition
point which will be performed in Sect.~\ref{sec:resDY}. We then see
that, if at least one of
the parton distributions is flavour singlet, the position of the
saddle point remains in the large--$N$ region down to fairly low values
of $\tau\sim0.01$. This extension of the region where resummation is
relevant to small $\tau$ is due to the rise of the anomalous dimension
related to the pole at $N=1$ Eq.~(\ref{polapprox}) in the singlet
sector.

Larger values of $\tau\sim0.1$ correspond to values of $N_0$ which are
anyway in the large--$N$ region. However, in this region the rapid drop
of the parton distribution, due both to its initial shape and to its
evolution, greatly increases the impact of resummation by raising the
position of the saddle, which at $\tau\sim0.2$ is already $N_0\sim 6$.

\subsection{The impact of PDFs: the Drell-Yan process}
\label{sec:ipdfdy}

We now assess the impact of parton distributions in the specific case
of Drell-Yan production. In this case, the quantity $\sigma(\tau,Q^2)$
which appears in Eq.~(\ref{fact}) is given by
\beq
\sigma(\tau,Q^2)=\frac{1}{\tau} \frac{d\sigma_{\rm DY}}{d Q^2}(\tau,Q^2),
\label{dyfact}
\eeq
where $\frac{d\sigma_{\rm DY}}{d Q^2}(\tau,Q^2)$ is
the invariant mass distribution of Drell-Yan pairs.
The corresponding partonic quantity $\hat\sigma(z,\as(Q^2))$ of
Eq.~(\ref{fact}) is then
\beq
\hat\sigma(z,\as(Q^2))=\frac{1}{z}\frac{d\hat\sigma_{DY}}{dQ^2}(z,\as(Q^2)).
\label{eq:dyxesc}
\eeq
It is further convenient to define a dimensionless coefficient
function $C(z,\as(Q^2))$ through
\beq
\hat\sigma(z,\as(Q^2))=\hat\sigma_0\, C(z,\as(Q^2)).
\label{cfdef}
\eeq
The coefficient function admits the perturbative expansion
\beq
C(z,\as)=
\left[\delta(1-z)+\frac{\as}{\pi}\,C_1(z)
+\(\frac{\as}{\pi}\)^2 C_2(z)+\dots\right];
\label{eq:exp}
\eeq
the Born term $\hat\sigma_0$ contains the electroweak coupling and
it is given e.g.~in Ref.~\cite{EHW}. 
Threshold logarithms only appear
in the quark--antiquark channel, while in other partonic channels they
are suppressed by powers of $1-z$.
In the quark--antiquark channel the
next-to-leading order contribution  is given by~\cite{DYNLO}
\beq
C_1(z)=C_F\left\{4\plus{\frac{\ln(1-z)}{1-z}}
-\frac{4}{1-z}\ln \sqrt{z}-2(1+z)\ln\frac{1-z}{\sqrt{z}}
+\(\frac{\pi^2}{3} -4\)\delta(1-z)
\right\}.
\label{eq:c1x}
\eeq
Note that of course  
at leading order $\hat \sigma(N)$
is just a constant. The  Mellin transform of the NLO term is
\begin{multline}
C_1(N)=C_F\Bigg\{ \frac{2\pi^2}{3} -4 +2\gammae^2+
2\psi_0^2(N)-\psi_1(N) + \psi_1(N+2)+ 4 \gammae \psi_0(N)  
\\
+ \frac{2}{N} \[ \gammae + \psi_0(N+1)\] + \frac{2}{N+1} 
\[ \gammae +\psi_0(N+2)\]\Bigg\},
\label{eq:c1}
\end{multline}
and was  shown in Fig.~\ref{fig:sd4}.

\begin{figure}[tb]
\begin{center}
\includegraphics[width=0.85\textwidth]{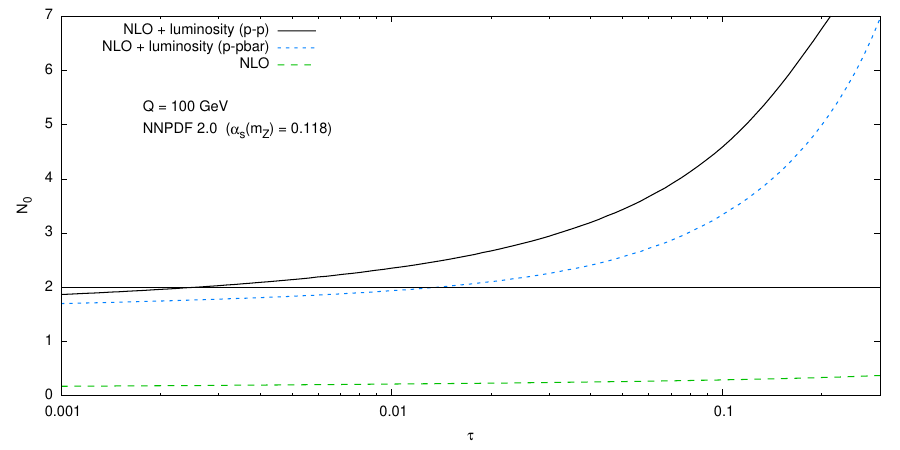}
\caption{
The position of the saddle point $N_0$ for the Mellin
  inversion
integral Eq.~(\ref{imt}) as a function of $\tau$ 
with the cross-section
Eq.~(\ref{factN}) determined  using the $\Ord(\as)$ Drell-Yan
cross-section Eq.~(\ref{eq:c1}) for neutral dileptons
and NNPDF2.0~\cite{Ball:2010de} parton
distributions, with  $Q=100$~GeV. The two upper curves refer, from the
top, to $pp$ and $p\bar p$ collisions. The lowest (dashed)
curve is the position of the saddle at the parton level, i.e.~omitting
the parton luminosity in Eq.~(\ref{factN}).}
\label{fig:sd1}
\end{center}
\end{figure}
We have then determined the position of the saddle point $N_0$ in a
realistic situation, i.e., using the
partonic cross-section Eq.~(\ref{eq:c1}) for 
Drell-Yan production of a neutral lepton pair of invariant mass
$Q=100$~GeV  at a $pp$ or $p\bar p$ collider, with a parton
luminosity determined  using
NNPDF2.0~\cite{Ball:2010de} parton distributions.  
Comparing the realistic curves of Fig.~\ref{fig:sd1} to those in
Fig.~\ref{fig:n0app}, which were determined using the toy initial PDFs 
Eqs.~(\ref{inpdfform},~\ref{pdfexps}) and neglecting the contribution
from the hard cross-section we see that the $pp$ curve of
Fig.~\ref{fig:sd1} agrees well with the sea--valence curve of
Fig.~\ref{fig:n0app}, as one would expect given that in $pp$
collisions one must always pick up at least a sea (antiquark) PDF. 
The case of $p\bar p$ is slightly more subtle: in this case, for
$\tau\gtrsim0.1$ the curve in  Fig.~\ref{fig:sd1} agrees well with the 
valence--valence curve of
Fig.~\ref{fig:n0app}. However, for smaller values of $\tau$ the
position of $N_0$ computed using the full luminosity decreases much
more slowly: this is due to the fact that as $N\lesssim 2$ the 
contribution $\gamma_+$ rapidly grows due to the pole
Eq.~(\ref{leadandims}) so that even the valence distribution is
dominated by it. As a consequence, even in the $p\bar p$ the
resummation region is further extended to somewhat lower $\tau$ values
than it would be the case for a pure valence--valence luminosity.

In Fig.~\ref{fig:sd1} we also show the position of the saddle that
is obtained if the parton luminosity is omitted, i.e.~for the
parton-level cross-section. It is clear that, as we argued in 
Sect.~\ref{sec:ipdfdy},
the position of the saddle is  determined by the PDFs, and it is much
larger than that found at the parton level, thereby supporting the
conclusion that the convolution with the parton luminosity greatly
enhances the importance of resummation, and it extends it to a much
wider kinematic region.

In summary, we conclude that the resummation region $N\gtrsim 2$ 
corresponds to $\tau \gtrsim 0.003$ for $pp$ collisions, and $\tau\gtrsim
0.02$  for $p\bar p$ collisions. For $\tau\lesssim0.1$ the position of
the saddle is determined by the pole in the anomalous dimension,
while for larger values of $\tau$ the large $x$ drop of PDFs, due
both to their initial shape and to perturbative evolution, very
substantially enhances the impact of resummation.
These values have been obtained for $\sqrt{Q^2}=100$~GeV.
We have checked that they depend very weakly on $Q^2$, which is
expected, because $Q^2$ enters the determination of the saddle point
only through the scale dependence of the parton densities.

\subsection{The resummation region for the Drell-Yan process}
\label{sec:resDY}

Having established that the PDFs drive the position of the saddle
point, and having explicitly determined this value, we now have to
determine quantitatively the value of $N$ at which
resummation becomes important. So far, we have taken conventionally
the momentum-conservation point $N= 2$ as the value at which  
logarithmically enhanced contributions give a sizable contribution to
the cross-section. We would now like to establish this in a
quantitative way.
 
To this purpose, we compare $C_1(N)$ Eq.~(\ref{eq:c1}) to its
logarithmic approximation.
The logarithmically-enhanced term in $C_1(z)$ Eq.~(\ref{eq:c1}) is
\beq
C_1^{\rm log}(z)=4C_F\plus{\frac{\ln(1-z)}{1-z}},
\label{logenhanced}
\eeq
whose Mellin transform is
\beq
C_1^{\rm log}(N)=
C_F\left[2\psi_0^2(N)-2\psi_1(N) + 
4 \gammae \psi_0(N)+\frac{\pi^2}{3}+2\gammae^2\right].
\label{eq:Mpsi}
\eeq
The function $C_1(N)$ was displayed as a function of $N$ along the real
positive axis in Fig.~\ref{fig:sd4}. The logarithmic approximation
Eq.~(\ref{eq:Mpsi}) is also shown in the same figure. It is clear that
the full result and the log approximation
agree at large $N$, up to a small constant shift. Indeed,
\beq
\lim_{N\to\infty}\left[C_1(N)-C_1^{\rm log}(N)\right]=
C_F\(\frac{\pi^2}{3} -4\).
\eeq
For $N$ slightly above 2 the logarithmic
contribution is already about 50\% of the full result, but it rapidly
deviates from it as $N$ decreases. This suggests that indeed the
logarithmic contribution is sizable for $N\gtrsim 2$.

It is interesting to observe that there is a certain arbitrariness in
the definition of the logarithmically enhanced contribution: indeed, 
the Mellin transform Eq.~(\ref{eq:Mpsi}) of the logarithmic
contribution Eq.~(\ref{logenhanced}) contains terms which are not
logarithmically enhanced in $N$--space: the Mellin transform of the
next$^k$-leading-log $x$ expression only coincides with the
next$^k$-leading-log $N$ expression up to subleading terms. For example, 
$\psi_1\toinf{N} \frac{1}{N}$, so that we could equally well identify
the logarithmically enhanced contribution to Eq.~(\ref{eq:c1}) with
\beq
C_1^{\rm log'}(N)=
C_F\[2\psi_0^2(N) + 4 \gammae \psi_0(N) +\frac{\pi^2}{3}+ 2\gammae^2\],
\label{eq:Mpsi1}
\eeq
which is the Mellin transform of
\beq
C_1^{\rm log'}(z)=4C_F
\left\{\plus{\frac{\ln(1-z)}{1-z}}
-\frac{\ln\sqrt{z}}{1-z}\right\}.
\label{logp}
\eeq
In other words, logarithmically enhanced contributions in $N$--space
also contain subleading terms when transformed to $z$--space, and
conversely (see Appendix~\ref{app:mellin} for details on the relevant
Mellin transforms).

The curve corresponding to $C_1^{\rm log'}(N)$ is also shown in
Fig.~\ref{fig:sd4}. It is clear that in the large $N\gtrsim 2$
region it
differs from $C_1^{\rm log}(N)$ by a negligible amount. However, at
small $N$ this particular logarithmic approximation turns out to be
much closer to the full result. Whereas of course the choice of
subleading terms to be included in the resummation is essentially
arbitrary, in practice some choices may be less natural than others: this
will be  discussed in Sect.~\ref{sec:subl} below.

\begin{figure}[htb]
\begin{center}
\includegraphics[width=0.75\textwidth]{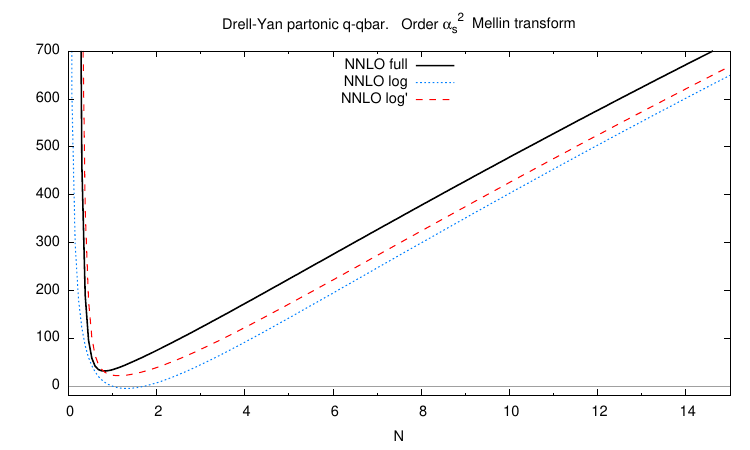}
\caption{Next-to-next-to-leading order Drell-Yan coefficient
function $C_2$ as a function of $N$, and its logarithmic approximations.}
\label{fig:sd5}
\end{center}
\end{figure}
In order to check that these conclusions are generic, we repeat the
comparison at the next perturbative order.  The order $\as^2$
Drell-Yan cross-section $C_2(z)$ has been computed in
Ref.~\cite{vanNeerven:1991gh}. In Fig.~\ref{fig:sd5} it is compared to
its logarithmic approximation $C_2^{\rm log}(N)$, defined as the
Mellin transform of all contributions to $C_2(z)$ which grow
logarithmically as $z\to1$, as well as another logarithmic
approximation $C_2^{\rm log'}(N)$ which differs by the former through
power-suppressed terms, analogous to  $C_1^{\rm log'}(N)$ and 
to be discussed in more detail in
Sect.~\ref{sec:subl} below.

It is apparent from Fig.~\ref{fig:sd5} that the conclusions drawn at
NLO are qualitatively unchanged, though there is some quantitative
difference: the logarithmic approximation, which differs by a constant
from the asymptotic large--$N$ behaviour of the coefficient function,
provides a sizable or even dominant contribution to it for
$N\gtrsim2-3$, according to the choice of subleading terms:
approximations which differ by subleading terms are all very close for
$N\gtrsim 3$, though they may differ substantially at small $N$.

As a final comment we note that the region in which we have found
logarithmic effects to be important is in fact rather wider than the
region in which $\as\ln^2 N\sim 1$. In this region even though
logarithmically enhanced terms may lead to a substantial contribution,
they behave in an essentially perturbative way, in that $(\as\ln^2
N)^{k+1}<(\as\ln^2 N)^k$, and therefore the all-order behaviour of the
resummation is irrelevant. In this intermediate region, the
resummation may have a significant impact, but with significant
ambiguities related to subleading terms.  \eject

\section{Resummation}
\label{sec:resummation}

After having provided an assessment of the kinematic region in which
logarithmically enhanced terms are significant, we now discuss their
all-order resummation. We will assume knowledge of the
resummed results in $N$--Mellin space, both at the level of inclusive
cross-sections~\cite{Sterman:1986aj,Catani:1989ne,fr} and rapidity
distributions~\cite{Laenen:1992ey,Mukherjee:2006uu,bolz}, and we will
concentrate on the ambiguities related to their definition. First, we
briefly review some prescriptions introduced in order to deal with the
divergent nature of the perturbative expansion for resummed
quantities. Next, we discuss the role and impact of subleading terms
in the resummation, with specific reference to the subleading terms
which are introduced by different resummation prescriptions.

\subsection{Resummation prescriptions}
\label{sec:MPBP}

Resummation of the partonic cross-section $\hat\sigma(z,\as(Q^2))$
Eq.~(\ref{fact}) is most naturally performed in $N$ space, because the
solution of the relevant Altarelli-Parisi
equation~\cite{Catani:1989ne}, the relevant factorization
theorem~\cite{Sterman:1986aj,Contopanagos:1996nh} and the relevant
renormalization group equation~\cite{Contopanagos:1996nh,fr} are all
naturally formulated in $N$--space. The underlying physical reason is
that, in the soft limit, while amplitudes factorize due to the eikonal
approximation both in $N$-- and in $z$--space, the relevant phase space
only factorizes in $N$--space but not in $z$--space~\cite{Catani:1996rb}.

Recently, resummation has also been performed using SCET
techniques both in $N$--space~\cite{scet} and in momentum
space~\cite{Becher:2006nr,Becher:2007ty}; in the latter approach the soft
scale whose logarithms are resummed is not the partonic scale
$Q^2(1-z)$ but rather a soft scale $\mu_s$, independent of
the parton momentum fraction $z$, but related to 
the hadronic scale $Q^2(1-\tau)$. 
As a consequence, the hard coefficient function depends on $\tau$
not only through the convolution variable, but also directly through
the soft scales: therefore, the
resummed result can no longer be factorized by Mellin transformation
into the product of a parton density and a hard
coefficient.
For this reason, a direct comparison between the result
obtained through the SCET approach and that based on standard
factorization Eq.~(\ref{fact}) is not possible at the parton level.
A phenomenological comparison is possible~\cite{Becher:2007ty}, but it
requires either assuming a specific form of
the parton distribution functions, or switching to a SCET formulation
in which $\mu_s$ depends on $N$, which however is not the Mellin transform
of the $\tau$--space SCET result, advocated for phenomenolgy in 
Ref.~\cite{Becher:2007ty}.

The general structure of $N$--space resummed expressions
can be understood by considering the case of the inclusive coefficient function 
$C(N,\as)$, obtained by Mellin transformation from
$C(z,\as)$ Eq.~(\ref{eq:exp}). At the resummed level,
\begin{align}
C^{\rm res}(N,\as(Q^2)) 
&= g_0(\as) \,\exp \S\(\ab \ln\frac{1}{N}, \ab\), \label{eq:Cres}\\
\S(\lambda, \ab) &= \frac{1}{\ab}\, g_1(\lambda) + g_2(\lambda) +
\ab \, g_3(\lambda) + \ab^2 \, g_4(\lambda) 
+ \dots,\label{eq:S}\\
\ab&= a\,\as(Q^2)\,\beta_0,\label{eq:ab_def}
\end{align}
where $a$ is a process-dependent constant ($a=2$ for Drell-Yan production).  
At the next$^k$-to-leading logarithmic (N$^k$LL) level 
functions up to $g_{k+1}$ must be included, and $g_0$ must be computed
up to order $\as^k$ (see Appendix~\ref{sec:app-resumm} for
explicit expressions). Expansion of the resummed coefficient function
Eq.~(\ref{eq:Cres}) in powers of
$\as(Q^2)$ up to order $n$ gives the logarithmically enhanced
contributions to the fixed-order coefficient functions $C_i(N)$
Eq.~(\ref{eq:exp}) up to the same order, with, at the N$^k$LL level, 
all terms of order $\ln^m\frac{1}{N}$  with  $2(i-k)\le m\le 2i$ correctly
predicted. 
The inclusion up to the relevant order of the function $g_0$  is
necessary, despite the fact that $g_0$ does not depend on $\ln N$,
because of its interference with the expansion of the exponentiated 
logarithmically enhanced functions $g_i$ with $i\ge 1$. 

Matched resummed coefficient functions are then obtained by combining
the resummed result Eq.~(\ref{eq:Cres}) with the fixed-order expansion
in power of $\as$, and subtracting double-counting terms, i.e.~the
expansion of $C^{\rm res}(N,\as(Q^2))$  in powers of $\as(Q^2)$ up to the
same order:
\beq
C^{\text{N$^k$LL+N$^p$LO}}(N,\as) =
C^{\rm res}(N,\as)
+ \sum_{j=0}^{p}\left(\frac{\as}{\pi}\right)^j C_j(N) 
- \sum_{j=0}^{p}\frac{\as^j}{j!}
\[\frac{d^jC^{\rm res}(N,\as)}{d\as^j}\]_{\as=0}.
\label{eq:match}
\eeq

However the
perturbative expansion of 
$C^{\rm res}(z,\as(Q^2))$ Eq.~(\ref{eq:Cres}) in powers of
$\as(Q^2)$ turns out to be
divergent.
This follows from the fact that the functions $g_i$ in Eq.~(\ref{eq:S})
depend on $N$ through 
\beq
\label{alpharun}
\as(Q^2/N^a)=\frac{\as(Q^2)}{1+L}\left(1+\Ord(\as(Q^2))\right); \qquad
L\equiv\ab\,\ln\frac{1}{N},
\eeq
with $\ab$ defined in Eq.~\eqref{eq:ab_def}.
As a consequence, the expansion of the resummed partonic cross-section
in powers of $\as(Q^2)$ has a
finite radius of convergence dictated by $\abs{L}<1$, and
$\hat\sigma(N,\as(Q^2))$ at fixed $\as(Q^2)$ has a branch 
cut in the complex $N$--plane along  the positive real axis from
$N_L=\exp(1/\ab)$ to $+\infty$.
But a Mellin transform always has a convergence abscissa, so  
$C^{\rm res}(N,\as(Q^2))$ cannot be a Mellin transform. 

On the
other hand,  any finite-order truncation of the series expansion
\be
C^{\rm res}(N,\as(Q^2))=
\sum_{i=0}^\infty\left(\frac{\as(Q^2)}{\pi}\right)^i C^{\rm res}_i(N)
\label{expsigman}
\ee
behaves as a power of $\ln N$ at large $N$ and it is thus free of
singularities for $N$ large enough. Hence, $C^{\rm res}(N,\as(Q^2))$ 
can be viewed as the
Mellin transform of the function $C^{\rm res}(z,\as(Q^2))$
\be
C^{\rm res}(z,\as(Q^2))=\sum_{i=0}^\infty\left(\frac{\as(Q^2)}{\pi}\right)^i
C^{\rm res}_i(z)
\label{expsigmaz}
\ee
such that 
\be
C^{\rm res}_i(z)\equiv\int_{\bar N-i\infty}^{\bar N+i\infty}
\frac{dN}{2\pi i}\,z^{-N}\, C^{\rm res}_i(N).
\label{sigpertexp}
\ee
It follows, by contradiction, that the series Eq.~(\ref{expsigmaz})
must diverge~\cite{frru}. 
It turns out~\cite{frru} that, if the Mellin inversion
Eq.~(\ref{sigpertexp}) is performed to finite logarithmic accuracy, 
the series
Eq.~(\ref{expsigmaz}) acquires a finite but nonzero radius of
convergence in $z$; however, this does
not help given that the convolution integral Eq.~(\ref{fact}) always
goes over the region where the series diverges.

Hence, any resummed definition must either explicitly or implicitly
deal with the divergence of the perturbative expansion
Eq.~(\ref{expsigmaz}). We now consider two prescriptions in which this
is done by constructing a resummed expression to which the divergent
series is asymptotic.

\subsubsection{Minimal prescription}

The minimal prescription (MP)~\cite{cmnt} defines the resummed
hadronic cross-section as
\beq
\sigma_{\rm MP}(\tau,Q^2)=
\hat\sigma_0\,\frac{1}{2\pi i}\int_{c-i\infty}^{c+i\infty}
dN\,\tau^{-N}{\cal L}(N,Q^2)\, C^{\rm res}(N,\as(Q^2)),
\label{eq:main1}
\eeq
where $\hat\sigma_0$ is defined in Eq.~\eqref{cfdef}.
The integration path in Eq.~\eqref{eq:main1} is 
taken to the left of the cut, but to the right of all other
singularities.
It is shown in Ref.~\cite{cmnt} that the cross-section obtained in this way
is finite, and that it is an asymptotic sum of the divergent series
obtained by substituting the expansion Eq.~(\ref{expsigman}) in
Eq.~(\ref{eq:main1}) and performing the Mellin inversion order by
order in $\as(Q^2)$. Of course, if the expansion is truncated to
any finite order the MP simply gives the exact inverse Mellin transform,
namely, the truncation of Eq.~(\ref{expsigmaz}) to the same finite order.

\begin{figure}[htb]
\begin{center}
\includegraphics[width=0.45\textwidth,page=1]{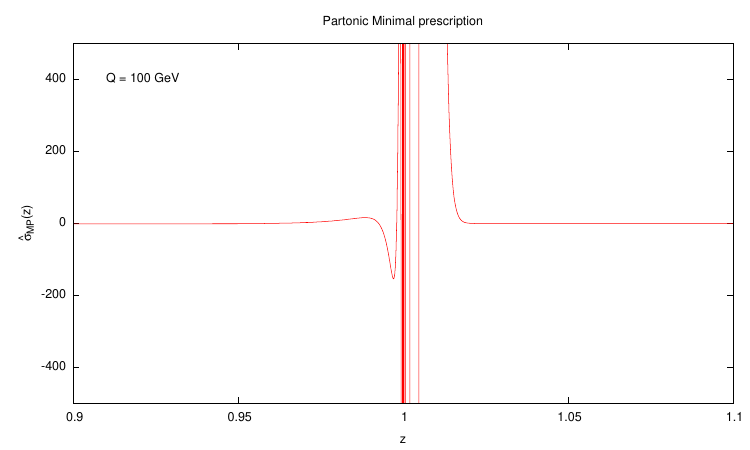}
\includegraphics[width=0.45\textwidth,page=2]{Minimal_partonic}
\caption{The  partonic cross-section $\hat\sigma_{\rm MP}(Q^2,z)$
  computed using the minimal prescription, i.e.~setting
  ${\cal L}(N,Q^2)=1$ in the 
  integral Eq.~(\ref{eq:main1}), evaluated at $\sqrt{Q^2}=8$~GeV and
  $\sqrt{Q^2}=100$~GeV. The curve shown is obtained  using on the
  r.h.s.~of Eq.~(\ref{eq:main1}) the NLL expression for the Drell-Yan
  coefficient function
  Eq.~(\ref{eq:Cres},~\ref{eq:S}) with only $g_1$ and $g_2$ included.} 
\label{fig:MP_osc}
\end{center}
\end{figure}
If the MP is applied  to the resummed partonic cross-section, i.e.~if
one omits the parton luminosity ${\cal L}(N,Q^2)$ in
Eq.~(\ref{eq:main1}) then, because of the branch cut to the right of
the integration contour, the ensuing integral gives a function
$\hat\sigma_{\rm MP}(z,Q^2)$, which does not vanish when
$z>1$~\cite{cmnt}.
However, as shown in Ref.~\cite{cmnt}, the contribution from
the $z\ge1$ region is exponentially suppressed in $\frac{\Lambda_{\rm QCD}}{Q}$.
In the vicinity of $z=1$ the integral oscillates strongly, as 
shown in Fig.~\ref{fig:MP_osc}, where $\hat\sigma_{\rm MP}(z,Q^2)$ is displayed
for the Drell-Yan resummed cross-section, evaluated at two scales
which are relevant for the phenomenological discussion of
Sect.~\ref{sec:pheno}. When folded with the luminosity,
the hadronic cross-section receives a
contribution from the unphysical region and the corresponding integral
thus does not have the form of
a convolution. The ensuing integral is finite, but
the  oscillatory behaviour of the partonic cross-section makes its
numerical computation difficult.
A technical solution to this problem
is provided in Ref.~\cite{cmnt}; here, we will propose in
Sect.~\ref{sec:rap} a different solution, and use it for phenomenology in
Sect.~\ref{sec:pheno}.

\subsubsection{Borel prescription}
\label{sec:borel}

An alternative prescription is based on the Borel
summation of the divergent series. This prescription was developed in
Refs.~\cite{frru,afr}; here we give an equivalent, but a somewhat
simpler presentation of it.
To this purpose,
it is convenient to rewrite the resummed coefficient function
$C^{\rm res}(N,\as(Q^2))$ Eq.~(\ref{eq:Cres}) as
\beq
\label{sigdef}
C^{\rm res}(N,\as(Q^2))=1+\Sigma(L,\as(Q^2)),
\eeq
where $C^{\rm res}_0(N,\as)=1$ is the ($N$--independent) Born result,
so that only logarithmically enhanced terms are included in $\Sigma$.
Using Eq.~\eqref{logminimal2} we see that
\begin{align}
&\Sigma(z,\as(Q^2))\equiv\frac{1}{2\pi i}
\int_{\bar N-i\infty}^{\bar N+i\infty} dN\,z^{-N}\Sigma(L,\as(Q^2))
=\left[\frac{R(z)}{\ln\frac{1}{z}}\right]_+
\nonumber\\
&R(z)=\sum_{k=1}^\infty h_k(\as(Q^2))\,\ab^k\,c_k(z),
\label{Rdef}
\end{align}
where 
\beq
c_k(z)=\left.\frac{d^k}{d\xi^k}\frac{\ln^\xi\frac{1}{z}}{\Gamma(\xi)}
\right|_{\xi=0}
\label{ckcoeffs}
\eeq
and $h_k(\as(Q^2))$ are the coefficients of the expansion
\be\label{SigmaL}
\Sigma(L,\as(Q^2))=\sum_{k=1}^\infty h_k(\as(Q^2))\,L^k,
\ee
whose $\as(Q^2)$ dependence will henceforth be omitted for
notational simplicity.
Using Eq.~(\ref{logminimal3})
Eq.~\eqref{Rdef} can be written as
\beq
R(z)=\frac{1}{2\pi i}\oint\frac{d\xi}{\xi}
\,\frac{\ln^\xi\frac{1}{z}}{\Gamma(\xi)}
\sum_{k=1}^\infty k!\,h_k\left(\frac{\ab}{\xi}\right)^k.
\label{RK}
\eeq

The Borel prescription can be now formulated. First, we note that the
Borel transform of $R(z)$ with respect to $\ab$, which is found replacing
\beq
\ab^k\to \frac{w^{k-1}}{(k-1)!}
\label{bt}
\eeq
in Eq.~(\ref{RK}), is convergent, and  can be summed in closed form:
\be
\hat R(w,z)=\frac{1}{2\pi i}\oint\frac{d\xi}{\xi}\,
\frac{\ln^\xi\frac{1}{z}}{\Gamma(\xi)}
\sum_{k=1}^\infty k\,h_k\,\frac{w^{k-1}}{\xi^k}
=\frac{1}{2\pi i}\oint\frac{d\xi}{\xi}\,
\frac{\ln^\xi\frac{1}{z}}{\Gamma(\xi)}\,
\frac{d}{dw}\Sigma\left(\frac{w}{\xi},\as(Q^2)\right).
\ee
The branch cut of $\Sigma(L,\as(Q^2))$, $-\infty<L\leq-1$, is mapped 
onto the range $-w\leq\xi\leq 0$ on the real axis
of the complex $\xi$ plane. Hence, the $\xi$ integration path
is any closed curve which encircles the cut.

Next, we observe that the inverse Borel transform of $\hat R(w,z)$ 
does not exist because it involves evaluation of 
the function $1/\Gamma(\xi)$ on the negative real axis where it is badly
behaved as $\xi\to-\infty$~\cite{frru}.
We may however define the inversion integral by  introducing a cutoff 
at some finite value $C$
of the Borel variable $w$: if we replace $R(z)$ Eq.~(\ref{RK}) with
\be
\bar R_C(z)
=\int_0^C dw\,e^{-\frac{w}{\ab}}\hat R(w,z)
=
\frac{1}{2\pi i}\oint\frac{d\xi}{\xi}\,
\frac{\ln^\xi\frac{1}{z}}{\Gamma(\xi)}
\int_0^C dw\,e^{-\frac{w}{\ab}}\frac{d}{dw}
\Sigma\left(\frac{w}{\xi},\as(Q^2)\right)
\label{RBP}
\ee
then the resummed function $\Sigma(z,\as(Q^2))$ becomes
\be
\bar\Sigma_C(z,\as(Q^2))
=
\frac{1}{2\pi i}\oint\frac{d\xi}{\xi}\,
\frac{1}{\Gamma(\xi)}
\int_0^C dw\,e^{-\frac{w}{\ab}}\frac{d}{dw}
\Sigma\left(\frac{w}{\xi},\as(Q^2)\right)
\[\ln^{\xi-1}\frac{1}{z}\]_+.
\label{SBP}
\ee
It is proved in Refs.~\cite{frru,afr} 
that the original divergent series Eq.~(\ref{Rdef}) for $\Sigma$ is
asymptotic to the function $\bar\Sigma_C(z,\as(Q^2))$, and furthermore, that for
any finite-order truncation of the divergent series, the full and
cutoff results differ by a 
twist--$\(2+\frac{2C}{a}\)$ term, where $C$ is the cutoff. The
parameter $C$ can be chosen freely in the range $C\ge a$ (with $a$ as
in Eq.~\eqref{eq:ab_def}), with different choices differing
by power suppressed terms. We will use henceforth the ``minimal''
choice $C=a$.

A somewhat simpler result is obtained if the Borel transform is
performed through the replacement
\be
\ab^k\to \frac{1}{\ab}\frac{w^k}{k!}
\ee
in Eq.~(\ref{RK}), instead of Eq.~\eqref{bt}. In this case, instead of
Eqs.~(\ref{RBP},~\ref{SBP}) one gets
\begin{align}
R_C(z)
&=\frac{1}{2\pi i}\oint\frac{d\xi}{\xi}
\,\frac{\ln^\xi\frac{1}{z}}{\Gamma(\xi)}
\int_0^C \frac{dw}{\ab}\,e^{-\frac{w}{\ab}}
\Sigma\left(\frac{w}{\xi},\as(Q^2)\right)
\label{RBP2}
\\
\Sigma_C(z,\as(Q^2))
&=\frac{1}{2\pi i}\oint\frac{d\xi}{\xi}\,\frac{1}{\Gamma(\xi)}
\int_0^C \frac{dw}{\ab}\,e^{-\frac{w}{\ab}}
\Sigma\left(\frac{w}{\xi},\as(Q^2)\right)
\[\ln^{\xi-1}\frac{1}{z}\]_+.
\label{SBP2}
\end{align}
which differs from Eq.~(\ref{RBP}) by higher-twist terms, and therefore
provides an equally good resummation prescription; the difference is
in practice very small.

If we only wish to retain terms which do not vanish as $z\to1$ we may
expand
\beq 
\ln\frac{1}{z}=
1-z+\Ord((1-z)^2)\label{logexp}
\eeq
 with the result 
\beq
\Sigma_{\rm BP}(z,\as(Q^2))=\frac{1}{2\pi i}
\oint\frac{d\xi}{\xi}
\,\frac{1}{\Gamma(\xi)}
\int_0^C\frac{dw}{\ab}\,e^{-\frac{w}{\ab}}
\Sigma\left(\frac{w}{\xi},\as(Q^2)\right)
\[(1-z)^{\xi-1}\]_+.
\label{sigmaBP}
\eeq
Equations~(\ref{SBP}),~(\ref{SBP2}), and (\ref{sigmaBP}) provide three
alternate definitions of the resummed $\Sigma$ function which differ by
terms suppressed by powers of $1-z$. The difference between the first
two is negligible and the prescription Eq.~(\ref{SBP}) will not be
discussed further, but
Eqs.~(\ref{SBP2}) and~(\ref{sigmaBP}) differ by a sizable amount, as we
shall see in somewhat greater detail in Sect.~\ref{sec:subl}. 
Equation~({\ref{sigmaBP}) with $C=a$ is the default
Borel prescription of Ref.~\cite{frru,afr}.

\begin{figure}[htb]
\begin{center}
\includegraphics[width=0.6\textwidth]{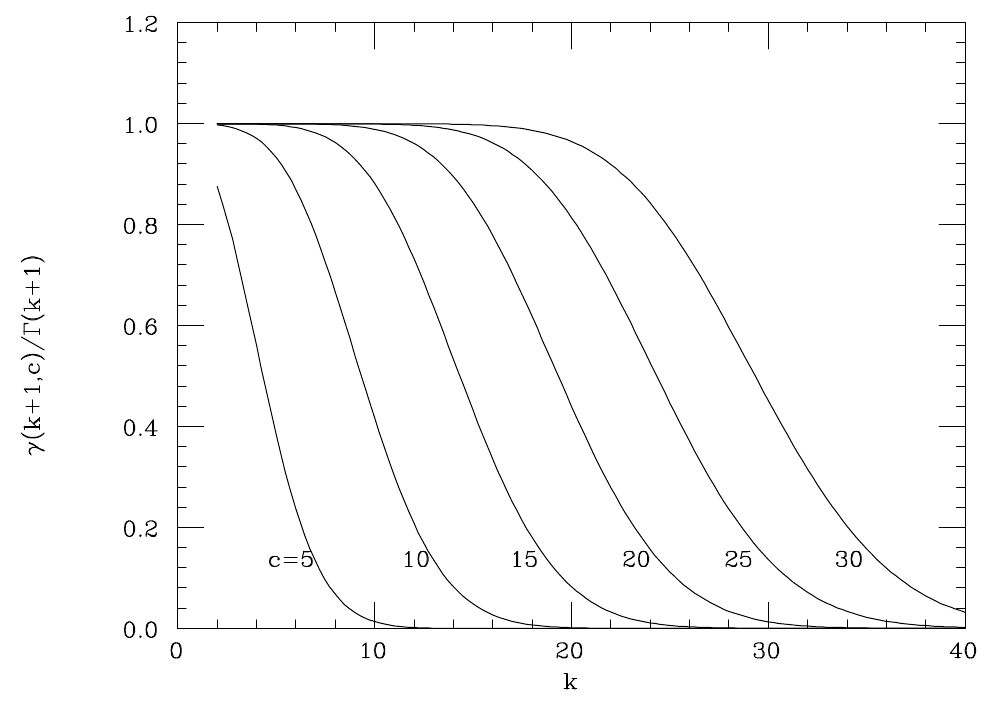}
\caption{\label{truncgamma}
The damping factor $f_k$ Eq.~(\ref{dampfac}) 
in the Borel inversion of $\ln^k\frac{1}{N}$, for various values of
$c\equiv\frac{C}{\ab}$.}
\end{center}
\end{figure}
When using the Borel prescription, the divergent series
Eq.~(\ref{expsigmaz}) is made convergent by cutting off its
high-order behaviour: indeed, using the Borel prescription
Eq.~(\ref{SBP2}) to
compute the inverse Mellin transform of $\ln^k\frac{1}{N}$ one gets
\be
\left[
\frac{1}{2\pi i}\int_{\bar N-i\infty}
^{\bar N+i\infty}dN\,z^{-N}\,\ln^k\frac{1}{N}
\right]_C
=f_k(C/\ab) \left[\frac{c_k(z)}{\ln\frac{1}{z}}
\right]_+,
\label{logborel}
\ee
and similarly if the Borel prescription
Eq.~(\ref{sigmaBP}) is adopted instead. 
Equation~\ref{logborel} differs
from the exact result Eq.~(\ref{logminimal1}) by the factor
$f_k(C/\ab)$~\cite{frru}, given by 
\beq
f_k(c)=\frac{\gamma(k+1,c)}{\Gamma(k+1)},
\label{dampfac}
\eeq
where $\gamma$ is the truncated Gamma function.
This factor, plotted in Fig.~\ref{truncgamma},
effectively truncates the series before the divergence sets in.

\subsection{Subleading terms}
\label{sec:subl}

The Borel and minimal prescriptions summarized in Sect.~\ref{sec:MPBP}
differ in the way the high-order behaviour of the divergent series is
handled. This, as discussed in Refs.~\cite{frru,afr}, makes in
practice a small difference unless the hadronic $\tau$ is close to the
Landau pole of the strong coupling, 
$\tau_L=1-\(\frac{\Lambda_{\rm QCD}}{Q}\)^{\frac{2}{a}}$,
which is seldom the case, and never for
applications at collider energies that we are mostly interested
in. This is a consequence of the fact that for values of $\as$ in the
perturbative region it is only at very high orders that the effect of
the various prescriptions kicks in. For example, with the Borel
prescription and the ``minimal'' choice $C=2$, if $\as=0.11$ then
$c\approx 15$, so from Fig.~\ref{truncgamma} one sees that the
perturbative expansion is truncated between the tenth and twentieth
order.

However, prescriptions may also differ in the subleading terms which are
introduced when performing the resummation. To understand this,
consider the result one gets applying the various prescriptions to any
finite truncation of the expansion of $\Sigma(L,\as(Q^2))$ in powers
of $\as(Q^2)$.  The minimal prescription then just gives the exact
Mellin inverse Eq.~(\ref{logminimal2}). Because this result depends on
$1-z$ through $\ln\frac{1}{z}$, in $z$ space it
generates a series of power suppressed terms Eq.~(\ref{logexp}).

If the Borel prescription is defined according to Eq.~(\ref{SBP2}),
then the result Eq.~(\ref{logborel}) is obtained, which for $k$ small
enough that $f_k\approx 1$ coincides with the minimal prescription,
including subleading terms.  However, with the Borel prescription the
$z$ dependence is under analytic control:  it is entirely
contained in the factor $\ln^{\xi-1}\frac{1}{z}$ in Eq.~\eqref{SBP2},
and it can thus be modified at will.  Indeed, while the original Borel
prescription Eq.~(\ref{SBP2}) treats subleading terms as the minimal
prescription, and thus it coincides with it for finite not too high
order truncations of the perturbative expansion, using our default
Borel prescription Eq.~(\ref{sigmaBP}) instead of Eq.~(\ref{SBP2})
only leading power contributions in $1-z$ are retained. However, this
implies that  power-suppressed contributions in $1/N$ are
introduced: indeed, the exact Mellin transforms of Eq.~(\ref{SBP2})
and Eq.~(\ref{sigmaBP}) are respectively given by
\begin{align}
\int_0^1 dz\, z^{N-1} \Sigma_C(z,\as(Q^2)) &=
\frac{1}{2\pi i} \oint\frac{d\xi}{\xi}\,
\left[N^{-\xi}-1\right]
\int_0^C \frac{dw}{\ab}\,e^{-\frac{w}{\ab}}\,
\Sigma\left(\frac{w}{\xi},\as(Q^2)\right)\nonumber\\
\int_0^1 dz\, z^{N-1} \Sigma_{\rm BP}(z,\as(Q^2)) &=
\frac{1}{2\pi i} \oint\frac{d\xi}{\xi}\,
\left[\frac{\Gamma(N)}{\Gamma(N+\xi)}-\frac{1}{\Gamma(1+\xi)}\right]
\int_0^C \frac{dw}{\ab}\,e^{-\frac{w}{\ab}}\,
\Sigma\left(\frac{w}{\xi},\as(Q^2)\right),
\label{BPcomp}
\end{align}
which differ by terms suppressed by powers of $\frac{1}{N}$.

Hence,  the BP Eq.~(\ref{sigmaBP}) on the one hand, 
and the MP (and the BP Eq.~(\ref{SBP})) on the other hand correspond
to two opposite extreme choices in the treatment of subleading terms: 
in the MP, all $1/N$ powers suppressed terms in $N$--space are set to zero,
but this leads to $1-z$ power suppressed terms in $z$ space, while in
the BP the opposite is true. The
difference between the MP and the default BP Eq.~(\ref{sigmaBP}) could
thus be taken as a maximal estimate of the impact of subleading terms.

With the Borel prescription it is also possible to construct
intermediate,  possibly optimized, choices of subleading terms.
Indeed, in Sect.~\ref{sec:resDY} we have noticed that
in the case of the NLO
coefficient function $C_1(N)$, the Mellin transform $C_1^{\rm log}(N)$
Eq.~(\ref{eq:Mpsi})  of the logarithmically enhanced $z$--space terms
can be brought in better agreement with the full result by inclusion
in it  
of some terms suppressed by powers of $z$ --- see $C_1^{\rm log'}(N)$, defined in
Eq.~\eqref{eq:Mpsi1}. This increased agreement can
be understood by inspection of the $z$--space expression of the
improved result, $C_1^{\rm log'}(z)$ Eq.~(\ref{logp}): the extra power-suppressed
terms which are introduced in it turn out to be 
present in the full result. These terms are in fact of
kinematical origin: soft resummation follows from the kinematic fact
that as $z\to1$ the dependence of partonic cross-sections on $z$ is
always in the combination $Q^2(1-z)^2$~\cite{fr}, essentially because
this is the upper limit of the integral over the energy of radiated
gluons for the Drell-Yan process. 
However, the upper integration limit is really
\beq
k^0_{\rm max}=\sqrt{\frac{Q^2(1-z)^2}{4z}},
\eeq
so that in fact the resummation produces logarithmic terms of the form
$\ln\frac{1-z}{\sqrt{z}}$.  

This suggests to define an all-order generalization of the
approximation $C_1^{\rm log'}(z)$, by simply letting 
\be
\ln(1-z)\to
\ln\frac{1-z}{\sqrt{z}}
\label{logreplace}
\ee
in all resummed expressions. In the Borel
prescription, this is easily done:
\beq
\Sigma_{\rm BP'}(z,\as(Q^2))=\frac{1}{2\pi i}
\oint\frac{d\xi}{\xi}
\,\frac{1}{\Gamma(\xi)}
\int_0^C\frac{dw}{\ab}\,e^{-\frac{w}{\ab}}
\Sigma\left(\frac{w}{\xi},\as(Q^2)\right)
\[(1-z)^{\xi-1}\]_+z^{-\frac{\xi}{2}}.
\label{sigmaBPp}
\eeq
With this choice, the kinematic correction Eq.~(\ref{logreplace}) is
automatically included to all orders. However, beyond $\Ord(\as)$
other subleading terms of the same form but not of
kinematic origin will in general be present.
This choice also arises in a natural way in the context of
soft-collinear effective
theories, and it was adopted in Ref.~\cite{Becher:2007ty}.

It turns out that the modified Borel prescription (BP$^\prime$)
Eq.~\eqref{sigmaBPp} is
closer to the MP than the default one Eq.~(\ref{sigmaBP}) because 
\beq
\ln\frac{1}{z}=\frac{1-z}{\sqrt{z}}\left(1+\Ord((1-z)^2)\right),
\label{logMPexp}
\eeq
so that
\beq
\frac{\ln^k\ln\frac{1}{z}}{\ln\frac{1}{z}}
=\frac{\sqrt{z}}{1-z}\ln^k\frac{1-z}{\sqrt{z}} \left(1+\Ord((1-z)^2)\right).
\label{allMPexp}
\eeq
Equation~(\ref{allMPexp}) shows that, amusingly, up to terms
suppressed by two powers of $1-z$, the minimal prescription
effectively also performs the kinematic
subleading replacement Eq.~(\ref{logreplace}), though at the cost of
also introducing an overall factor
$\sqrt{z}$ which is absent in the known perturbative contributions.

\begin{figure}[htb]
\begin{center}
\includegraphics[width=0.48\textwidth,page=1]{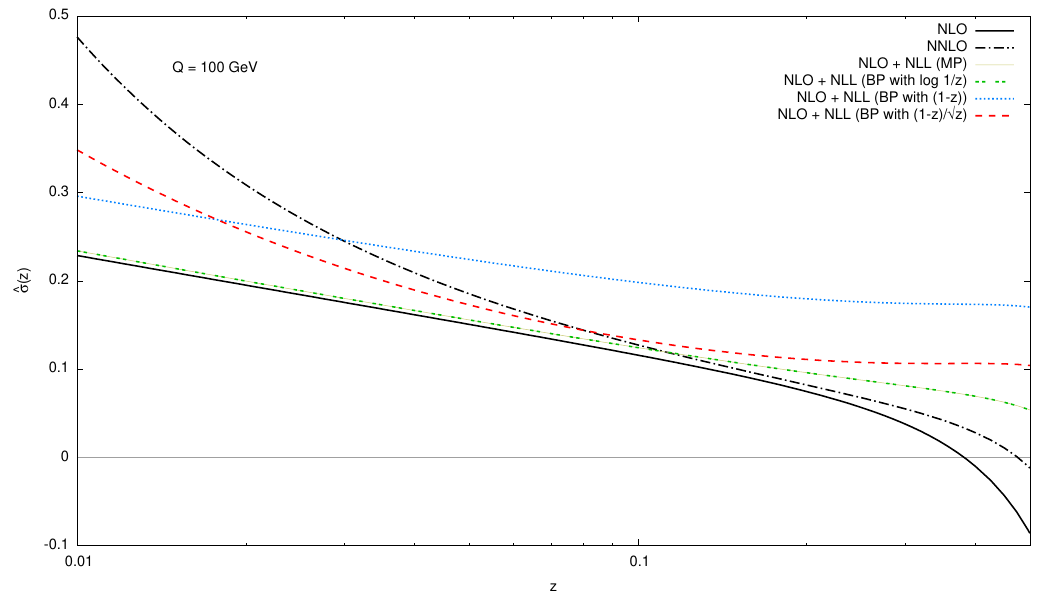}
\includegraphics[width=0.48\textwidth,page=2]{graph_partonic}\\
\includegraphics[width=0.48\textwidth,page=3]{graph_partonic}
\includegraphics[width=0.48\textwidth,page=4]{graph_partonic}
\caption{The next-to-leading log (NLL) resummed  Drell-Yan matched to
  the fixed next-to-leading order (NLO, top row) or
  next-to-next-to-leading order (NNLO bottom row), with various
  resummation prescription. The right plots show the very large $z$
  region. The three alternate versions Eq.~(\ref{SBP2}), Eq.~(\ref{sigmaBP})
and Eq.~(\ref{sigmaBPp}) of the Borel prescription are respectively
denote BP with $\log 1/z$, BP with $(1-z)$ and BP with $(1-z)/\sqrt{z}$.
}
\label{fig:subl}
\end{center}
\end{figure}
In order to assess the impact of these different prescriptions, in
Fig.~\ref{fig:subl} we compare the matched result Eq.~(\ref{eq:match})
for the Drell-Yan coefficient function in the quark--antiquark channel
obtained by including terms up to NLL in the resummed expression, and
either up to $\Ord(\as)$ (NLO) or $\Ord(\as^2)$ (NNLO) in the
fixed-order expansion, with various resummation prescriptions.  First,
we note that the minimal prescription and the Borel prescription
Eq.~(\ref{SBP2}) (denoted as BP with $\log 1/z$ in the figure) are
essentially indistinguishable (for values of $z$ less than about 0.9,
where the oscillatory behaviour of the minimal prescription sets in).
This is what one expects, since they contain the same subleading terms
and they only differ in the treatment of the high-order divergence.

However, the Borel prescription Eq.~(\ref{sigmaBP}) is seen to differ
by a non-negligible amount from the minimal prescription: at large
$z\gtrsim 0.3$ where the resummation kicks in it is small in
comparison to the size of the resummation itself, while at small
$z\lesssim 0.03$, where the resummation just leads to unreliable
subleading contributions, it is smaller than the typical higher order
correction, as it is seen by comparing results matched to the NLO and
NNLO. However, in the intermediate $z$ region the subleading terms
introduced by this prescription are uncomfortably large.

The improved BP$^\prime$ Borel prescription
Eq.~\eqref{sigmaBPp} (denoted as BP with $(1-z)/\sqrt{z}$ in the
figure) 
as expected 
differs less from the MP
both at large and at small $z$; also, it does not introduce
unnaturally large subleading terms for any value of $z$. The
difference between MP and this last version of the Borel prescription
is not negligible but small in comparison to the size of resummation
effects in the region where the resummation is relevant, and it is
smaller than typical higher-order terms 
in the region in which the resummation is not relevant. It can thus be
taken as a reliable estimate of the ambiguity in the resummation.
It is interesting to observe that the subleading terms which are
introduced by the replacement Eq.~(\ref{logreplace}) grow at small
$z$, and in fact are thus reproducing part of the small--$z$ growth of
perturbative coefficient functions. The fact that inclusion of these
terms is important in keeping the ambiguities of large $z$ resummation
under control suggests that the large $z$ and small $z$ resummation
regions are not well separated, and that there  
might be an interplay between small-- and
large--$z$ resummation.
\section{Rapidity distribution}
\label{sec:rap}

The resummed expression for rapidity distributions was only derived
relatively recently in
Refs.~\cite{Mukherjee:2006uu,bolz}, confirming a conjecture of
Ref.~\cite{Laenen:1992ey}. In this Section, after introducing rapidity
distributions at fixed perturbative order, we will briefly review
this result, and also compare it to a  somewhat different expression
later  derived using SCET methods in Ref.~\cite{Becher:2007ty}. We
will then discuss the numerical implementation of resummed results
that will be used for phenomenology in Sect.~\ref{sec:pheno}.

The hadronic rapidity $Y$ distribution for a Drell-Yan pair of
invariant mass $Q$ produced in hadronic collisions at center-of-mass
energy $\sqrt{s}$ is given by
\beq
\frac{d\sigma}{dQ^2 dY}(\tau,Y,Q^2) =
\sum_{i,j} \int_{x_1^0}^1 dx_1 \int_{x_2^0}^1 dx_2\,
f_i^1(x_1,\muf^2)\, f_j^2(x_2,\muf^2)\,
\frac{d\hat\sigma_{ij}}{dQ^2 dy}
\(\frac{\tau}{x_1x_2},y,\as(\mur^2),\frac{Q^2}{\muf^2},\frac{Q^2}{\mur^2}\),
\eeq
where
\beq
y = Y-\frac{1}{2}\ln\frac{x_1}{x_2},\qquad
x_1^0=\sqrt{\tau}e^Y,\qquad
x_2^0=\sqrt{\tau}e^{-Y},\qquad
\tau=\frac{Q^2}{s}.
\eeq
The sum runs over all partons in hadrons $1$ and $2$.
To simplify notations, in the following we suppress the explicit
dependence of the partonic cross-sections $d\hat\sigma_{ij}$ on $\as$
and on the factorization and renormalization scales $\muf,\mur$.
We define
\beq
C_{ij}(z,y)=
\frac{1}{z}\frac{d\hat\sigma_{ij}}{dQ^2 dy}(z,y),
\eeq
so that
\beq\label{eq:rap_distr}
\frac{1}{\tau}\frac{d\sigma}{dQ^2 dY} =
\sum_{i,j} \int_{x_1^0}^1 \frac{dx_1}{x_1} \int_{x_2^0}^1 
\frac{dx_2}{x_2} \, f_i^1(x_1)\, f_j^2(x_2) \,
C_{ij} \( \frac{\tau}{x_1x_2},y\).
\eeq

Especially in fixed-target experiments, distributions sometimes are
given in terms of the Feynman $x_F$ variable instead of rapidity. 
The variable $x_F$ is defined by
\beq
x_F=\frac{2q_L}{\sqrt{s}},
\label{xfdef}
\eeq
where $q_L$ is the longitudinal momentum of the Drell-Yan pair,
and it is related to the rapidity $Y$ and the transverse momentum $q_T$
by
\beq
Y=\frac{1}{2}\log\frac
{\sqrt{x_F^2+4\tau(1+\hat q_T^2)}+x_F}
{\sqrt{x_F^2+4\tau(1+\hat q_T^2)}-x_F}
\label{xfy}
\eeq
where $\hat q_T^2=q_T^2/Q^2$.
At leading order $q_T=0$, and at NLO $q_T$ is fixed uniquely in terms
of $x_F$ by the kinematics, so up to this order the  
$x_F$ and rapidity distributions are
simply proportional:
\beq
\left.\frac{d\sigma}{dQ^2 dY}\right|_{\rm NLO}=
\sqrt{x_F^2 + 4\tau(1+\hat q_T^2)}
\left.\frac{d\sigma}{dQ^2 dx_F}\right|_{\rm NLO},
\eeq
though at higher orders they will be different.

\subsection{Fixed-order results}

At next-to-leading order,
the rapidity distribution receives contributions from
quark--antiquark and quark--gluon subprocesses:
\beq
\frac{d\sigma^{\rm NLO}}{dQ^2\,dY} =
\frac{d\sigma^{\rm NLO}_{q\bar q}}{dQ^2\,dY} +
\frac{d\sigma^{\rm NLO}_{qg+gq}}{dQ^2\,dY}.
\eeq
The result is conveniently expressed in terms of new
variables $z,u$, defined by
\beq
x_1 = \sqrt{\frac{\tau}{z}}  e^Y \sqrt{\frac{z+(1-z)u}{1-(1-z)u}}, \qquad
x_2 = \sqrt{\frac{\tau}{z}}  e^{-Y} \sqrt{\frac{1-(1-z)u}{z+(1-z)u}}
\label{x1x2tozu}
\eeq
with inverse
\beq
z = \frac{\tau}{x_1 x_2}, \qquad
u = \frac{e^{-2y}-z}{(1-z)(1+e^{-2y})}
\label{zutox1x2}
\qquad
\(e^{-2y} = \frac{x_1}{x_2}e^{-2Y}\).
\eeq
The partonic threshold region $Q^2\to \hat s=x_1x_2s$ corresponds to $z\to 1$.

At order $\as$ the $q\bar q$ contribution is given by
\beq
\frac{1}{\tau}\frac{d\sigma^{\rm NLO}_{q\bar q}}{dQ^2\,dY} =
\int_\tau^1 \frac{dz}{z} \int_0^1 du \, L_{q\bar q}(z,u) 
\left[\delta(1-z) + \frac{\as}{2\pi}C_F F(z,u)\right]
\label{sigmaNLOqqbar}
\eeq
where we have defined
\beq
L_{q\bar q}(z,u)= \sum_q c_{q\bar q}\, f_q^1(x_1)\, f_{\bar q}^2(x_2),
\label{qqbarlum}
\eeq
the constants $c_{q\bar q}$ are suitable combinations 
of the coupling constants of partons to vector boson, and
\begin{align}
F(z,u) =&\; \Big( \delta(u)+\delta(1-u) \Big) \Bigg[ \delta(1-z) 
\(\frac{\pi^2}{3} -4\)
+2\,(1+z^2) \plust{\frac{\ln(1-z)}{1-z}} \nonumber \\
&\qquad\qquad\qquad\qquad\quad
+\ln\frac{Q^2}{\muf^2}\plust{\frac{1+z^2}{1-z}}
-\frac{1+z^2}{1-z}\ln z +1-z \Bigg] 
\nonumber \\
&\;
+\frac{1+z^2}{1-z}\[\plust{\frac{1}{u}}+\plust{\frac{1}{1-u}}\]-2(1-z).
\label{Fzu}
\end{align}
The $qg+gq$ contribution is given by
\beq
\frac{1}{\tau}\frac{d\sigma^{\rm NLO}_{qg+gq}}{dQ^2\,dY} =\frac{\as}{2\pi}T_F 
\int_\tau^1 \frac{dz}{z} \int_0^1 du \,
\big[ L_{qg}(z,u) \,G(z,u) + L_{gq}(z,u) \,G(z,1-u) \big]
\eeq
where 
\beq
L_{qg}(z,u)=\sum_q c_{qg}\,f_q^1(x_1)\, f_g^2(x_2),
\qquad(q\leftrightarrow g),
\eeq
and
\begin{multline}
G(z,u)=\delta(u)\left[\(z^2+(1-z)^2\)\(\ln\frac{(1-z)^2}{z} 
+ \ln\frac{Q^2}{\muf^2} \) +2z(1-z) \right] \\
+\(z^2+(1-z)^2\)\plust{\frac{1}{u}} +2z(1-z) +(1-z)^2 u.
\end{multline}

The inclusive coefficient functions, whose form in the
quark--antiquark channel and for $\muf^2=Q^2$  has already
been discussed in Sect.~\ref{sec:ipdfdy},
are found integrating over the hadronic rapidity $Y$.
We get
\beq
\frac{d\sigma^{\rm NLO}}{dQ^2} =
\frac{d\sigma^{\rm NLO}_{q\bar q}}{dQ^2} +
\frac{d\sigma^{\rm NLO}_{qg+gq}}{dQ^2}
\eeq
with
\begin{align}
\frac{1}{\tau}\frac{d\sigma^{\rm NLO}_{q\bar q}}{dQ^2} &=
\int_\tau^1 \frac{dz}{z}\,{\cal L}_{q\bar q}\(\frac{\tau}{z}\) \left[
\delta(1-z) + \frac{\as}{2\pi}\, C_F \, F_{\rm int}(z)
\right] \label{eq:NLO}\\
\frac{1}{\tau}\frac{d\sigma^{\rm NLO}_{qg+gq}}{dQ^2} &=
\frac{\as}{2\pi}T_F
\int_\tau^1 \frac{dz}{z} \, 
\left[{\cal L}_{qg}\(\frac{\tau}{z}\) 
+{\cal L}_{gq}\(\frac{\tau}{z}\) \right]G_{\rm int}(z)
\end{align}
and
\bea
F_{\rm int}(z) &=& 2\, C_1(z) + 2\ln\frac{Q^2}{\muf^2} \plust{\frac{1+z^2}{1-z}}\\
G_{\rm int}(z) &=& \(z^2+(1-z)^2\) \( \ln\frac{(1-z)^2}{z} 
+ \ln\frac{Q^2}{\muf^2} \) +\frac{1}{2} + 3z -\frac{7}{2}z^2,
\eea
where $C_1(z)$ was given in Eq.~(\ref{eq:c1x})
and
\beq\label{eq:lum_z_tot}
{\cal L}_{ij}(z) = \int_z^1 \frac{dw}{w} \, 
\sum_{i,j} c_{ij}\, f_i^1(w)\, f_j^2\(\frac{z}{w} \).
\eeq

The NNLO rapidity distribution has been computed in Ref.~\cite{admp},
where its lengthy analytic expression can be found.

\subsection{Resummation}
\label{subs:resummation}

As already mentioned, large threshold logs only appear in the
quark--quark channel, while the contributions from other channels are
suppressed by at least one more power of $1-z$
as $z \to 1$ in comparison to  the $q\bar q$ contribution.
We therefore consider the $q\bar q$ term of Eq.~\eqref{eq:rap_distr},
\beq 
\frac{1}{\tau}\frac{d\sigma_{q\bar q}}{dQ^2 dY}=
\int_{x_1^0}^1 \frac{dx_1}{x_1} \int_{x_2^0}^1 
\frac{dx_2}{x_2}  \sum_q c_{q\bar q}\, f_q^1(x_1)\, f_{\bar q}^2(x_2) \,
C \( \frac{\tau}{x_1x_2},Y-\frac{1}{2}\ln\frac{x_1}{x_2} \),
\eeq
where $C(z,y)=C_{q\bar q}(z,y)$ is the quark-antiquark coefficient function,
which is independent of the flavour of the colliding quarks.

Threshold resummation of rapidity distributions is based on the
observation~\cite{Mukherjee:2006uu,bolz} (conjectured in
Ref.~\cite{Laenen:1992ey}) that at large $z$
the coefficient function $C(z,y)$ factorizes as
\beq
C(z,y)=C(z)\,\delta(y)\[1+\Ord(1-z)\],
\eeq
where $C(z)$ is the rapidity-integrated coefficient. This is easily
proved by rewriting
$C(z,y)$ in terms of its
Fourier transform with respect to $y$:
\beq
\tilde C(z,M) = \int_{-\infty}^{+\infty}dy\,e^{iMy}\,C(z,y).
\label{eq:fourier}
\eeq
The integration range in Eq.~\eqref{eq:fourier}
is restricted by kinematics to 
$\ln\sqrt{z}\leq y\leq-\ln\sqrt{z}$.
Hence, one may expand the exponential $e^{iMy}$ in powers
of $y$,
\beq
\tilde C(z,M)
=\int_{\ln\sqrt{z}}^{-\ln\sqrt{z}}dy\,C(z,y)\[1+\Ord(y)\]
=C(z)\[1+\Ord(1-z)\],
\eeq
where $C(z)$ is the
rapidity-integrated coefficient function. Hence,
$\tilde C(z,M)$ is independent
of $M$ up to terms which as $z\to1$ are suppressed by powers of 
$\abs{\ln z}=1-z+\Ord((1-z)^2)$, and one immediately gets the desired
factorized form:
\beq
C(z,y)=
\int_{-\infty}^{+\infty}\frac{dM}{2\pi}\,e^{-iMy}\,\tilde C(z,M)
=C(z)\,\delta(y)\[1+\Ord(1-z)\].
\eeq

Up to power-suppressed terms we can thus write
\beq
\label{eq:sigma_step2}
\frac{1}{\tau}\frac{d\sigma^{\rm res}}{dQ^2 dY} = 
\int_{x_1^0}^1 \frac{dx_1}{x_1} \int_{x_2^0}^1\frac{dx_2}{x_2}\,
\sum_q c_{q\bar q}\, f_q^1(x_1)\, f_{\bar q}^2(x_2)\,
\delta\(Y-\frac{1}{2}\ln\frac{x_1}{x_2}\)\,
C^{\rm res}\(\frac{\tau}{x_1x_2}\).
\eeq
Because Eq.~(\ref{eq:sigma_step2}) only depends on the 
rapidity-integrated coefficient
function, threshold resummation is simply performed by using for the
latter the resummed expressions discussed in Sect.~\ref{sec:resummation}.

Note that resummation of the $x_F$ distributions can be 
performed in the same way.
This is because the kinematical bounds on $Y$,
\beq
\abs{Y}\leq\log\frac{1}{\sqrt{\tau}}
\eeq
translates into 
\beq
x_F^2\leq(1-\tau)^2,
\eeq
and a similar relation holds at the partonic level. Hence the 
the above argument holds for the $x_F$ distribution as well.

The dependence on parton distributions in Eq.~(\ref{eq:sigma_step2}) 
can be easily rewritten in terms of the differential parton luminosity
Eq.~\eqref{qqbarlum} with fixed values of the momentum fractions
$x_i$. To this purpose, perform the change of variables
\beq
z=\frac{\tau}{x_1 x_2}\;,\qquad Y_{\rm cm}=\frac{1}{2}\ln\frac{x_1}{x_2},
\eeq
with the inverse
\beq
x_1=\sqrt{\frac{\tau}{z}}\,e^{Y_{\rm cm}}\;,\qquad 
x_2=\sqrt{\frac{\tau}{z}}\,e^{-Y_{\rm cm}}.
\eeq
The line $Y_{\rm cm}=Y$, selected by the delta function,
is contained in the integration region for all values of $z$ in
\beq
\tau e^{2\abs{Y}} \leq z \leq 1.
\eeq
Hence,
\begin{align}
\frac{1}{\tau}\frac{d\sigma^{\rm res}}{dQ^2 dY} 
&= \int_{\tau e^{2\abs{Y}}}^1 \frac{dz}{z}\,C^{\rm res}(z)
\int dY_{\rm cm}\,\delta(Y-Y_{\rm cm})
\sum_q c_{q\bar q}\, f_q^1\(\sqrt{\frac{\tau}{z}}e^{Y_{\rm cm}}\)\,
f_{\bar q}^2\(\sqrt{\frac{\tau}{z}}e^{-Y_{\rm cm}}\)
\nonumber\\
&= \int_{\tau e^{2\abs{Y}}}^1 \frac{dz}{z}\,C^{\rm res}(z)\,
L_{q\bar q}\(z,\frac{1}{2}\)
\label{eq:main_conv}
\end{align}
where we have used Eq.~\eqref{qqbarlum}. 
The integration range in Eq.~\eqref{eq:main_conv} can be extended
down to $z=\tau$ because  $\Lrap(z,1/2)$ vanishes for $z<\tau e^{2\abs{Y}}$:
\beq
\frac{1}{\tau}\frac{d\sigma^{\rm res}}{dQ^2 dY} 
= \int_\tau^1 \frac{dz}{z}\,C^{\rm res}(z)\,\Lrap\(z,\frac{1}{2}\).
\label{eq:main_conv2}
\eeq
By inspection of Eq.~\eqref{x1x2tozu}, we see that, for $u=1/2$, $x_1,x_2$
depend on $z$ through the ratio $\tau/z$.
Therefore, Eq.~\eqref{eq:main_conv2} has the form of a convolution
product: this greatly simplify its treatment, which is then analogous
to that of the resummed integrated cross-section, with the replacement 
\beq
{\cal L}_{q\bar q}\(\frac{\tau}{z}\)\to L_{q\bar q}\(z,\frac{1}{2}\).
\label{lumireplacement}
\eeq

The matched resummed expression for the 
Drell-Yan cross-section can finally be written in analogy to
Eq.~(\ref{eq:match}):
\begin{align}
\frac{d\sigma^{\text{N$^k$LL+N$^p$LO}}}{dQ^2 dY} 
&=\frac{d\sigma^{\text{N$^p$LO}}}{dQ^2 dY}
+\frac{d\sigma^{\rm res}}{dQ^2 dY} -
\sum_{j=0}^{p}\frac{\as^j}{j!}\[\frac{d^j}{d\as^j}\frac{d\sigma^{\rm res}}{dQ^2 dY}
\]_{\as=0}.\label{eq:difresum}
\end{align}

Before turning to the numerical implementation of the resummed result,
we discuss briefly a different way of relating the resummation of
rapidity distribution to that of the inclusive cross-section which
has been more recently presented in 
Ref.~\cite{Becher:2007ty}. This is based on writing the rapidity
distribution in terms of the variables $z,u$ defined in
Eqs.~(\ref{x1x2tozu},~\ref{zutox1x2}):
\beq
\frac{1}{\tau}\frac{d\sigma_{q\bar q}}{dQ^2\,dY} =
\int_\tau^1 \frac{dz}{z} \int_0^1 du \, \Lrap(z,u) \,
\bar C(z,u),
\eeq
with
\beq
\bar C(z,u)=\abs{\frac{\partial(\ln x_1,\ln x_2)}{\partial(\ln z,u)}}
C\(z,y(z,u)\).
\eeq
It is then observed that at NLO the logarithmically enhanced 
terms in the partonic threshold limit $z\to 1$ appear as coefficients of
the combination $\delta(u)+\delta(1-u)$, as one can check by inspection
of Eq.~\eqref{Fzu}.
At higher orders, logarithmic terms in general multiply non-trivial functions
of $u$, but Eq.~\eqref{x1x2tozu} implies that
the $u$ dependence of $x_1,x_2$ is of
order $1-z$, so
\beq
\bar C(z,u)
=\abs{\frac{\partial(\ln x_1,\ln x_2)}{\partial(\ln z,u)}}
C\(z,y(z,u)\)
=\[\delta(u)+\delta(1-u)\]\bar C(z)\[1+\Ord(1-z)\].
\label{Cbar}
\eeq
Hence
\beq
\frac{1}{\tau}\frac{d\sigma^{\rm res}}{dQ^2\,dY} =
\int_\tau^1 \frac{dz}{z}\, \bar C^{\rm res}(z)\[\Lrap(z,0) +\Lrap(z,1) \],
\eeq
where
\beq
\bar C^{\rm res}(z)=\frac{1}{2}\, C^{\rm res}(z)
\eeq
as it is easily seen integrating 
Eq.~\eqref{Cbar}
with respect to $u$ between 0 and 1 and noting that
\beq
\abs{\frac{\partial(\ln x_1,\ln x_2)}{\partial(\ln z,u)}}
\abs{\frac{\partial u}{\partial y}}=
\frac{z^2}{\tau}\abs{\frac{\partial u}{\partial y}
\(
\frac{\partial x_1}{\partial z}\frac{\partial x_2}{\partial u}
-
\frac{\partial x_2}{\partial z}\frac{\partial x_1}{\partial u}\)}
=1.
\eeq
It follows that
\beq
\frac{1}{\tau}\frac{d\sigma^{\rm res}}{dQ^2\,dY} =
\int_\tau^1 \frac{dz}{z}\, C^{\rm res}(z)\, \frac{\Lrap(z,0) +\Lrap(z,1)}{2}.
\label{eq:main_convn}
\eeq

Equation~(\ref{eq:main_convn}) differs 
by power suppressed terms from the resummed result 
previously derived Eq.~\eqref{eq:main_conv2}, as it is easy to check explicitly.
Indeed, using Eq.~\eqref{x1x2tozu}
and expanding
$L_{q\bar q}(z,0)$, $L_{q\bar q}(z,1)$ and $L_{q\bar q}(z,1/2)$ 
in powers of $z$ about $z=1$ it is easy to check that
\beq
\frac{\Lrap(z,0)+\Lrap(z,1)}{2}
=\Lrap\(z,\frac{1}{2}\)+\Ord((1-z)^2).
\eeq
Because the difference is suppressed by two powers of $1-z$ one
expects it to be small, and indeed we have checked that (using the
 Borel prescription) the difference between
Eqs.~\eqref{eq:main_conv2} and \eqref{eq:main_convn} is negligible, and
specifically much smaller than the difference between Borel and
minimal prescription.

We note that, because $\Lrap(z,0)$ and $\Lrap(z,1)$ are not functions
of $\tau/z$, the form  Eq.~\eqref{eq:main_convn} of the resummed
result does not have the structure of a convolution and thus would
require a separate numerical implementation. 
For the same reason, a comparison between
Eqs.~(\ref{eq:main_conv2}) and~\eqref{eq:main_convn} using the minimal
prescription cannot be performed, because 
Eq.~\eqref{eq:main_convn} cannot be expressed in terms
of the Mellin transform of $C^{\rm res}(z)$.
We will disregard  the
form Eq.~\eqref{eq:main_convn} of rapidity distributions henceforth.

\subsection{Numerical implementation}

For the sake of phenomenology, an efficient numerical implementation 
of resummed results using the various prescriptions is necessary.
Such an implementation was hitherto not available and it will be
discussed here.

The minimal prescription
involves the numerical evaluation of the complex integral
\beq
\label{eq:main2}
\frac{1}{\tau}\frac{d\sigma^{\rm res}}{dQ^2 dY}
=\frac{1}{2\pi i} \int_{c-i\infty}^{c+i\infty}dN\,\tau^{-N} 
L_{q\bar q}\(N,\frac{1}{2}\)\,C^{\rm res}(N,\as)
\eeq
where the integration path is usually chosen as in Fig.~\ref{fig:MP_path}
in order to make the integral absolutely convergent.
\begin{figure}[tbp]
\centering
\includegraphics[scale=0.3]{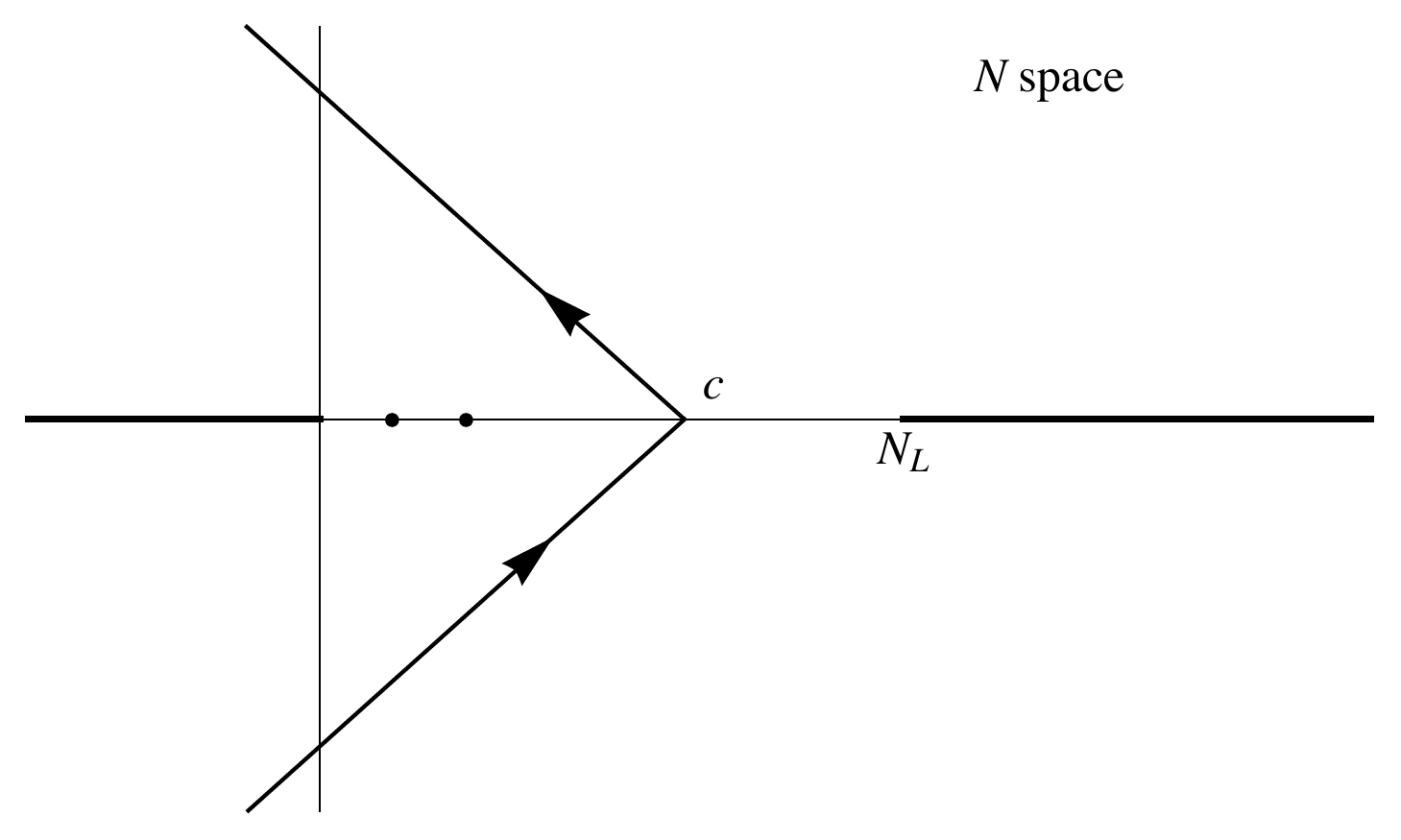}
\caption{The analytic structure of the integrand of \eqref{eq:main2} 
and the minimal prescription path.}
\label{fig:MP_path}
\end{figure}
However, parton densities obtained from data analysis are commonly
available  as functions of $z$ in
interpolated form through common interfaces such as
LHAPDF~\cite{Bourilkov:2006cj}, and the numerical 
evaluation of their Mellin transform does not converge
along the path of integration (specifically, for 
${\rm Re}\;N<0$) and must be defined by analytic
continuation.
The option of  applying the MP
to the partonic cross-section, and then convoluting the result with
the parton luminosity in momentum space is not viable, because
the MP does not have the structure of a convolution: the
partonic cross-section does not vanish for $z\ge1$, and it oscillates wildly
in the region $z\sim 1$. 
This problem is discussed in Ref.~\cite{cmnt}, where it is handled by
adding and subtracting the results of the minimal prescription
evaluated with a fake luminosity which allows for analytic integration.

Another possibility, adopted for example in
Ref.~\cite{bolz}, is to use parton distributions whose Mellin
transform can be computed exactly at the initial scale. This, however,
greatly restricts the choice of parton distributions, and specifically
it prevents the use of current state-of-the-art PDFs from global fits.
It is  thus not
suitable for precision phenomenology.

The method adopted here, based on an idea suggested long
ago~\cite{Furmanski:1981ja}, consists of expanding the function 
$\Lrap(z,1/2)$ (or $\Lum_{q\bar q}(z)$ for the inclusive cross-section)
on a basis of polynomials whose Mellin transform can be computed
analytically. We have chosen Chebyshev polynomials, for which
efficient algorithms for the computation of the expansion coefficients
are available.  The details of the procedure are illustrated in
Appendix~\ref{app:cheb}. The obvious drawback of this
procedure is that it must be carried on for each value of the scale
$\muf$ and, in the case of rapidity distribution, for each value of
$\tau$ and $Y$.

We now turn to the discussion of an implementation issue which is
specific to the Borel prescription (BP), and has to do with the choice
of the cutoff $C$. As discussed in Sect.~\ref{sec:borel} the 
minimal choice is $C=2$; however
when $C\ge 1$
the $\xi$ integration path in Eq.~(\ref{sigmaBP}) 
includes values of $\xi$
with ${\rm Re}\;\xi<-1$, for which the convolution integral diverges.
As discussed  in Ref.~\cite{afr}, the integral can be nevertheless
defined by analytic continuation, by subtracting and adding back from
${\cal L}_{q\bar q}(\tau/z)/z$ its Taylor expansion in $z$ around $z=1$: the
subtracted  integrals converge, and the compensating terms can be
determined analytically and continued in the desired region.

Here we propose a different method which is numerically much
more efficient. The idea is to perform the convolution integral 
with the luminosity analytically, before 
the complex $\xi$ integral of Eq.~\eqref{sigmaBPp}. 
When ${\rm Re\;}\xi>0$ we can use the identity
\beq\label{eq:BP_dependence_z}
\plus{(1-z)^{\xi-1}} = (1-z)^{\xi-1} - \frac{1}{\xi}\delta(1-z),
\eeq
under the integral, 
because the two ensuing integrals are separately convergent. We get
\beq
\int_{\tau}^1 \frac{dz}{z}\,\plus{(1-z)^{\xi-1}}z^{-\frac{\xi}{2}}
{\cal L}_{q\bar q}\(\frac{\tau}{z}\)
= \int_{\tau}^1 dz \,
\(\frac{1-z}{\sqrt{z}}\)^\xi  g(z,\tau) 
-{\cal L}_{q\bar q}(\tau)\[\frac{1}{\xi}-\int_\tau^1dz\,
\frac{(1-z)^{\xi-1}}{z^{\frac{\xi}{2}}}\]
\label{eq:conv_borel}
\eeq
where we have defined
\beq\label{g_zx}
g(z,\tau)= \frac{1}{1-z}
\left[ 
\frac{1}{z}\,{\cal L}_{q\bar q}\(\frac{\tau}{z}\)-{\cal L}_{q\bar q}(\tau)\right].
\eeq

The function $g(z,\tau)$ is more easily approximated by
an expansion on the basis
of Chebyshev polynomials than ${\cal L}_{q\bar q}\(\frac{\tau}{z}\)$.
Proceeding as in  Appendix~\ref{sec:cheb_borel} we find
\beq
g(z,\tau) = \sum_{p=0}^n b_p(1-z)^p,
\eeq
where $n$ is 
the order of the approximation, and the coefficients
$b_p = b_p(\tau,\muf^2)$ can be computed by numerical methods.
We have
\begin{multline}\label{eq:BPnew_final}
\frac{1}{\tau}\frac{d\sigma^{\rm res}}{dQ^2} = 
\frac{1}{2\pi i}
\oint \frac{d\xi}{\xi}\frac{1}{\Gamma(\xi)}
\int_0^C \frac{dw}{\ab} \, e^{-\frac{w}{\ab}}\Sigma\left(\frac{w}{\xi}\right)\\
\left[\sum_{p=0}^n b_p\mathrm{B}\(\xi+p+1,1-\frac{\xi}{2};1-\tau\) -
{\cal L}_{q\bar q}(\tau) \left[\frac{1}{\xi}-\mathrm{B}
\(\xi, 1-\frac{\xi}{2}; 1-\tau\) \right] \right],
\end{multline}
where
\beq
\mathrm{B}(b,a;1-\tau)=\int_\tau^1 dz\, z^{a-1}(1-z)^{b-1}
\eeq
is the incomplete Beta function.
The function $\mathrm{B}(b,a,1-\tau)$ is singular at $b=0$.
In Eq.~\eqref{eq:BPnew_final}, the first argument of the B functions
in the integrand vanishes for non positive integer values of
$\xi=0,-1,\dots$, in correspondence of zeros of
$1/\Gamma(\xi)$. Thus, the $\xi$ integrand has only a branch cut 
in $-w\leq\xi\leq 0$.
This expression is in fact valid for
arbitrarily large values of $C$, hence it can also be used in the case $C>2$,
which requires multiple subtractions if the method of Ref.~\cite{afr}
is used.

The rapidity distribution is obtained by repeating the above procedure
with
\beq
{\cal L}_{q\bar q}\(\frac{\tau}{z}\)\to L_{q\bar q}\(z,\frac{1}{2}\).
\eeq
Since $L(z,\frac{1}{2})$
vanishes for $z<\tau e^{-2\abs{Y}}$, the convolution integral is
actually restricted to the region $\tau e^{2\abs{Y}} \leq z \leq 1$.
This difference amounts to replacing $\tau$ by $\tau e^{2\abs{Y}}$:
\begin{multline}
\frac{1}{\tau}\frac{d\sigma^{\rm res}_k}{dQ^2 dY} = 
\frac{1}{2\pi i}
\oint \frac{d\xi}{\xi}\,\frac{1}{\Gamma(\xi)}
\int_0^C \frac{dw}{\ab} \, e^{-\frac{w}{\ab}}\Sigma\left(\frac{w}{\xi}\right)\\
\Bigg[\sum_{p=0}^n b_p\mathrm{B}\(\xi+p+1,1-\frac{\xi}{2};1-\tau e^{2\abs{Y}}\)\\
- L_{q\bar q}\(1,\frac{1}{2}\) \left[\frac{1}{\xi}-\mathrm{B}
\(\xi, 1-\frac{\xi}{2}; 1-\tau e^{2\abs{Y}}\) \right] \Bigg].
\end{multline}

\eject
\section{Phenomenology}
\label{sec:pheno}

We will now turn to a phenomenological assessment of the impact of
resummation for Drell-Yan invariant mass and rapidity
distributions. As discussed in Sect.~\ref{sec:intro}, threshold
resummation is not currently included in the calculations of
fixed-target and collider Tevatron Drell-Yan data which are included
in global parton fits such as MSTW08~\cite{Martin:2009iq},
NNPDF2.0~\cite{Ball:2010de}, or CT10~\cite{Lai:2010vv}; neither it has
been so far included in predictions for the LHC, such as those of
Ref.~\cite{Balossini:2009sa}.

We will therefore consider three cases: $pp$ collisions at a
center-of-mass energy of 38.76~GeV, which corresponds to the case of
the experiment E866/NuSea, taken as representative of Tevatron
fixed-target experiments; $p\bar p$ collisions at the Tevatron
collider energy of $\sqrt{s}=1.960$~TeV; and the LHC case, $pp$
collisions at the intermediate energy of $\sqrt{s}=7$~TeV and at the design
energy $\sqrt{s}=14$~TeV. For the Tevatron and LHC configurations, we will
consider both charged ($\ell\bar\nu$) and neutral ($\ell\bar\ell$)
Drell-Yan pairs, taking into account in the latter case the
interference between virtual $Z$ and $\gamma$. Lepton masses will
always be neglected. We will show results for both the invariant mass
distribution as a function of $\tau=Q^2/s$, and for the
doubly-differential distribution in invariant mass and rapidity as a
function of the rapidity for fixed values of $\tau$.
Specifically, for invariant mass distributions we will show results for the
$K$-factor defined as
\beq
K\(\tau, \frac{\muf^2}{Q^2}, \frac{\mur^2}{Q^2}\) =
\frac{\dfrac{d\sigma}{dQ}\(\tau,\dfrac{\muf^2}{Q^2}, 
\dfrac{\mur^2}{Q^2}\)}{\dfrac{d\sigma^{\rm LO}}{dQ}(\tau,1,1)},
\eeq
where $\mu_F$ and $\mu_R$ are the factorization and renormalization
scale, respectively. Since we will be considering
different experimental configurations, the results for the $K$-factors
will be shown for fixed value of $s$, with $Q^2$ determined by the value
of $\tau$, $Q^2=\tau s$.
The Born cross-section ${\dfrac{d\sigma^{\rm LO}}{dQ}(\tau,1,1)}$,
which provides the scale for these plots, is
shown in Fig.~\ref{fig:born} for LHC at $7$~TeV
and Tevatron at $1.96$~TeV.
\begin{figure}[htb]
\begin{center}
\includegraphics[width=0.49\textwidth]{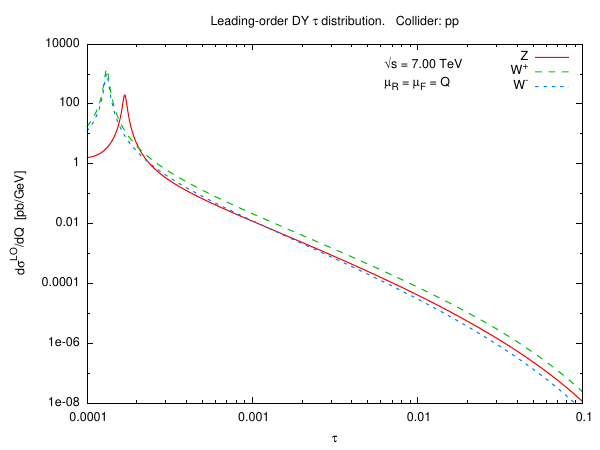}
\includegraphics[width=0.49\textwidth]{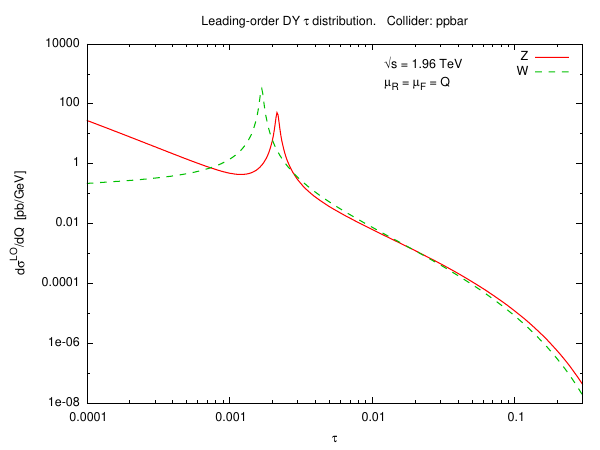}
\caption{The  
invariant mass distribution of charged and neutral Drell-Yan pairs
in $pp$ collisions at $\sqrt{s}=7$~TeV (left) and in $p\bar p$
collisions at $\sqrt{s}=1.96$~TeV (right) computed at leading order with
NNPDF2.0 parton distributions at $\as(M_Z^2)=0.118$.
}
\label{fig:born}
\end{center}
\end{figure}

Our aim will be to assess the potential impact of inclusion of
resummation effects on cross-sections and their associate perturbative
uncertainty, both on experiments which are already used for PDF
determinations (and thus, potentially, on the PDF extraction itself),
as well as on future LHC measurements, both for real $W$ and $Z$
production as well as for high-mass 1~TeV Drell-Yan pair
(relevant for instance
as a background to hypothetical $Z^\prime$ production). 
For each observable, we will show fixed-order predictions at leading,
next-to-leading and next-to-next-to-leading order, and,
correspondingly, leading, next-to-leading and next-to-next-to-leading
log resummed curves.

All curves will be computed using a fixed (NLO)
set of parton distributions. In a realistic situation, parton
distributions would be refitted each time at the corresponding
perturbative order; the effect of the perturbative corrections on the
hard cross-section is then partly reabsorbed in the PDFs (with fixed
experimental data), and the effect on the Drell-Yan process 
gets tangled with the effect on other processes which are used for PDF
fitting. Hence, a comparative assessment of size of various
perturbative corrections on cross-sections and uncertainties, which is
our main aim here, can only be done with fixed PDF. It should be born
in mind, however, that our predictions will only be fully realistic
when considering the NLO case.

In order to assess perturbative uncertainties, we will perform
standard variations of factorization and renormalization scales, and
furthermore in order to assess the ambiguities related to the
resummation procedure we will compare results obtained with the
minimal and Borel prescriptions, as discussed in
Sect.~\ref{sec:resummation}: specifically, for the Borel prescription
we will use the modified BP$^\prime$ Eq.~(\ref{sigmaBPp}), which
provides a more moderate but realistic estimate of ambiguities as
discussed in Sect.~\ref{sec:subl}, and take $C=2$ (see
Sect.~\ref{sec:borel}). Note that we have checked explicitly that also
at the hadronic level curves obtained using the Borel
prescription~\eqref{SBP2}, which contains logarithms of
$\frac{1}{z}$, are indistinguishable from those obtained with the
minimal prescription, in agreement with our discussion in
Sect.~\ref{sec:subl}.

Other sources of uncertainty will be discussed briefly in
Sect.~\ref{sec:thunc}, where we will provide an overall assessment of
uncertainties related to the value of the strong coupling and to the
parton distributions, and then present and evaluate critically the use
of scale variation to assess perturbative uncertainties. In the
remainder of this Section detailed predictions for the three cases of
interest will be presented.

All the numerical results of this Section are obtained using the \texttt{C++}
code \texttt{ReDY} (Resummed Drell-Yan)~\cite{ReDY}.
The computation of fixed-order cross-sections in this code relies on the
\texttt{Vrap} code by L.~Dixon~\cite{Vrap}, supplemented by 
soft-gluon resummation described in the previous Sections.

\subsection{Uncertainties}
\label{sec:thunc}

Theoretical predictions for the Drell-Yan process are affected by a
number of uncertainties, related to the treatment of both the strong
and electroweak interactions. 
Of course, in a realistic experimental situation further
uncertainties arise because of the need to introduce kinematic cuts,
which in turns requires comparing to fully exclusive
calculations~\cite{Catani:2009sm}.
Here, we will make no attempt to
estimate the latter, nor electroweak uncertainties 
and their interplay with strong corrections
(see e.g. Ref.~\cite{Balossini:2009sa} for a
recent discussion), and we will concentrate on uncertainties related to the treatment of the strong
interactions. 
 Before turning to an assessment of the way higher order
corrections can be estimated from scale variation, we discuss
uncertainties related to the value of the strong coupling and to the
choice of parton distributions (PDFs).

\subsection{Uncertainties due to the parton distributions and $\as$}
\label{alphapdf}

The uncertainty due to PDFs is usually dominant in hadron
collider processes. Tevatron Drell-Yan data are used for PDF
determination, so PDF uncertainties here reflect essentially the
current theoretical uncertainty in knowledge of this process, as
well as possible tension between Drell-Yan data 
and other data which go into global PDF
fit (which however seems~\cite{Ball:2010de} to be very moderate).
Predictions for the LHC are affected by sizable PDF uncertainties
because of the need to extrapolate to a new kinematical region, and
also, in the case of Drell-Yan, because at the LHC, unlike at the
Tevatron, one of the two PDFs which enter the leading-order process
is sea-like. 

PDF uncertainties for the invariant mass distribution of 
neutral Drell-Yan pairs at $\sqrt{s}=7$~TeV are shown in
Fig.~\ref{fig:totLHCpdf} as a function of $\tau=Q^2/s$. We use NNPDF2.0
PDFs with $\as(M_Z^2)=0.118$; other PDF sets are expected to give
similar results~\cite{PDF4LHC}. Because we are using a fixed PDF set,
the  uncertainty does not depend
significantly on the perturbative order, or the inclusion of resummation.
\begin{figure}[htb]
\begin{center}
\includegraphics[width=0.7\textwidth]{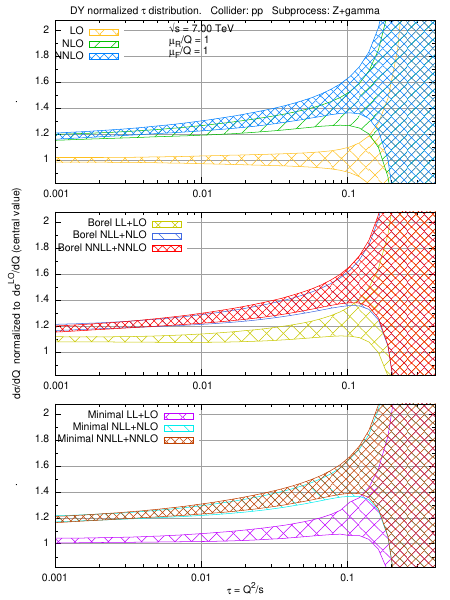}
\caption{
Invariant mass distribution of neutral Drell-Yan pairs
in $pp$ collisions at $\sqrt{s}=7$~TeV. The
band corresponds to the PDF uncertainty, using the
NNPDF2.0 set with $\as(M_Z^2)=0.118$.
}
\label{fig:totLHCpdf}
\end{center}
\end{figure}
It ranges between 5\% and 15\% for $\tau\lesssim
0.1$. For larger values of $\tau$ the cross-section becomes
essentially undetermined, because there are no data in PDF global fits
to constrain PDFs in that region: the few available large--$x$ data
are at much lower scale, and the uncertainty due to lack of
information at very large $x\gtrsim0.5$ contaminates PDFs down to
$x\gtrsim0.1$ when evolving up to the LHC scale.
Note however that the Drell-Yan cross-section at large
$\tau\gtrsim0.1$ rapidly drops to unmeasurably small values (see
Fig.~\ref{fig:born}). The fact that PDF uncertainties blow up for
$\tau\gtrsim0.1$ implies that data in this region would allow a
determination of PDFs in a region where they are currently almost
unknown; conversely, any signal of new physics in this region would
have to be validated by measurements in an independent channel (such
as for example jet production) which provides an independent
constraint on the relevant PDFs.

In Fig.~\ref{fig:nnpdf} we show the PDF uncertainties for the rapidity
distribution of neutral Drell-Yan pairs with $Q=1$~TeV at
$\sqrt{s}=7$~TeV, again using the NNPDF2.0 set with $\as(M_Z^2)=0.118$. As in
the previous case, the PDF uncertainty does not depend significantly
on the perturbative order, and it is typically larger than 5\%.
\begin{figure}[htb]
\begin{center}
\includegraphics[width=0.7\textwidth]{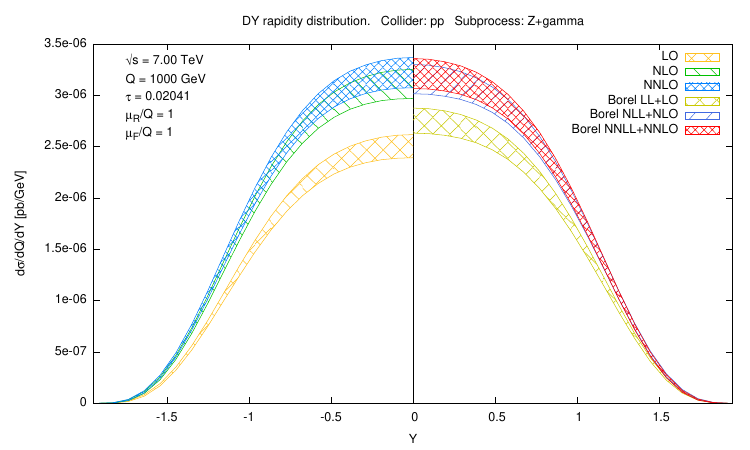}
\caption{
  Rapidity distribution of neutral Drell-Yan pairs with invariant
  mass $Q=1$ TeV in $pp$ collisions at $\sqrt{s}=7$~TeV. Unresummed
  results are shown for negative $Y$ and resummed results for positive
  $Y$.  The band corresponds to the PDF uncertainty, obtained by the
  NNPDF2.0 set with $\as(M_Z^2)=0.118$.}
\label{fig:nnpdf}
\end{center}
\end{figure}

We now turn to the uncertainty due to the value of $\as$.
The current PDG~\cite{PDG} value for $\as(M_Z^2)$ is taken from
Ref.~\cite{Bethke:2009jm} and it is $0.1184\pm 0.0007$. However, this
uncertainty seems quite small, especially when taking into account the
fluctuation in central values between the determinations that go into
it, and the dependence on the perturbative order of some of them: indeed,
current recommendations for precision LHC studies from the PDF4LHC
group~\cite{PDF4LHC} advocate a rather more conservative uncertainty
estimate. We thus take
\beq
\label{alphaband}
\as(M_Z^2)=0.118\pm 0.002
\eeq
as a reasonable current range.

The impact of this uncertainty on $\as$ on the Drell-Yan cross-section
at $\sqrt{s}=7$~TeV can be estimated from Fig.~\ref{fig:tot7alpha},
\begin{figure}[htb]
\begin{center}
\includegraphics[width=0.7\textwidth]{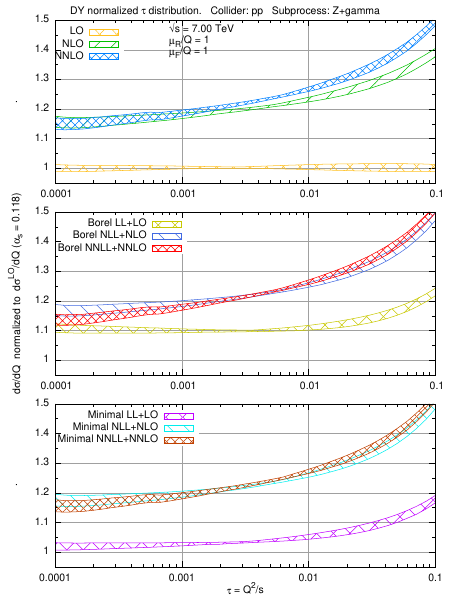}
\caption{
  Invariant mass distribution of neutral Drell-Yan pairs in $pp$
  collisions at $\sqrt{s}=7$~TeV.  The uncertainty bands corresponds
  to a variation of  $\as(M_Z^2)$ in the range $0.116$ to $0.120$ in the hard
  matrix element and in the parton distributions;  NNPDF2.0 parton
  distributions are used.}
\label{fig:tot7alpha}
\end{center}
\end{figure}
where we show the effect on the inclusive Drell-Yan cross-section due
to variation in the range Eq.~(\ref{alphaband}) of the value of
$\as(M_Z^2)$ used both in the computation of the hard matrix element and
the PDF evolution. Results are only shown for $\tau<0.1$, because for
larger value the PDF uncertainty blows up and results loose
significance,  as discussed above.
Note that this full dependence of the physical
cross-section on $\as$ is in general somewhat different from
the dependence of the hard matrix element alone, because of the
dependence on $\as$ of the relevant parton luminosity. This total
dependence might be larger or smaller according to whether the
luminosity is correlated or anticorrelated to the value of $\as$,
either of which might be the case for a quark luminosity, according to
the kinematic region~\cite{Ball:2010de}.  A priori, the size of the
uncertainty due to variation of $\as$ in the matrix element and
that due to the dependence on $\as$ of the PDFs are likely to
be comparable: after all, the Drell-Yan rapidity distribution plays a
significant role in the determination of the PDFs themselves.

It appears from Fig.~\ref{fig:tot7alpha} that the $\as$ uncertainty
increases with the perturbative order, but it is of similar size at
the resummed and unresummed level; at NNLO it is of order of $\sim
1.5$\% at LHC energies; we have checked that it is about a factor two
larger at Tevatron fixed-target experiments. The uncertainty due to
$\as$ on rapidity distributions is clearly of comparable size.

Noting that within the approximation of linear error propagation the
PDF and $\as$ uncertainties should be combined 
in quadrature~\cite{Lai:2010nw}, we conclude that PDF uncertainties
are somewhat larger than $\as$ uncertainties and the combined
effect of PDF and $\as$ uncertainties is likely to be smaller
than about 10\% but not much smaller, at least in the region in which
PDFs are constrained by presently available data. 
Once PDF uncertainties will be
reduced due to LHC data, it should be possible to keep the combined
effect of these uncertainties at the level of few percent. Therefore,
perturbative accuracies at the percent level are relevant for precision
phenomenology.

\subsection{Perturbative uncertainties: scale variations}
\label{scales}

A standard way of estimating unknown higher order perturbative
corrections is to vary factorization and renormalization scales. We
perform this variation 
by writing the factorized expression Eq.~(\ref{factN}) as
\beq
\label{factNscal}
\sigma(N,Q^2)={\cal L}(N,\muf^2)\;
\hat\sigma\(N,\as(\mur^2),\frac{Q^2}{\muf^2},\frac{Q^2}{\mur^2}\),
\eeq
which is independent of $\muf^2$ and $\mur^2$ at the
order at which the partonic coefficient 
$\hat\sigma\(N,\as(\mur^2),\frac{Q^2}{\muf^2},\frac{Q^2}{\mur^2}\)$
is computed.
The residual scale dependence is therefore of the first neglected
order, and can be used as an estimate of the higher order terms in
the perturbative expansion.
We vary the two scales in the range
\beq\label{eq:scale_variation}
\abs{\ln\frac{\muf}{Q}} \leq \ln 2 ,\qquad
\abs{\ln\frac{\mur}{Q}} \leq \ln 2 ,\qquad
\abs{\ln\frac{\mur}{\muf}} \leq \ln 2 ,
\eeq
depicted in Fig.~\ref{fig:scales}, which guarantees that both
higher-order corrections to the partonic cross-section and to perturbative
QCD evolution are generated, with the last condition ensuring that no
artificially large scale ratios are introduced.
\begin{figure}[tbp]
  \centering
  \includegraphics[scale=1]{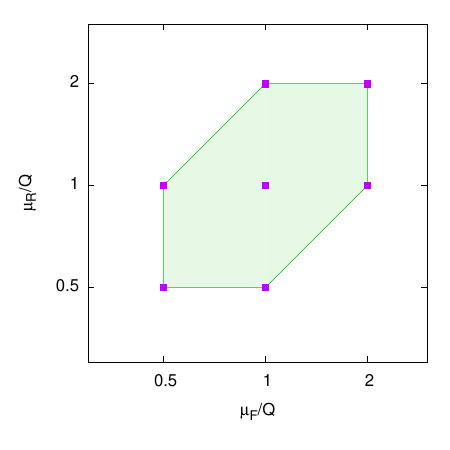}
  \caption{Scale variation grid; the cross-section is computed
    in correspondence of the purple dots.}
  \label{fig:scales}
\end{figure}

In the sequel, we will perform scale variation of both unresummed and
resummed cross-sections. The interpretation of results deserves a
comment. When performing scale variation of a result determined at
fixed $\Ord(\as^k)$, one generates terms of $\Ord(\as^{k+1})$:
consequently, the scale uncertainty is reduced as
one increases the perturbative order. However, terms generated by
scale variation are proportional to those which are present at the
given order: therefore, scale variation underestimates the size of higher
order corrections when these are enhanced by higher logarithmic
powers. For instance, scale variation of  the $\Ord(\as)$
Drell-Yan coefficient function $C_1(N)$ Eq.~(\ref{eq:c1}) produces terms which
at large $N$ grow at most as $\ln^2 N$, whereas the actual 
$\Ord(\as^2)$ $C_2(N)$ coefficient function at large $N$ grows as
$\ln^4 N$. Hence, if $N$ is so large that these terms dominate the
coefficient functions and must be resummed to all orders their impact
might be rather larger than the scale variation of the fixed-order
result may suggest. Nevertheless, in this case the scale dependence of the resummed
result will still be smaller than that of the fixed-order result
because the resummed result includes the dominant contributions to the
cross-section to all orders.

However, in Sect.~\ref{sec:resDY} we have seen that there is an
intermediate kinematic region in which logarithmically enhanced
contributions may provide a sizable fraction of the coefficient
function even though $\as\ln^2 N\ll 1$: in this case, the resummation
improves the fixed-order result in that it includes a sizable fraction
of the higher order correction, but it still behaves in a perturbative
way, i.e.~terms of higher order in $\as$ included through the
resummation give an increasingly small contribution. If so, 
the scale dependence of the resummed result may well be
comparable to or even larger than that of the fixed-order
result, because the resummation amounts to the inclusion of large
terms in the next few higher orders, which are not necessarily seen when
performing the scale variation of the lower orders. Furthermore, 
resummation
only affects the quark channel, while fixed-order scale variation
mixes the quark and gluon channels: in an intermediate region, the
logarithmic terms in the quark channel may be sizable, but with the
gluon channel not being entirely negligible.  In such case, the scale
variation is dominated by subleading terms and thus we expect the
residual scale dependence of the resummed result to differ according to the
resummation prescription. We will see that this is indeed the case for
resummation of Tevatron rapidity distributions, with scale variation
of resummed results different according to whether the Borel or
minimal prescription is used.

\subsection{Tevatron at fixed target: NuSea}
\label{fixedtarget}

\begin{figure}[htb]
\begin{center}
\includegraphics[width=0.7\textwidth]{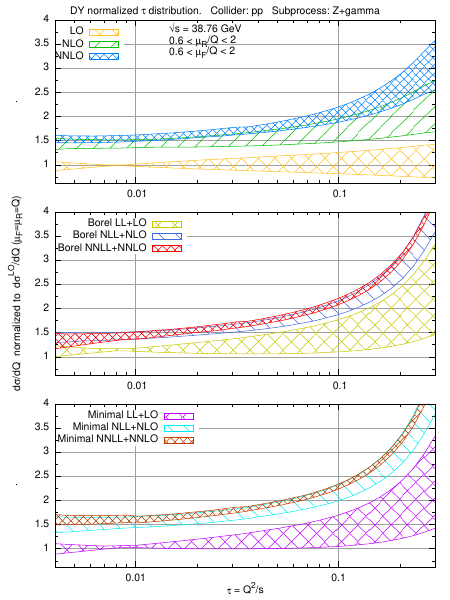}
\caption{Invariant mass distribution of neutral Drell-Yan pairs
in $pp$ collisions at $\sqrt{s}=38.76$~GeV.}
\label{fig:totft}
\end{center}
\end{figure}
We begin by studying the invariant mass distribution of lepton pairs
produced by collisions of a proton beam of energy
$E=800$~GeV on a proton or deuteron target, at rest in the laboratory
($\sqrt{s}=38.76$~GeV). This is the experimental
configuration of the experiment E866/NuSea.
We first consider the inclusive invariant mass distribution. Results
are shown in Fig.~\ref{fig:totft}. All uncertainties shown here and
henceforth are due to scale variation as described in Sect.~\ref{scales}.
As expected, the width of the error bands decreases
with increasing perturbative order. Note that for sufficiently small
$\tau$ the uncertainty blows up, due to the fact that for fixed $s$
the small $\tau$ limit corresponds to low scale: for example, at this energy
$\tau=10^{-3}$ corresponds to $Q\approx 1.2$~GeV, and varying the scales
as in Eq.~\eqref{eq:scale_variation} the values $\mur,\muf\approx 0.6$~GeV
are reached.

Turning now to the
resummed results, we note that the numerical impact of
resummation is large for $\tau \gtrsim 0.1$, while for
$0.03\lesssim\tau\lesssim0.1$ it is moderate but sill not negligible.
Furthermore, starting with the NLL level,
the scale uncertainty band for resummed results is dramatically
smaller than in the case of fixed-order results. This is because scale
variation of the LL result produces NLL terms which beyond the first
few orders are not contained in the fixed-order result; starting with
NLL these terms are already included in the resummed result.
It is interesting to
note that in the case of the resummed cross-section (with both
prescriptions) the NNLL band is almost entirely contained in the NLL
band, while the fixed-order NLO and NNLO error bands are only
marginally compatible with each other.  The ambiguity in the
resummation, estimated from the difference
between Borel
and minimal prescription, is not negligible,
but smaller than the scale uncertainty;
moreover, it is more evident at small $\tau$,
since the different subleading terms
give a larger contribution in that region.

\begin{figure}[htb]
\begin{center}
\includegraphics[width=0.7\textwidth,page=1]{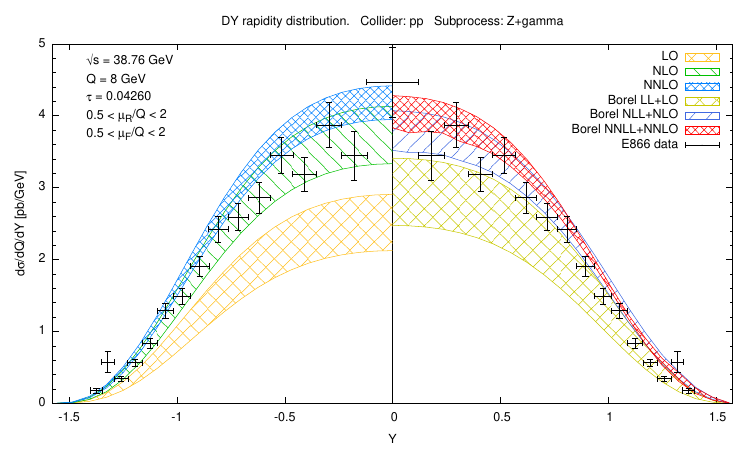}
\\
\includegraphics[width=0.7\textwidth,page=2]{pheno_plots/graph_rap_pp_nusea}
\caption{Rapidity distribution of neutral Drell-Yan pairs of invariant
mass $Q=8$~GeV in $pp$ collisions
at $\sqrt{s}=38.76$~GeV; E866 data are also shown.}
\label{fig:rapft}
\end{center}
\end{figure}
The experiment E866/NuSea~\cite{e866} has measured the distribution in
$x_F$ Eq.~(\ref{xfdef}) of lepton pairs with an invariant mass
$Q=8$~GeV. 
The E866 data are displayed in Fig.~\ref{fig:rapft}, superimposed to
the QCD prediction and the corresponding scale uncertainty.
The distribution is symmetric about $Y=0$; the curves shown for $Y<0$ refer
to fixed-order calculations, and those with $Y>0$ to resummed results. 
The data agree with the NLO calculation because these data were
included in the determination of the  NLO PDFs that we are using.

The impact of the resummation is small but not negligible: for
instance the difference between NNLO and NNLL is about half of the
difference between NNLO and NLO. Furthermore, the scale uncertainty of
resummed results is somewhat smaller than that of the unresummed ones.
This is consistent with the
observation that for this experiment $\tau=0.04$, which, as 
discussed in Sect.~\ref{sec:saddlepoint}, is in the region
in which resummation is relevant. 
However, the difference between resummed results obtained using the
Borel and the minimal prescription is almost as large as the size of
the resummation itself: in fact the NLL Borel results is somewhat
lower than the NNLO one, while the NLL minimal prescription
result is a bit higher.
Hence we conclude that the overall impact of the resummation on these
data is essentially negligible. A large and negative NLL resummed
correction to the NLO result was claimed in
Ref.~\cite{bolz}, using the minimal prescription, but we do not confirm it:
we find a positive and rather smaller correction.
The result of Ref.~\cite{bolz} was first criticized in 
Ref.~\cite{Becher:2007ty}.
Our result with the Borel prescription is
in good quantitative agreement with Ref.~\cite{Becher:2007ty};
however,
the minimal prescription gives a somewhat larger correction, though still
positive.

\subsection{Tevatron collider}

We now turn to Drell-Yan production in $p\bar p$ collisions
at a center-of-mass energy $\sqrt{s}=1.96$~TeV. 
Results for the invariant mass distribution  of neutral and
charged Drell-Yan pairs in this configuration are shown
in Fig.~\ref{fig:totTev}.
\begin{figure}[htbp]
\begin{center}
\includegraphics[width=0.496\textwidth]{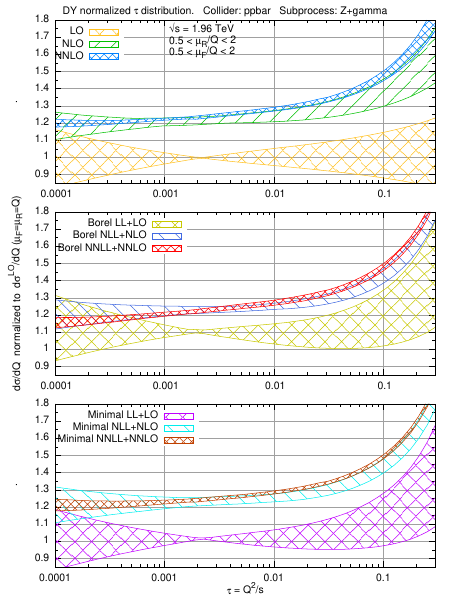}
\includegraphics[width=0.496\textwidth]{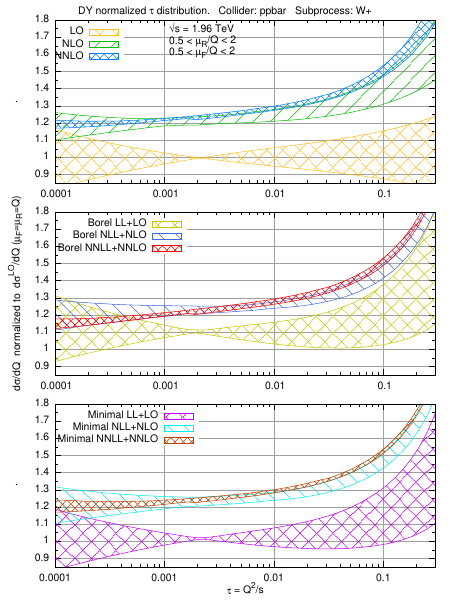}
\caption{Invariant mass distribution of neutral (left) and charged 
(right) Drell-Yan pairs
in $p\bar p$ collisions at $\sqrt{s}=1.960$~TeV.}
\label{fig:totTev}
\end{center}
\end{figure}

The behaviour of these curves is similar to that seen in the case of NuSea,
Fig.~\ref{fig:totft}, but with the impact of the resummation yet a
bit smaller, as one would expect both because of the higher energy and
because of the collider configuration, as discussed in
Sect.~\ref{sec:ipdfdy} (in particular Fig.~\ref{fig:sd1}).
Interestingly, even when the resummation has a very small impact, it
still leads to a non-negligible reduction of the uncertainty: this 
is consistent with the expectation
that for these medium-small values of $\tau$ 
resummation is in fact a perturbative
correction, as discussed in the end of Sect.~\ref{sec:resDY}.
Note that in these plots the smallness of leading-order 
uncertainty bands  when $\tau\approx 0.002$
(i.e.~$Q^2\approx 100$~GeV$^2$)
is due to the
fact that the scale dependence of the parton luminosity, to which the
leading-order cross-section is proportional (see Eq.~(\ref{sigmaNLOqqbar})), is
stationary at this scale.

\begin{figure}[htbp]
\begin{center}
\includegraphics[width=0.7\textwidth,page=1]{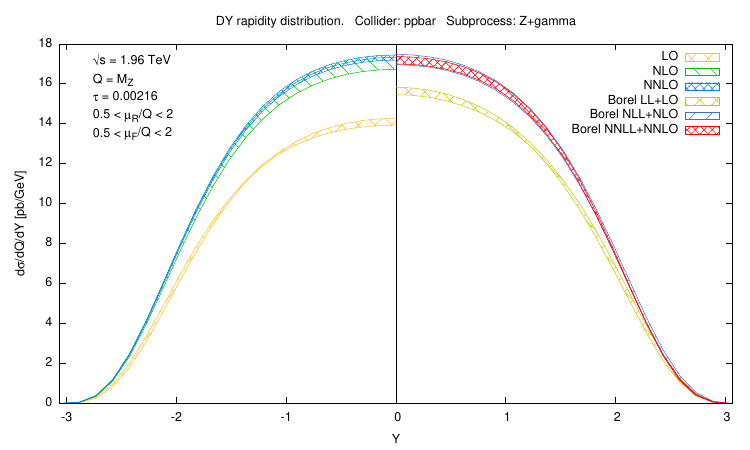}
\\
\includegraphics[width=0.7\textwidth,page=2]{pheno_plots/graph_rap_ppbar_196_Z}
\caption{Rapidity distribution of neutral Drell-Yan pairs of invariant
  mass $Q=M_Z$ in $p\bar p$ collisions at $\sqrt{s}=1.96$~TeV
  (the contribution of virtual $\gamma$ at the $Z$ peak is included).}
\label{fig:rapTEVZ}
\end{center}
\end{figure}
\begin{figure}[htbp]
\begin{center}
\includegraphics[width=0.7\textwidth,page=1]
{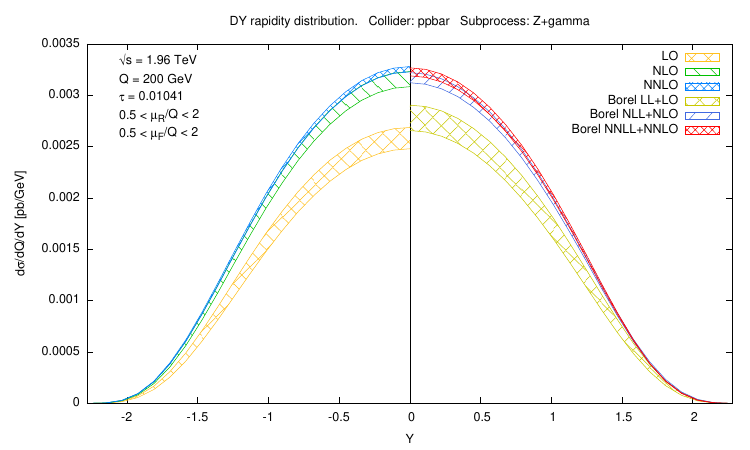}
\\
\includegraphics[width=0.7\textwidth,page=2]
{pheno_plots/graph_rap_ppbar_196_200}
\caption{Rapidity distribution of neutral Drell-Yan pairs of invariant
  mass $Q=200$~GeV in $p\bar p$ collisions at $\sqrt{s}=1.96$~TeV.}
\label{fig:raptevg}
\end{center}
\end{figure}
\begin{figure}[htbp]
\begin{center}
\includegraphics[width=0.67\textwidth,page=1]{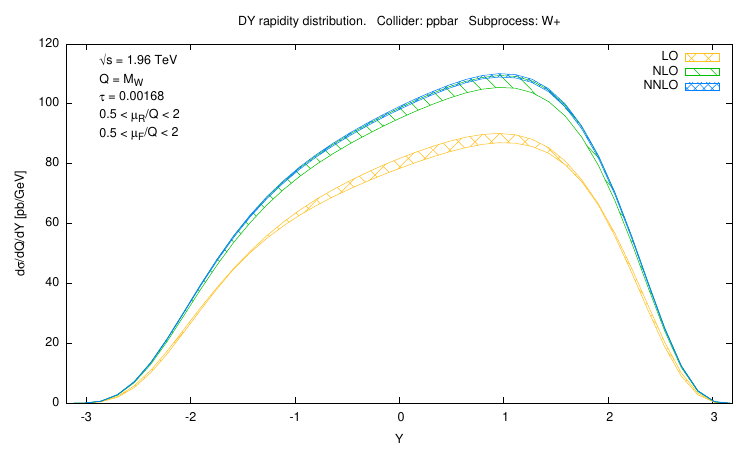}
\\
\includegraphics[width=0.67\textwidth,page=2]{pheno_plots/graph_rap_ppbar_196_Wp}
\\
\includegraphics[width=0.67\textwidth,page=3]{pheno_plots/graph_rap_ppbar_196_Wp}
\caption{Rapidity distribution of charged Drell-Yan pairs of invariant
  mass $Q=M_W$ in $p\bar p$ collisions at $\sqrt{s}=1.96$~TeV.}
\label{fig:rapTEVWp}
\end{center}
\end{figure}

We now turn to rapidity distributions, shown in
Figs.~\ref{fig:rapTEVZ} and~\ref{fig:raptevg} for neutral Drell-Yan
pairs of invariant mass $Q=M_Z$ and $Q=200$~GeV respectively, and in
Fig.~\ref{fig:rapTEVWp} for charged Drell-Yan pairs of invariant mass
$Q=M_W$. The impact of the resummation is now very small, as one would
expect given the smallness of the relevant values of $\tau$. However,
interestingly, resummed uncertainty bands are systematically smaller
than the unresummed one, with the resummation ambiguity (i.e.~the
difference between minimal and Borel results) now essentially
negligible. Hence, even in this small $\tau$ region the resummation
leads to perturbative improvement, while behaving of course as a
perturbative correction.  Again, note that the smallness of
leading-order uncertainty bands is due to the fact that the scale here
is close to the stationary point of the scale dependence of the parton
luminosity already seen in Fig.~\ref{fig:totTev}.

We can compare these uncertainties to that of typical current data,
thanks to recent measurements at the Tevatron. Specifically,
the rapidity distribution of $e^+ e^-$ pairs with
invariant mass in the range $66$~GeV $\leq Q\leq 116$~GeV has
been recently measured by the CDF collaboration~\cite{Aaltonen:2010zz}.
In principle, the data should be
compared with the theoretical prediction for the full process
\beq
p+\bar p \to e^+ + e^- +X.
\eeq
For values of $Q$ close to the
$Z$ mass, however, a good approximation is provided by the Breit-Wigner
approximation, which amounts to assuming that the amplitude is dominated by 
$Z$ exchange, and takes into account the finite width $\Gamma_Z$ of the $Z$
boson:
\beq
\label{BW}
\frac{d\sigma(\tau,Y,Q^2)}{dQ^2dY}
=\frac{2M_Z\Gamma_{e^+ e^-}}{(Q^2-M_Z^2)^2+M_Z^2\Gamma^2_Z}
\frac{1}{2\pi}\frac{d\sigma_Z}{dY}
\eeq
where $\Gamma_{e^+e^-}$ is the $Z$ decay width into
a lepton pair, and $d\sigma_Z$ is the
differential cross-section for the production of a real on-shell $Z$ boson.
Eq.~\eqref{BW} gives
\beq
\label{BW2}
\frac{d\sigma(\tau,Y,Q^2)}{dQ^2dY}
=\frac{M_Z^2\Gamma_Z^2}{(Q^2-M_Z^2)^2+M_Z^2\Gamma^2_Z}
\frac{d\sigma(\tau,Y,M_Z^2)}{dQ^2dY},
\eeq
and therefore
\beq
\int_{Q^2_{\rm min}}^{Q^2_{\rm max}}dQ^2\,\frac{d\sigma(\tau,Y,Q^2)}{dQ^2 dY}
= N(Q^2_{\rm min},Q^2_{\rm max})
\frac{d\sigma(\tau,Y,M_Z^2)}{dQ^2dY},
\eeq
where
\beq
N(Q^2_{\rm min},Q^2_{\rm max})=
M_Z^2 \Gamma_Z^2\int_{Q^2_{\rm min}}^{Q^2_{\rm max}}dQ^2\,
\frac{1}{(Q^2-M_Z^2)^2+M_Z^2\Gamma^2_Z}
\eeq
is just a $Y$--independent multiplicative factor.

\begin{figure}[htb]
\begin{center}
\includegraphics[width=0.9\textwidth,page=2]
{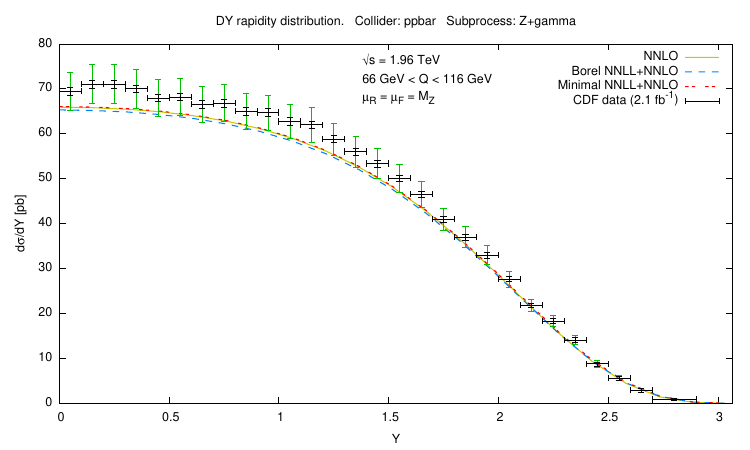}
\caption{Rapidity distribution of $Z$ bosons in $p\bar p$ collisions
at $\sqrt{s}=1.96$~TeV (the contribution of virtual $\gamma$ at the $Z$ peak
is included). Data are taken from~\cite{Aaltonen:2010zz}. The smaller black
uncertainty bands are statistical only, while the wider green bands also
include normalization uncertainties.}
\label{fig:rapTEVZdata}
\end{center}
\end{figure}

In Fig.~\ref{fig:rapTEVZdata} we show the CDF
data~\cite{Aaltonen:2010zz}, corresponding to an integrated luminosity
of $2.1 \text{ fb}^{-1}$, compared to the NNLO QCD prediction with
the inclusion of threshold resummation at NNLL, using both Borel
and minimal prescriptions.
The comparison shows an excellent agreement in shape between the data
and the theoretical curves; there is clearly a mismatch in
normalization of the total cross-section, which is however consistent
with the sizable 6\% normalization uncertainty on the cross-section.
Also in this case, as for the NuSea experiment, this simply
reflects the fact that these data are used in the determination of the
PDFs that we are using, and the difference between the NLO expression
used in PDF fitting and the NNLO one shown here is much smaller than
the experimental uncertainties.

A similar comparison can be performed for the $W^\pm$ asymmetry, defined as
\beq
A_W(Y)=\frac
{\dfrac{d\sigma_{W^+}}{dY}-\dfrac{d\sigma_{W^-}}{dY}}
{\dfrac{d\sigma_{W^+}}{dY}+\dfrac{d\sigma_{W^-}}{dY}},
\eeq 
also measured by CDF~\cite{Aaltonen:2009ta}. In this case, normalization 
uncertainties cancel in the ratio.
In Fig.~\ref{fig:rapTEVWdata}
we show the measured CDF data~\cite{Aaltonen:2009ta} compared to the
QCD prediction at NNLO and resummed NNLO+NNLL (Borel and minimal
prescriptions).
\begin{figure}[htbp]
\begin{center}
\includegraphics[width=0.9\textwidth,page=2]
{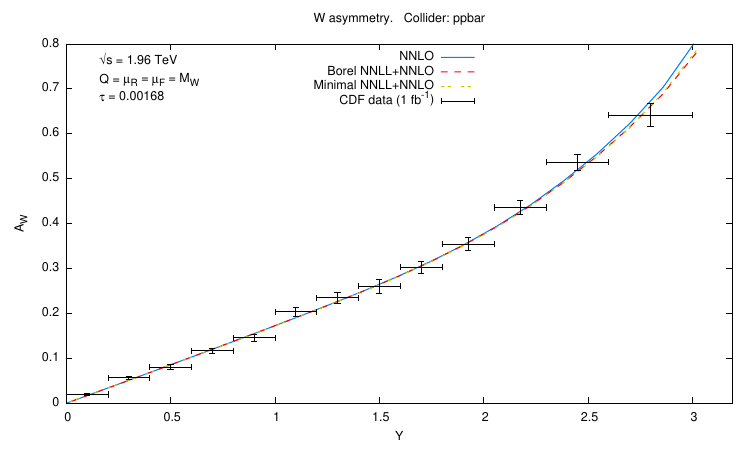}
\caption{$W^\pm$ asymmetry in $p\bar p$ collisions at $\sqrt{s}=1.96$~TeV.
Data are taken from~\cite{Aaltonen:2009ta}.}
\label{fig:rapTEVWdata}
\end{center}
\end{figure}

Clearly, the accuracy of present-day data is insufficient to
appreciate the effect of NNLO or resummation correction, and it is
rather comparable to the difference between LO and NLO predictions,
which can thus be barely appreciated. However, an improvement of
statistical accuracy by an order of magnitude would be sufficient for
NNLO and resummation corrections 
to become significant. The normalization uncertainty 
has a negligible effect on the shape of the distribution
and therefore it does not affect this conclusion

\subsection{LHC}
\label{sec:LHC7}

We now consider predictions for Drell-Yan production at the  LHC, both at
7~TeV and 14~TeV.
Invariant mass distributions for both neutral and charged Drell-Yan
pairs are shown in
Figs.~\ref{fig:totLHC7},~\ref{fig:totLHC7W}. While the impact of
fixed-order perturbative corrections is unsurprisingly similar to
that at the Tevatron collider shown in
Fig.~\ref{fig:totTev}, interestingly the reduction
in uncertainty obtained thanks to the resummation is more marked at
the LHC, consistent with the expectation (recall
Sect.~\ref{sec:saddlepoint}) that the effect of the resummation is
somewhat more significant at a $pp$ than at a $p\bar p$ collider. 
Moreover, the consistency of the NLO error band
with the NNLO prediction is improved by the inclusion of
resummation.
Of course, as in the case of Tevatron, for realistic values of $\tau\lesssim
0.1$ the impact of the resummation is mostly on the uncertainty but
very small or negligible on central values, so the resummation is
behaving as a perturbative correction.
\begin{figure}[htbp]
\begin{center}
\includegraphics[width=0.7\textwidth]{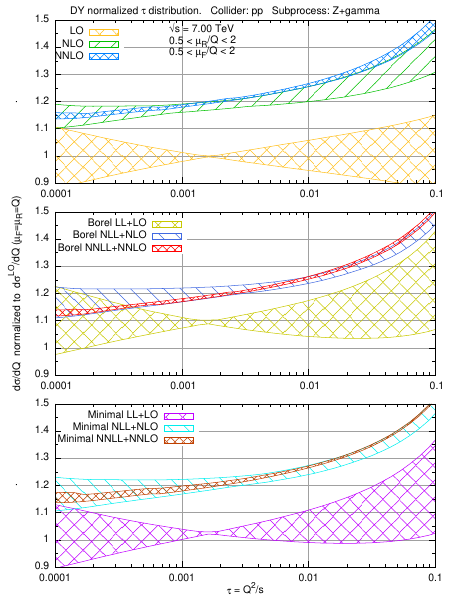}
\caption{Invariant mass distribution of neutral Drell-Yan pairs
in $p p$ collisions at $\sqrt{s}=7$~TeV.}
\label{fig:totLHC7}
\end{center}
\end{figure}
\begin{figure}[htbp]
\begin{center}
\includegraphics[width=0.496\textwidth]{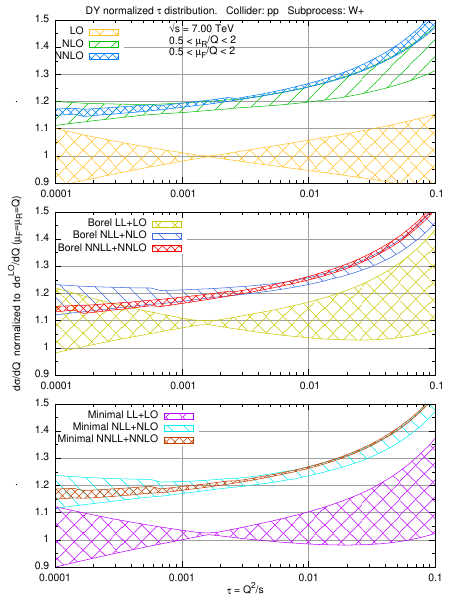}
\includegraphics[width=0.496\textwidth]{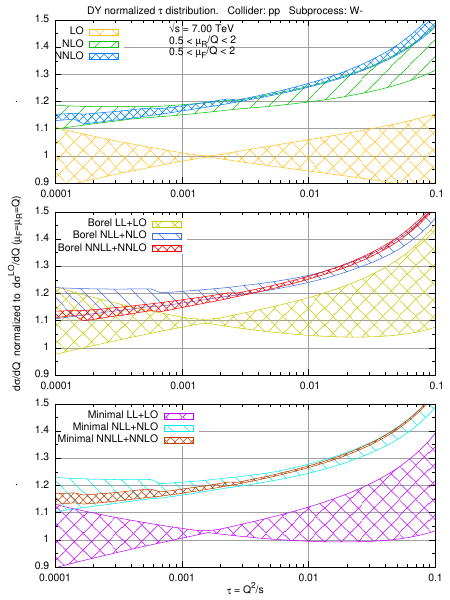}
\caption{Invariant mass distribution of positively (left) and negatively (right)
  charged Drell-Yan pairs in $p p$ collisions at $\sqrt{s}=7$~TeV.}
\label{fig:totLHC7W}
\end{center}
\end{figure}

Turning to rapidity distributions, we present results for the following
observables:
\begin{itemize}
\item[--] 
neutral Drell-Yan pairs with invariant mass $Q=1$~TeV, Fig.~\ref{fig:rapLHC7g}
\item[--] 
neutral Drell-Yan pairs with invariant mass $Q=M_Z$, Fig.~\ref{fig:rapLHC7Z}
\item[--]
positively charged Drell-Yan pairs with invariant mass $Q=M_W$, 
Fig.~\ref{fig:rapLHC7Wp}
\item[--]
negatively charged Drell-Yan pair with invariant mass $Q=M_W$,
Fig.~\ref{fig:rapLHC7Wm}
\end{itemize}

The first case corresponds to $\tau\sim 0.02$, comparable to
the case of a final state with $Q=200$~GeV at the Tevatron shown in 
Fig.~\ref{fig:raptevg}. As in that case, we clearly see an improvement
in uncertainty (with small resummation ambiguities)
when going to the resummed result, though also in that
case the effect on central value is moderate.
On the other hand, the other cases correspond to very small values of
$\tau$ and indeed in this case the uncertainty on resummed results is
larger than that on unresummed ones, indicating that whatever effect
is induced by the resummation is related to the inclusion of terms
which are not dominant in this region. This is also reflected in a
sizable difference between Borel and minimal results.
\begin{figure}[htbp]
\begin{center}
\includegraphics[width=0.7\textwidth,page=1]{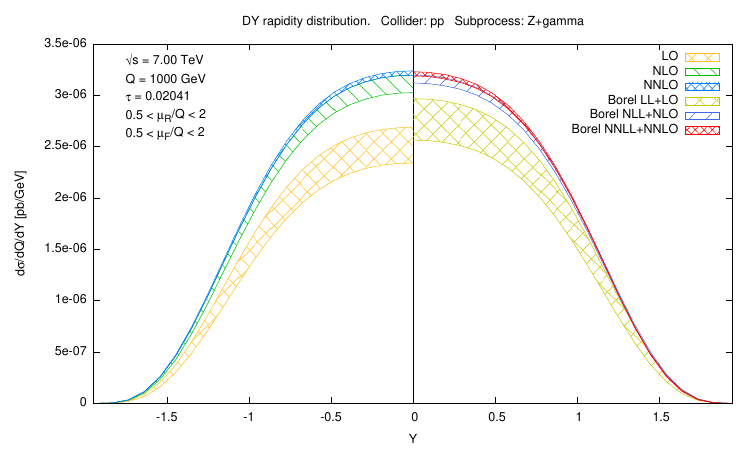}
\\
\includegraphics[width=0.7\textwidth,page=2]{pheno_plots/graph_rap_pp_7_1000}
\caption{Rapidity distribution of neutral Drell-Yan pairs of invariant
  mass $Q=1$~TeV in $pp$ collisions at $\sqrt{s}=7$~TeV.}
\label{fig:rapLHC7g}
\end{center}
\end{figure}

\begin{figure}[htbp]
\begin{center}
\includegraphics[width=0.7\textwidth,page=1]{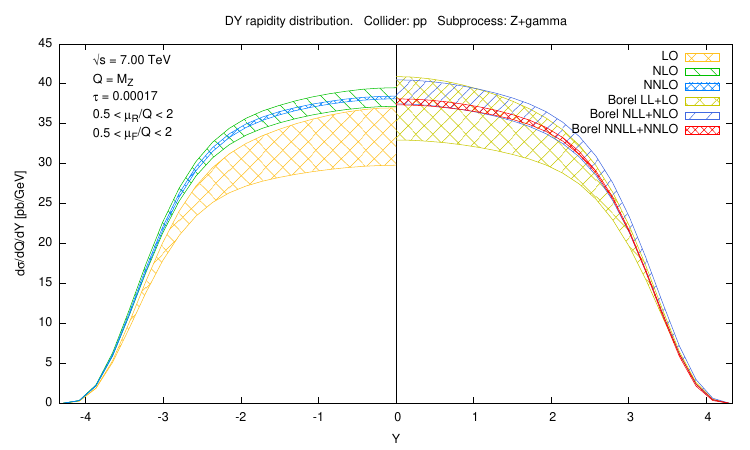}
\\
\includegraphics[width=0.7\textwidth,page=2]{pheno_plots/graph_rap_pp_7_Z}
\caption{Rapidity distribution of neutral Drell-Yan pairs of invariant
  mass $Q=M_Z$ in $pp$ collisions at $\sqrt{s}=7$~TeV
  (the contribution of virtual $\gamma$ at the $Z$ peak is included).}
\label{fig:rapLHC7Z}
\end{center}
\end{figure}

\begin{figure}[htbp]
\begin{center}
\includegraphics[width=0.7\textwidth,page=1]{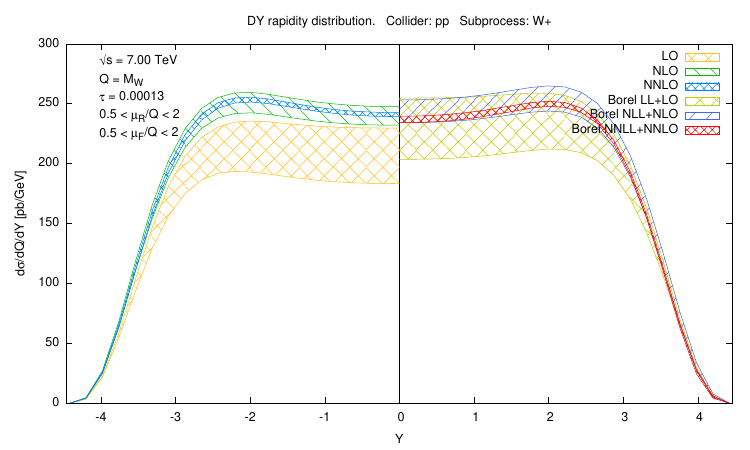}
\\
\includegraphics[width=0.7\textwidth,page=2]{pheno_plots/graph_rap_pp_7_Wp}
\caption{Rapidity distribution of positively charged Drell-Yan pairs of invariant
  mass $Q=M_W$ in $pp$ collisions at $\sqrt{s}=7$~TeV.}
\label{fig:rapLHC7Wp}
\end{center}
\end{figure}

\begin{figure}[htbp]
\begin{center}
\includegraphics[width=0.7\textwidth,page=1]{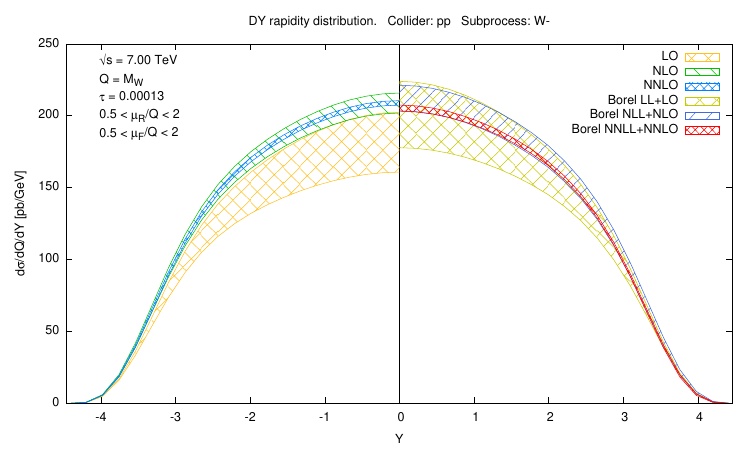}
\\
\includegraphics[width=0.7\textwidth,page=2]{pheno_plots/graph_rap_pp_7_Wm}
\caption{Rapidity distribution of negatively charged Drell-Yan pairs of invariant
  mass $Q=M_W$ in $pp$ collisions at $\sqrt{s}=7$~TeV.}
\label{fig:rapLHC7Wm}
\end{center}
\end{figure}

Finally, in Figs.~\ref{fig:totLHC14}--\ref{fig:rapLHC14Wm} results for
this same invariant mass and rapidity distributions for the LHC at
14~TeV are shown. The general behaviour is the same as that at the
lower energy.

\begin{figure}[htbp]
\begin{center}
\includegraphics[width=0.7\textwidth]{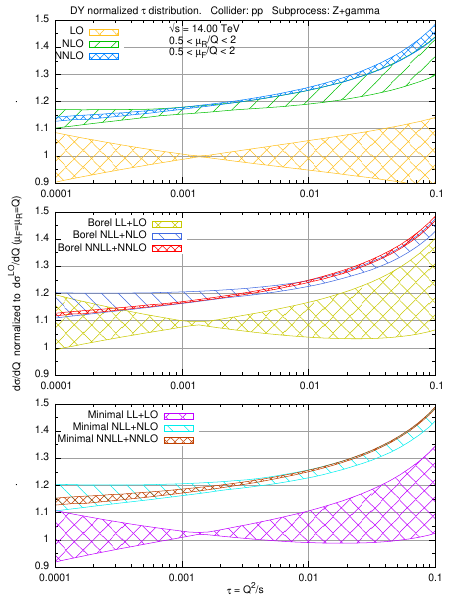}
\caption{Invariant mass distribution of neutral Drell-Yan pairs
in $p p$ collisions at $\sqrt{s}=14$~TeV.}
\label{fig:totLHC14}
\end{center}
\end{figure}
\begin{figure}[htbp]
\begin{center}
\includegraphics[width=0.496\textwidth]{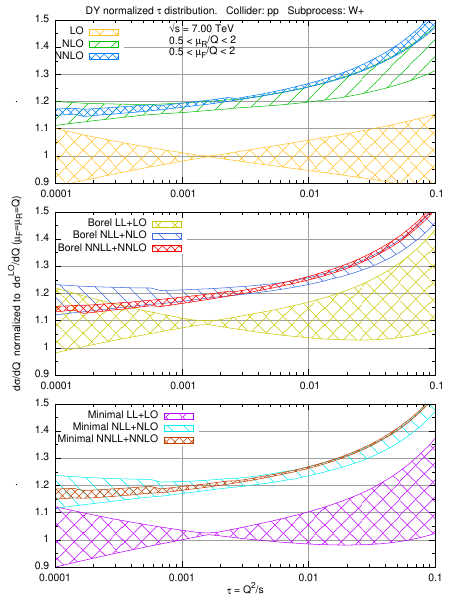}
\includegraphics[width=0.496\textwidth]{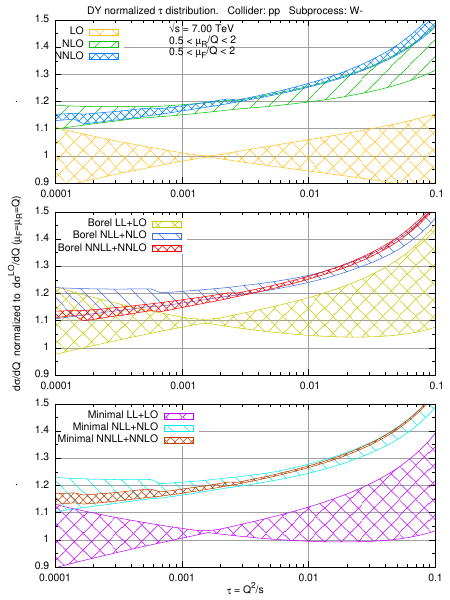}
\caption{Invariant mass distribution of positively (left) and negatively (right)
  charged Drell-Yan pairs in $p p$ collisions at $\sqrt{s}=14$~TeV.}
\label{fig:totLHC14Wpm}
\end{center}
\end{figure}

\begin{figure}[htbp]
\begin{center}
\includegraphics[width=0.7\textwidth,page=1]{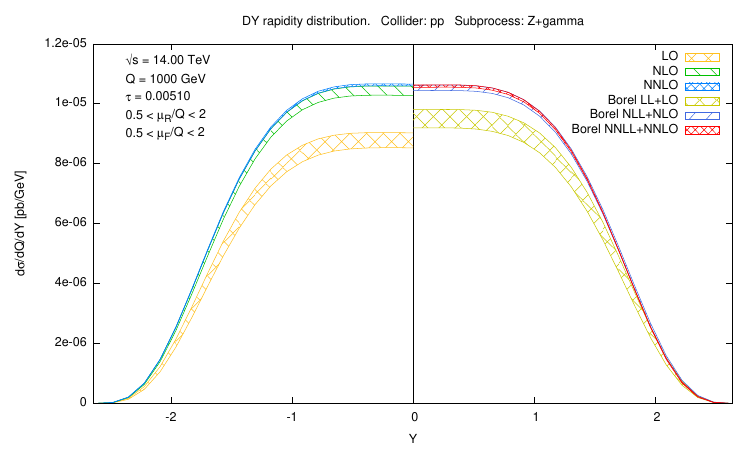}
\\
\includegraphics[width=0.7\textwidth,page=2]{pheno_plots/graph_rap_pp_14_1000}
\caption{Rapidity distribution of neutral Drell-Yan pairs of invariant
  mass $Q=M_Z$ in $pp$ collisions at $\sqrt{s}=14$~TeV
  (the contribution of virtual $\gamma$ at the $Z$ peak is included).}
\label{fig:rapLHC14g}
\end{center}
\end{figure}

\begin{figure}[htbp]
\begin{center}
\includegraphics[width=0.7\textwidth,page=1]{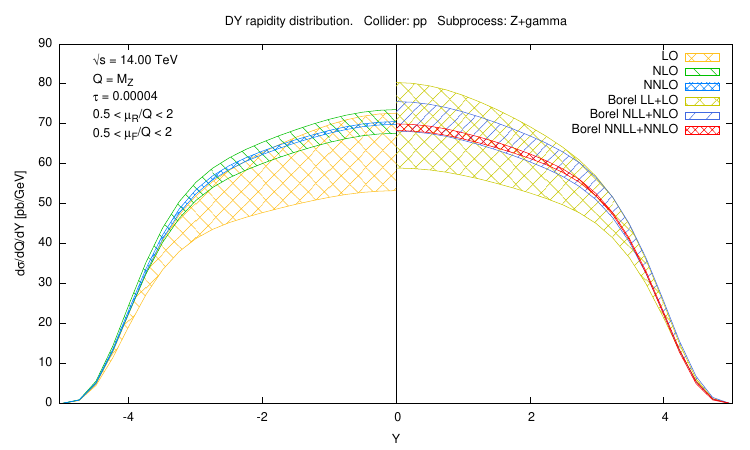}
\\
\includegraphics[width=0.7\textwidth,page=2]{pheno_plots/graph_rap_pp_14_Z}
\caption{Rapidity distribution of $Z$ bosons in $pp$ collisions
at $\sqrt{s}=14$~TeV (the contribution of virtual $\gamma$ at the $Z$ peak
is included).}
\label{fig:rapLHC14Z}
\end{center}
\end{figure}

\begin{figure}[htbp]
\begin{center}
\includegraphics[width=0.7\textwidth,page=1]{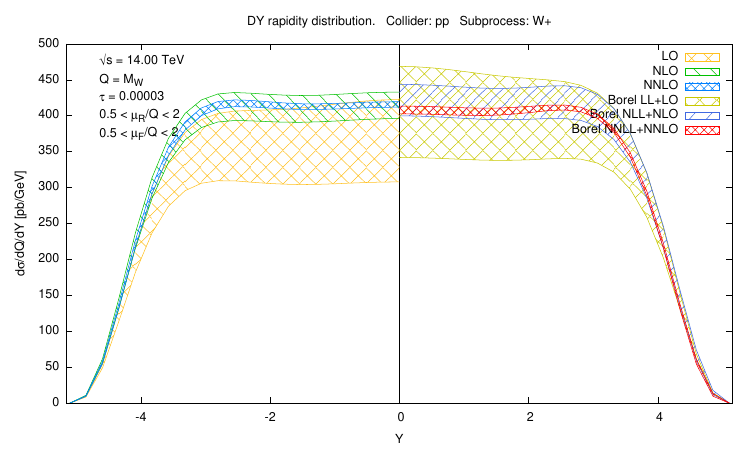}
\\
\includegraphics[width=0.7\textwidth,page=2]{pheno_plots/graph_rap_pp_14_Wp}
\caption{Rapidity distribution of positively charged Drell-Yan pairs
of invariant mass $Q=M_W$ in $pp$ collisions at $\sqrt{s}=14$~TeV.}
\label{fig:rapLHC17Wp}
\end{center}
\end{figure}

\begin{figure}[htbp]
\begin{center}
\includegraphics[width=0.7\textwidth,page=1]{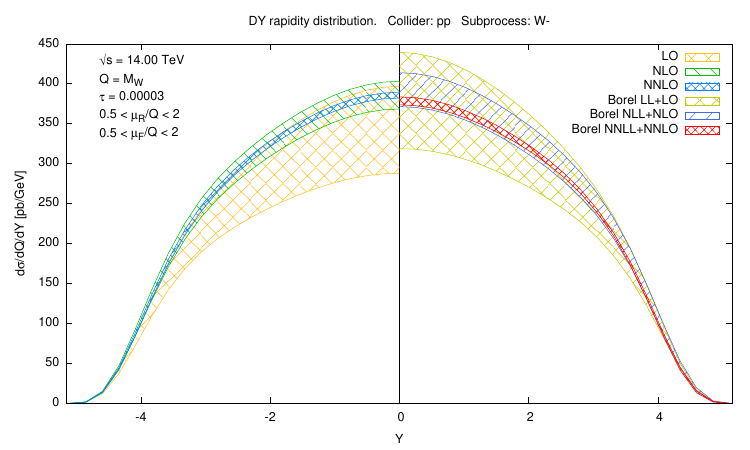}
\\
\includegraphics[width=0.7\textwidth,page=2]{pheno_plots/graph_rap_pp_14_Wm}
\caption{Rapidity distribution of negatively charged Drell-Yan 
pairs of invariant mass $Q=M_W$ in $pp$ collisions at $\sqrt{s}=14$~TeV.}
\label{fig:rapLHC14Wm}
\end{center}
\end{figure}

\section{Conclusions}

We have presented a discussion of the theory and phenomenology of
threshold resummation of rapidity distributions, with special regard
to present-day hadron colliders. On the theoretical side, our main
result is that we have provided a way to estimate the value of
$\tau=Q^2/s$ at which resummation of threshold logarithms is expected
to provide an improvement over fixed-order calculations. This result
has been accomplished through a determination of 
the relevant partonic
center-of-mass energy, whose distance from threshold 
determines the impact of resummation. This
estimate relies on the singularity structure of the anomalous
dimensions which drive $Q^2$ evolution of parton distributions in
perturbative QCD.  Using this technique, we have shown that
resummation is expected to be relevant down to fairly small values of
$\tau\sim 0.003$ at $pp$ colliders, and $\tau\sim 0.02$ at $p\bar p$
colliders.

At the phenomenological level, we have shown how to use different
versions of the Borel resummation prescription and their comparison to
the minimal prescription as a means to assess the ambiguities related
to the resummation, and in particular to the treatment of the
subleading terms. The application of these tools to the Tevatron and
LHC has shown that resummation is relevant for the production of
states of mass as light as $W$ and $Z$ vector bosons at the Tevatron,
and for the production of heavy dileptons of mass in the TeV range at
the LHC: in all these cases threshold resummation leads to a
significant improvement in perturbative accuracy. The impact of
resummation on Tevatron fixed-target rapidity distributions is less
clear, in that, despite being larger, the effect of resummation may be
marred by its ambiguities.

Our general conclusion is that the impact of threshold resummation at
hadron colliders for $\tau\gsim 0.01$ is comparable to that of NNLO
corrections, and it should thus be included both in the determination
of parton distributions and in precision phenomenology, though care
should be taken in also estimating carefully the ambiguity which is
intrinsic in the resummation procedure.

\bigskip
\noindent{\bf Acknowledgements:} We thank 
M.~Grazzini and A.~Vicini for discussions and S.~Catani for
correspondence. This work was supported in part 
by the European network HEPTOOLS under contract
MRTN-CT-2006-035505 and by an Italian PRIN-2008 grant.

\eject

\appendix

\section{Saddle point}
\label{app:saddle}
We collect here some results used in Sect.~\ref{sec:saddlepoint} on the
saddle-point approximation to Mellin transform integrals of typical
parton distributions.

First, we use the saddle-point method to show 
that the large--$N$ behaviour of a Mellin transform describes
the original function of $x$ in the large--$x$ region.
To see this, consider the Mellin transform of a generic function $f(x)$,
\be
\tilde f(N)=\int_0^1dx\,x^{N-1} f(x);
\label{mtt}
\ee
we assume that $f(x)\geq 0$ in the range $0\leq x\leq1$.
For $N$ real, $\tilde f(N)$ is a decreasing function of $N$, because the area 
below the curve $x^{N-1} f(x)$ obviously decreases as $N$ increases. 

The inverse Mellin transform is given by
\be
f(x)=\frac{1}{2\pi i}\int_{\bar N-i\infty}^{\bar N+i\infty}dN\,x^{-N}\tilde f(N),
\label{imtt}
\ee
with $\bar N$ larger than the real part of the rightmost singularity
of $\tilde f(N)$.
We estimate the range of $N$ which gives the dominant contribution
by saddle-point:
we rewrite Eq.~(\ref{imtt}) as
\be
f(x)=\frac{1}{2\pi i}\int_{\bar N-i\infty}^{\bar N+i\infty}dN\,e^{E(x,N)}
\label{imt2}
\ee
where
\be
E(x,N)=N\ln\frac{1}{x}+\ln\tilde f(N).
\ee
For appropriate $\tilde f(N)$, as in the case of the physical cross-section,
$E(x,N)$ has a unique minimum on the real $N$ axis at $N=N_0$, with
\be
E'(x,N_0)=\ln\frac{1}{x}+\frac{\tilde f'(N_0)}{\tilde f(N_0)}=0,
\label{saddlepoint}
\ee
where a prime denotes differentiation with respect to $N$.
The inversion integral is
dominated by the region of $N$ around $N_0$,
and can be approximated by 
\begin{align}
f(x)&\approx
\frac{1}{2\pi i}\int_{N_0-i\infty}^{N_0+i\infty}dN\,
e^{E(x,N_0)+\frac{E''(x,N_0)}{2}(N-N_0)^2}
\nonumber\\
&=\frac{1}{\sqrt{2\pi}}\frac{e^{E(x,N_0)}}{\sqrt{E''(x,N_0)}}
\label{imt3}
\end{align}
(after the variable change $N=N_0+it$ and gaussian integration).  When
$x\to 1$, the slope of the straight line $N\ln\frac{1}{x}$
decreases, and the position $N_0$ of the minimum is pushed to larger
values. We conclude that the behaviour of $f(x)$ for $x$ close to 1 is
determined by the large--$N$ behaviour of $\tilde f(N)$, as we set out
to prove.

As an explicit example, we consider
\be
f(x)=x^\alpha\ln^\beta\frac{1}{x},
\label{ffunct}
\ee
with $\alpha$ and $\beta$ positive constants. The function
$f(x)$ Eq.~(\ref{ffunct}) vanishes both at $x=0$ 
and $x=1$, and has a maximum at
\be
x=e^{-\frac{\beta}{\alpha}}
\ee
which approaches $1$ as $\beta/\alpha\to 0$. 
We find
\be
\tilde f(N)=\frac{\Gamma(\beta+1)}{(N+\alpha)^{\beta+1}}.
\label{largeN}
\ee
Let us now compute $f(x)$ using the saddle-point
approximation to invert the Mellin transform Eq.~\eqref{largeN}.
We have
\bea
&&E(x,N)=N\ln\frac{1}{x}+\ln\Gamma(\beta+1)-(\beta+1)\ln(N+\alpha)
\\
&&E'(x,N)=\ln\frac{1}{x}-\frac{\beta+1}{N+\alpha}
\label{Eprime}
\\
&&E''(x,N)=\frac{\beta+1}{(N+\alpha)^2}.
\eea
From Eq.~\eqref{Eprime} we find that the saddle point is at
\beq
N_0=\frac{\beta+1}{\ln\frac{1}{x}}-\alpha,
\eeq
which is, as expected, an increasing function of $x$. Furthermore
\bea
&&e^{E(x,N_0)}=x^{-N_0}\tilde f(N_0)
=\frac{\Gamma(\beta+1)e^{\beta+1}}{(\beta+1)^{\beta+1}}
x^\alpha\ln^{\beta+1}\frac{1}{x}
\\
&&E''(x,N_0)=\frac{1}{\beta+1}\ln^2\frac{1}{x}.
\eea
Therefore, Eq.~\eqref{imt3} gives
\beq
f(x)\approx K(\beta)x^\alpha\ln^\beta\frac{1}{x},
\eeq
where 
\beq
K(\beta)=\frac{\Gamma(\beta+2)}
{\sqrt{2\pi(\beta+1)}\exp\[(\beta+1)\ln(\beta+1)-(\beta+1)\]}.
\label{k}
\eeq
We see that the saddle-point approximation in this case gives the exact result,
Eq.~\eqref{ffunct}, up to a normalization factor, which however
approaches 1 for large values of $\beta$ (the denominator
of Eq.~\eqref{k} is just $\Gamma(\beta+2)$ in the Stirling approximation).
Indeed, one can check that the Taylor expansion of $E(x,N)$ to second order
around $N_0$ is increasingly accurate as $\beta\to\infty$.

It is important to observe that the argument which shows that a Mellin
transform $f(N)$ Eq.~(\ref{mtt}) is a decreasing function of $N$ does
not apply if $f(x)$ in Eq.~(\ref{imtt}) is a distribution rather than
an ordinary function.  An obvious counter-example is the Dirac delta
function, which has an $N$--independent Mellin transform:
\beq
\int_0^1dx\,x^{N-1}\delta(1-x)=1.
\eeq
It is not difficult to find examples of distributions whose Mellin transforms
even increase as $N$ increases.
Let us consider for instance
\be
d(x)=\left[D(x)\right]_+,
\ee
where $D(x)$ is a positive function, at most as singular as $(1-x)^{-1}$
in $x=1$. With the usual definition of the plus prescription, we
have
\be
\tilde d(N)=\int_0^1dx\,\left(x^{N-1}-1\right)D(x),
\ee
which is finite under the above assumptions. The area below
$(1-x^{N-1})D(x)$ increases as $N$ increases,
and therefore $|\tilde d(N)|$ grows with $N$.

In fact (see also Appendix~\ref{app:mellin}) this is the typical
situation for the distributions
$\left[\frac{\ln^k\frac{1}{1-x}}{1-x}\right]_+$, which
characterize the large--$x$ and
thus large--$N$ behaviour of soft-gluon emission cross-sections after
the cancellation of infrared poles, and whose Mellin transforms 
behave as $\ln^{k+1}N$
when $N\to\infty$. The coefficient of this
logarithmic rise need not be positive, given that these are partonic
cross-sections obtained after subtraction of infrared poles and
factorization of collinear singularity.

Of course, the physical cross-section must be an ordinary function, so
in the factorized expression Eq.~(\ref{fact}) 
the decrease of the parton luminosity at large $N$ always offsets
the increase of the partonic cross-section $\hat \sigma$. Indeed,
because the partonic cross-section rises at most as a power of $\ln N$
as $N\to\infty$, it is easy to show that a sufficient condition 
for the Mellin inversion integral Eq.~(\ref{imt}) of the factorized
cross-section Eq.~(\ref{fact}) to exist is that the parton luminosity ${\cal
L} (z)$ 
vanishes at large $z$ at least as a positive power of $(1-z)$.

\section{Mellin transformation}
\label{app:mellin}

We collect here some results on the calculation of Mellin
transforms of functions and distributions which typically appear in
the perturbative coefficients.  We start from the identity
\beq
\int_0^1dz\,z^{N-1}\,\plus{\ln^{\xi-1}\frac{1}{z}}
=\int_0^1dz\,(z^{N-1}-1)\,\ln^{\xi-1}\frac{1}{z}
=\Gamma(\xi)\left(N^{-\xi}-1\right)
\label{logzlogn}
\eeq
(the plus prescription is not necessary if ${\rm Re\,}\xi>0$,
but its presence makes the l.h.s.~well defined even when $\xi=0$).
It follows that
\beq
N^{-\xi}=1+\frac{1}{\Gamma(\xi)}\int_0^1dz\,z^{N-1}
\plus{\ln^{\xi-1}\frac{1}{z}}
\eeq
and therefore
\beq
\frac{1}{2\pi i}\int_{\bar N-i\infty}^{\bar N+i\infty}dN\,
z^{-N}\,N^{-\xi}=\delta(1-z)+
\plus{\frac{\ln^{\xi-1}\frac{1}{z}}{\Gamma(\xi)}}
\eeq
with $\bar N>0$. Taking $k$ derivatives with respect to $\xi$ at $\xi=0$ we get
\be
\frac{1}{2\pi i}\int_{\bar N-i\infty}^{\bar N+i\infty}
dN\,z^{-N}\,\ln^k\frac{1}{N}
=\delta_{k0}\delta(1-z)
+\plus{
\left.\frac{d^k}{d\xi^k}\frac{\ln^{\xi-1}\frac{1}{z}}{\Gamma(\xi)}
\right|_{\xi=0}
}.
\label{logminimal1}
\eeq
Note that the result Eq.~(\ref{logminimal1}) is a distribution,
consistently with the fact that $\ln^k N$ is an increasing function
of $N$. In other words, the
inversion integral does not exist in the ordinary sense, but only in
the sense of distributions.

An equivalent form of Eq.~(\ref{logminimal1}) is
\beq
\frac{1}{2\pi i}\int_{\bar N-i\infty}^{\bar N+i\infty}
dN\,z^{-N}\,\ln^k\frac{1}{N}
=\delta_{k0}\delta(1-z)
+\frac{k!}{2\pi i}\plus{\oint\frac{d\xi}{\xi^{k+1}}\,
\frac{\ln^{\xi-1}\frac{1}{z}}{\Gamma(\xi)}
},
\label{logminimal3}
\eeq
where the integration contour is any closed curve in the complex plane
$\xi$ with the origin $\xi=0$ inside. This form proves useful in
the formulation of the Borel prescription for the Mellin inversion of
divergent series in $\ln N$.

A second equivalent form of Eq.~(\ref{logminimal1}) is obtained by
computing the derivative in the r.h.s.~explicitly:
\beq
\frac{1}{2\pi i}\int_{\bar N-i\infty}^{\bar N+i\infty}dN\,z^{-N}\,\ln^k\frac{1}{N}
=\delta_{k0}\delta(1-z)
+\plus{\frac{1}{\ln\frac{1}{z}}\sum_{j=1}^k\binom{k}{j}\Delta^{(j)}(0)\,
\ell^{k-j}},
\label{logminimal2}
\eeq
where
\beq
\Delta(\xi)=\frac{1}{\Gamma(\xi)};\qquad \ell=\ln\ln\frac{1}{z},
\eeq
We see that the terms in the sum over $j$
are decreasingly important as $z\to 1$.

Threshold logarithms typically appear in perturbative
coefficients through the Mellin transform
\beq
I_k(N)=\int_0^1 dz\,z^{N-1}\,\left[\frac{\ln^k(1-z)}{1-z}\right]_+.
\eeq
We have
\beq
I_k(N)=\left.\frac{d^k}{d\xi^k}F(N,\xi)\right|_{\xi=0},
\eeq
where
\beq
F(N,\xi)=\int_0^1dz\,(z^{N-1}-1)(1-z)^{\xi-1}
=\frac{1}{\xi}\left[
\frac{\Gamma(N)\Gamma(1+\xi)}{\Gamma(N+\xi)}-1\right].
\label{Fdef}
\eeq
Hence
\beq
I_k(N)=\frac{\Gamma(N)}{k+1}\sum_{j=0}^{k+1}\binom{k+1}{j}
\Gamma^{(j)}(1)\,\Delta^{(k+1-j)}(N).
\eeq

\section{Chebyshev polynomials}
\label{app:cheb}

In this Appendix we recall the definition and the main properties
of Chebyshev polynomials. The Chebyshev polynomials
\beq 
T_i(z) = \sum_{k=0}^i T_{ik}\, z^k,
\eeq
are defined in the range $z\in[-1,1]$ recursively by
\begin{subequations}\label{eq:cheb}
\begin{align}
T_0(z) &= 1 \\
T_1(z) &= z \\
T_2(z) &= 2z^2-1 \\
T_i(z) &= 2z\,T_{i-1}(z) -T_{i-2}. \label{eq:recursive}
\end{align}
\end{subequations}

A generic function $G(u)$ can be approximated
in the range $u\in[u_{\rm min}, u_{\rm max}]$ by its
expansion on the basis of Chebyshev
polynomials~\eqref{eq:cheb}, truncated at some order $n$:
\beq
G(u) \simeq -\frac{c_0}{2} +\sum_{i=0}^n c_i\, T_i(Au+B),
\eeq
where
\beq\label{eq:AB}
A=\frac{2}{u_{\rm max}-u_{\rm min}}, 
\qquad B=-\frac{u_{\rm max}+u_{\rm min}}{u_{\rm max}-u_{\rm min}}.
\eeq
Simple numerical algorithms for the calculation of the coefficients $c_i$
are available (we have used the routines of the \texttt{gsl}).

Simple algebra leads to
\beq
G(u)\simeq \sum_{k=0}^n \tilde c_k \, (Au+B)^k
\eeq
where
\beq
\tilde c_k = -\frac{c_0}{2}\delta_{k0} + \sum_{i=k}^n c_i\, T_{ik}.
\eeq

\subsection{Minimal Prescription}
\label{sec:cheb_minimal}

We have shown in Sect.~\ref{sec:MPBP} that the minimal prescription
can be conveniently implemented by means of an analytic expression
for the Mellin transform of the luminosity 
${\cal L}(z)$ (which can be either $\Lum_{q\bar q}(z)$ in the case of
invariant mass distributions or $\Lrap(z,1/2)$ in the case
of rapidity distributions). This can be obtained by expanding
${\cal L}(z)$ on the basis of Chebyshev polynomials, truncated at
some finite order $n$, and then taking its analytical Mellin transform.
The luminosity itself, however, is very badly behaved in the range
$(0,1)$: it is singular at $z=0$, and varies by orders of magnitude in the
range
\beq
0\leq z\leq z_{\rm max}
;\qquad z_{\rm max}
=
\begin{cases}
1 &\text{for the rapidity-integrated cross-section}\\
e^{-2\abs{Y}} & \text{for the rapidity distribution}
\end{cases}
\eeq
It is therefore convenient to expand a regularized function
\beq
F(z) = \frac{z^\beta}{(z_{\rm max}-z)^\delta} \, {\cal L}(z).
\eeq 
Values of $\beta$ and
$\delta$ in the range $3\div 7$ are normally suited to
make $F(z)$ smooth enough to be approximated by a reasonably
small number of Chebyshev polynomials. Equation~\eqref{eq:AB} gives
\beq 
A=\frac{2}{z_{\rm max}},\qquad B=-1.
\eeq
and the approximation is
\beq
F(z)= \sum_{k=0}^n \tilde c_k\, \(2\frac{z}{z_{\rm max}}-1\)^k
= \sum_{p=0}^n \hat c_p \, z^p,
\eeq
where
\beq
\hat c_p = \(\frac{2}{z_{\rm max}}\)^p \sum_{k=p}^n \binom{k}{p} 
(-1)^{k-p}\, \tilde c_k.
\eeq
The luminosity is easily recovered:
\beq\label{eq:lum_cheb_app_1}
{\cal L}(z)
=(z_{\rm max}-z)^\delta \sum_{p=0}^n \hat c_p \, z^{p-\beta}
= \sum_{j=0}^\delta \binom{\delta}{j}\,z_{\rm max}^{\delta-j}\,(-1)^j
\sum_{p=0}^n \hat c_p \, z^{p+j-\beta}\;,
\eeq
where the last equality holds for $\delta$ integer.
It is now immediate to obtain the Mellin transform:
\beq
{\cal L}(N) = \int_0^{z_{\rm max}} dz\, z^{N-1} \, \Lum(z) =
\sum_{p=0}^n\sum_{j=0}^\delta \hat c_p \, \binom{\delta}{j}\,(-1)^j \,
\frac{z_{\rm max}^{N+p+\delta-\beta}}{N+p+j-\beta}.
\eeq

Alternatively, one may introduce the variable
\beq
z= z_{\rm max}e^u
\eeq
and appoximate the function
\beq
F(u) = z{\cal L}(z)=z_{\rm max}\, e^u {\cal L}
\( z_{\rm max} e^u \)
\eeq
by an expansion on the basis of Chebyshev polynomials.
The variable $u$ ranges from $-\infty$ to 0 when $0\leq z\leq z_{\rm max}$;
however,
for practical purposes one only needs the luminosity 
for $z\geq z_{\rm  min}=z_{\rm max}e^{u_{\rm min}}$. We have therefore
\beq
A=-\frac{2}{u_{\rm min}}, \qquad B=1,
\eeq
and
\beq
F(u) = \sum_{k=0}^n \tilde c_k \(1-2 \frac{u}{u_{\rm min}}\)^k.
\eeq
We can now reconstruct ${\cal L}(z)$ through the replacement
$u=\ln\frac{z}{z_{\rm max}}$. We get
\beq
{\cal L}(z) = \frac{1}{z} \sum_{p=0}^n (-2)^p\, u_{\rm min}^{-p}\,
\ln^p\frac{z}{z_{\rm max}}\sum_{k=p}^n \binom{k}{p} \tilde c_k.
\eeq
The Mellin transform is computed using the result
\beq
\int_0^{z_{\rm max}} dz \, z^{N-2} \ln^p\frac{z}{z_{\rm max}}
= z_{\rm max}^{N-1} \frac{(-1)^p\, p!}{(N-1)^{p+1}}.
\eeq
We obtain
\beq
{\cal L}(N) = \int_0^{z_{\rm max}} dz \, z^{N-1}\Lum(z) 
= z_{\rm max}^{N-1} \sum_{p=0}^n \frac{\bar c_p}{(N-1)^{p+1}},
\eeq
where
\beq
\bar c_p= \frac{2^p}{u_{\rm min}^p} \sum_{k=p}^n \frac{k!}{(k-p)!}\,\tilde c_k.
\eeq
In practice, we have found that the second method is to be preferred
for small values of $\tau$, $\tau\lesssim 0.1$, while the previous one
works better for $\tau\gtrsim 0.1$.

\subsection{Borel Prescription}
\label{sec:cheb_borel}

In this case, we look for an approximation of the function
$g(z,\tau)$, Eq.~\eqref{g_zx}, as a function of $z\in[\tau,1]$. We have
\beq
g(z,\tau) = \sum_{k=0}^n \tilde c_k\, (Az+B)^k 
= \sum_{p=0}^n b_p\, (1-z)^p
\eeq
where
\beq
b_p = (-A)^p \sum_{k=p}^n \binom{k}{p} (A+B)^{k-p}\, \tilde c_k
\eeq
and
\beq
A=\frac{2}{1-\tau}, \qquad B= -\frac{1+\tau}{1-\tau}.
\eeq
Note that $A+B=1$ in this case. Therefore
\beq
b_p = \(\frac{-2}{1-\tau}\)^p \sum_{k=p}^n \binom{k}{p}\, \tilde c_k .
\eeq
In the case of the rapidity distributions, the variable $z$
is in the range $z\in[\tau e^{2\abs{Y}},1]$; therefore
\beq
b_p = \(\frac{-2}{1-\tau e^{2\abs{Y}}}\)^p \sum_{k=p}^n
\binom{k}{p}\, \tilde c_k.  
\eeq

\section{Resummed cross-section for the Drell-Yan process}
\label{sec:app-resumm}
\allowdisplaybreaks

In this Appendix we give the explicit expressions of the
functions $g_i$ which appear in the resummed Drell-Yan cross-section,
Eq.~\eqref{eq:S}.
We have~\cite{mvv}
\begin{subequations}\label{eq:g_i}
\begin{align}
g_0(\as) =& \, 1 + \as\, g_{01} + \as^2\, g_{02} + \Ord(\as^3)\\
g_1(\lambda) =&\, \frac{2A_1}{\beta_0} \left[ (1+\lambda)\ln(1+\lambda) -\lambda \right] \\
g_2(\lambda) =&\, \frac{A_2}{\beta_0^2} \left[ \lambda - \ln(1+\lambda) \right]
   + \frac{A_1}{\beta_0} \left[ \ln(1+\lambda) \( \ln\frac{Q^2}{\mur^2} -2\gammae \) -\lambda \ln\frac{\muf^2}{\mur^2} \right] \nonumber\\
       &\,+ \frac{A_1\beta_1}{\beta_0^2} \left[ \frac{1}{2}\ln^2(1+\lambda) +\ln(1+\lambda) -\lambda \right]\\
g_3(\lambda) =&\, \frac{1}{4\beta_0^3} \( A_3 - A_1  \beta_2 + A_1  \beta_1^2 - A_2  \beta_1 \) \frac{\lambda^2}{1+\lambda} \nonumber\\
&\, + \frac{A_1 \beta_1^2}{2\beta_0^3} \; \frac{\ln(1+\lambda)}{1+\lambda} \left[ 1 + \frac{1}{2} \ln(1+\lambda)\right]
      + \frac{ A_1  \beta_2 - A_1  \beta_1^2 }{2\beta_0^3} \; \ln(1+\lambda) \nonumber\\
&\, +  \( \frac{A_1  \beta_1}{\beta_0^2}  \gammae 
          + \frac{A_2  \beta_1 }{2\beta_0^3}
          \) \left[ \frac{\lambda}{1+\lambda} 
- \frac{\ln(1+\lambda)}{1+\lambda} \right]
\nonumber\\
&\, - \left(
            \frac{A_1  \beta_2 }{2\beta_0^3}
          + \frac{A_1}{\beta_0}   (\gammae^2 + \zeta_2) 
          + \frac{A_2}{\beta_0^2}  \gammae
          - \frac{D_2}{4\beta_0^2} 
         \right) 
           \frac{\lambda}{1+\lambda} 
\nonumber\\
&\,
       +   \left[
         \(
            \frac{A_1}{\beta_0}  \gammae
          + \frac{A_2 - A_1  \beta_1 }{2\beta_0^2}
          \) 
           \frac{\lambda}{1+\lambda} 
       +  \frac{A_1  \beta_1}{2\beta_0^2} \;
          \frac{\ln(1+\lambda)}{1+\lambda}
          \right] \ln\frac{Q^2}{\mur^2}
\nonumber\\
&\,
       -  \frac{A_2}{2\beta_0^2} \, \lambda \, \ln\frac{\muf^2}{\mur^2}
       + \frac{A_1}{4\beta_0} \left[
         \lambda\, \ln^2\frac{\muf^2}{\mur^2}
          - \frac{\lambda}{1+\lambda} \, \ln^2\frac{Q^2}{\mur^2}
          \right]
\end{align}
\end{subequations}
with
\beq
\frac{d}{d\ln\mu^2}\as(\mu^2) = -\beta_0 
\as^2 \( 1+\beta_1\as+\beta_2 \as^2+\ldots \) \;.
\eeq
The coefficients $g_{0k}$ can be found in~\cite{mv}, but without 
scale-dependent terms. Their full expression is given by
\begin{subequations}
\begin{align}
g_{01} =&\, \frac{C_F}{\pi}\left[ 4\zeta_2 -4 + 2\gammae^2 
+\(\frac{3}{2}-2\gammae\)\ln\frac{Q^2}{\muf^2} \right]\\
g_{02} =&\, \frac{C_F}{16\pi^2} \Bigg\{
          C_F \( \frac{511}{4} - 198\,\zeta_2 - 60\,\zeta_3 
+ \frac{552}{5}\,\zeta_2^2 
              - 128\,\gammae^2 + 128\,\gammae^2 \zeta_2 
+ 32\,\gammae^4 \) \nonumber\\
       &\qquad + C_A \( - \frac{1535}{12} + \frac{376}{3}\,\zeta_2 
+ {604 \over 9}\,\zeta_3 
                  - \frac{92}{5}\,\zeta_2^2 + \frac{1616}{27}\,\gammae
                  - 56\,\gammae \zeta_3 + {536 \over 9}\,\gammae^2
                  - 16\,\gammae^2 \zeta_2 + {176 \over 9}\,\gammae^3 \) 
\nonumber\\
&\qquad + n_f \( {127 \over 6} - {64 \over 3}\,\zeta_2 
+ {8 \over 9}\,\zeta_3 
                - {224 \over 27}\,\gammae - {80 \over 9}\,\gammae^2 
                - {32 \over 9}\,\gammae^3 \) \nonumber\\
&\qquad + \ln^2\frac{Q^2}{\muf^2} \left[ C_F \(32\gammae^2-48\gammae+18\)
 +C_A \(\frac{44}{3}\gammae-11\) +n_f\(2-\frac{8}{3}\gammae\) \right]
\nonumber\\
&\qquad + \ln\frac{Q^2}{\muf^2} 
\Bigg[ C_F \(48\zeta_3 +72\zeta_2 -93 -128\gammae \zeta_2 +128\gammae + 48\gammae^2 -64\gammae^3 \) \nonumber\\
       &\qquad\qquad\qquad +C_A \(\frac{193}{3} -24\zeta_3 -\frac{88}{3}\zeta_2 +16\gammae\zeta_2 -\frac{536}{9}\gammae -\frac{88}{3}\gammae^2 \) \nonumber\\ 
       &\qquad\qquad\qquad +n_f \(\frac{16}{3}\zeta_2 -\frac{34}{3} +\frac{80}{9}\gammae +\frac{16}{3}\gammae^2 \) \Bigg]
       \Bigg\} \nonumber\\
       &\,- \frac{\beta_0 C_F}{\pi}\left[ 4\zeta_2 -4 + 2\gammae^2 +\(\frac{3}{2}-2\gammae\)\ln\frac{Q^2}{\muf^2} \right] \ln\frac{\muf^2}{\mur^2} \; .
\end{align}
\end{subequations}
The coefficients appearing in the previous functions are
\begin{align}
\beta_0 &= \frac{11 C_A - 4T_F\, n_f}{12\pi} = \frac{33-2n_f}{12\pi}
\\
\beta_1 &= \frac{17C_A^2-(10C_A+6C_F)T_F\,n_f}{24\pi^2 \beta_0}
= \frac{153-19n_f}{(33-2n_f)2\pi}
\\
\beta_2 &= \frac{1}{(4\pi)^3\beta_0} \left[
  \frac{2857}{54}\,C_A^3 + \left( 2 C_F^2 - \frac{205}{9}\,C_F C_A
 - \frac{1415}{27}\,C_A^2 \right) T_F n_f
  + \left( \frac{44}{9}\,C_F + \frac{158}{27}\,C_A \right) T_F^2 n_f^2 \right]
\nonumber\\
& = \frac{3}{(33-2n_f)32\pi^2}\left[ 2857-\frac{5033}{9}n_f 
+\frac{325}{27}n_f^2 \right]
\\
A_1 &= \frac{C_F}{\pi} = \frac{4}{3\pi}\\
A_2 &= \frac{C_F}{2\pi^2}
\left[ C_A \(\frac{67}{18} - \frac{\pi^2}{6}\) - \frac{10}{9}T_F\, n_f \right] 
= \frac{201-10n_f}{27\pi^2}-\frac{1}{3}\\
A_3 &= \frac{C_F}{4\pi^3} \left[ C_A^2 
\left( \frac{245}{24} - \frac{67}{9}\,\zeta_2
+ \frac{11}{6}\,\zeta_3 + \frac{11}{5}\,\zeta_2^2 \right) 
+ \left( -  \frac{55}{24}  + 2\,\zeta_3 \right)C_F \,n_f  \right. 
\nonumber\\ & 
\left. \mbox{} \qquad
+ \left( - \frac{209}{108} + \frac{10}{9}\,\zeta_2 -\frac{7}{3}\zeta_3 \right)
C_A \,n_f - \frac{1}{27} \, n_f^2 \right]
\\
D_2 &= \frac{C_F}{16\pi^2}
\left[ C_A\(-\frac{1616}{27}+\frac{88}{9}\pi^2+56\zeta_3\) 
+ \(\frac{224}{27}-\frac{16}{9}\pi^2\) n_f \right].
\end{align}

\subsection{Matching}\label{sec:matching}
Here we compute the terms to be subtracted in the resummed result in order
to avoid double counting.
Start from Eqs.~\eqref{eq:g_i} and expand them in powers of their argument
$\lambda$:
\begin{subequations}
\begin{align}
g_1(\lambda) 
=&\, \frac{A_1}{\beta_0} \left[ \lambda^2-\frac{1}{3}\lambda^3 
+ \Ord(\lambda^4) \right] \\
g_2(\lambda) 
=&\, \frac{A_1}{\beta_0} \(\ln\frac{Q^2}{\muf^2}-2\gammae\) \lambda
+ \( \frac{A_2}{2\beta_0^2} + \frac{A_1}{2\beta_0} 
\(2\gammae - \ln\frac{Q^2}{\mur^2}\) \) \lambda^2
            + \Ord(\lambda^3) \\
g_3(\lambda) 
=&\, \(-\frac{A_1}{\beta_0}(\gammae^2+\zeta_2)
-\frac{A_2}{\beta_0^2}\gammae+\frac{D_2}{4\beta_0^2}\) \lambda 
\nonumber\\
&\,+\( \frac{A_1}{\beta_0}\gammae\ln\frac{Q^2}{\mur^2} 
+\frac{A_2}{2\beta_0^2}\ln\frac{Q^2}{\muf^2}
 - \frac{A_1}{4\beta_0}\(\ln^2\frac{Q^2}{\mur^2}-\ln^2\frac{\muf^2}{\mur^2}\)
            \) \lambda + \Ord(\lambda^2).
\end{align}
\end{subequations}
The Taylor expansion of $g_0(\as)\,\exp\S(\lambda,\ab)$ is
\begin{align}
g_0(\as)\,\exp\S(\lambda,\ab) &= (1 + \as\, g_{01} + \as^2\, g_{02} 
+ \ldots)e^{\as\,\S_1+\as^2\,\S_2+\ldots}
\nonumber \\
&=1 + (\S_1+g_{01})\as + \(\frac{\S_1^2}{2} + \S_2 + \S_1\,g_{01} + g_{02}\)
 \as^2 + \Ord(\as^3).
\end{align}
where (note that, since $\lambda=\ab\ln\frac{1}{N}$,
it can be seen as an expansion in $\lambda$ with $\lambda/\ab$ fixed)
\begin{align}
\as\, \S_1 
&=\left[\frac{A_1}{\beta_0} \(\frac{\lambda}{\ab} + \ln\frac{Q^2}{\muf^2}
 - 2\gammae \)\right] \lambda\\
\as^2\, \S_2 &= \Bigg[ - \frac{A_1}{3\beta_0} \frac{\lambda}{\ab}
+ \( \frac{A_2}{2\beta_0^2} + \frac{A_1}{2\beta_0}
\(2\gammae - \ln\frac{Q^2}{\mur^2}\) \)  
              +\bigg\{-\frac{A_1}{\beta_0}(\gammae^2+\zeta_2)
-\frac{A_2}{\beta_0^2}\gammae+\frac{D_2}{4\beta_0^2} \nonumber\\
&\qquad\qquad
            +\frac{A_1}{\beta_0}\gammae\ln\frac{Q^2}{\mur^2} 
+\frac{A_2}{2\beta_0^2}\ln\frac{Q^2}{\muf^2}
            - \frac{A_1}{4\beta_0}\(\ln^2\frac{Q^2}{\mur^2}-
\ln^2\frac{\muf^2}{\mur^2}\)\bigg\} \frac{\ab}{\lambda} \Bigg] \lambda^2.
\end{align}

\eject

\end{document}